\documentclass[preprint,12pt]{elsarticle}

\usepackage[utf8]{inputenc}
\usepackage{setspace}
\usepackage[margin=1.25in]{geometry}
\usepackage{graphicx}
\usepackage{subfigure}
\graphicspath{ {./figures/} }
\usepackage{subcaption}
\usepackage{float}
\usepackage{amsmath}
\usepackage{amssymb}
\usepackage{physics}
\usepackage{lineno}
\usepackage{hyperref}

\journal{Computer Physics Communications}

\begin{document}

\begin{frontmatter}



\title{Verification of a hybrid gyrokinetic model using the advanced semi-Lagrange code ssV}


\author[maxplanck]{Sreenivasa chary Thatikonda\corref{cor1}} 
\ead{sreenivasa.thatikonda@ipp.mpg.de}
\cortext[cor1]{Corresponding author:}
\author[maxplanck,kuluven]{F. N. De Oliveira-Lopes}
\author[maxplanck]{A. Mustonen}
\author[maxplanck]{K. Pommois}
\author[maxplanck]{D. Told}
\author[maxplanck]{F. Jenko}
\affiliation[maxplanck]{organization={Max-Planck Institute for Plasma
Physics},
            addressline={Boltzmannstrasse 2}, 
            city={Garching},
            postcode={85748}, 
            state={Bavaria},
            country={Germany}}
\affiliation[kuluven]{organization={Centre for mathematical Plasma Astrophysics, KU Leuven},
            addressline={Celestijnenlaan 200B}, 
            city={Leuven},
            country={Belgium}}
\begin{abstract}
The super simple Vlasov (ssV) code was developed to study instabilities, turbulence, and reconnection in weakly magnetized plasmas, such as the solar wind in the dissipation range and the edge of fusion plasmas. The ssV code overcomes the limitations of standard gyrokinetic theory by using a hybrid model that incorporates fully kinetic ions and gyrokinetic electrons. This hybrid gyrokinetic model enables accurate modeling in regimes characterized by steep gradients and high-frequency dynamics. To achieve this, ssV implements a set of semi-Lagrangian numerical schemes, including Positive Flux Conservative (PFC), Flux Conservative fifth-order (FCV), FCV with Umeda limiters, and a Semi-Lagrangian Monotonicity-Preserving fifth-order scheme (SLMP5). Benchmark problems such as Landau damping, ion-acoustic waves, ion Bernstein waves, and kinetic Alfven waves were employed to evaluate the schemes. The SLMP5 scheme consistently delivered the best overall accuracy and numerical stability performance. The code also addresses well-known electromagnetic gyrokinetic simulation issues, such as the Ampere cancellation problem, using carefully chosen velocity-space resolutions and accurate integral evaluation.
\end{abstract}




\end{frontmatter}
\newpage

\section{Introduction}

The solar wind is a turbulent plasma composed primarily of ions and electrons interacting in a nearly collisionless environment~\cite{Bruno2013, Marsch2006, Chen2016}. Energy is injected at large scales as Alfvenic fluctuations and cascades toward smaller scales along the turbulent spectrum~\cite{Bruno2013,Chen2016,Cerri2016}. Understanding how this energy dissipates at sub-ion scales remains a major scientific challenge~\cite{Sahraoui2009,Howes2008}.

To tackle this problem, researchers have extensively studied processes such as plasma instabilities, magnetic reconnection, turbulent transport, and energy dissipation~\cite{Boldyrev2013,Yamada2010,Drake2003}. These studies span laboratory plasmas (e.g., tokamaks) and natural systems such as the solar wind, solar corona, and planetary magnetospheres~\cite{Schekochihin2009,Kivelson1995}. These environments are nearly collisionless and exhibit kinetic effects that govern their large-scale behavior~\cite{Gary1993,Matteini2020}.

At the sub-ion scale, electron dynamics become critical~\cite{TenBarge2013}. To address the associated complexities, precise computational models are used to identify fluctuations that drive energy transfer and uncover dissipation mechanisms~\cite{Howes2011}. Such models play a key role in understanding how microphysical processes shape macroscopic phenomena like solar wind propagation and planetary magnetospheres~\cite{Matteini2020}.

Computational plasma physics employs a hierarchy of models to balance the physical fidelity and computational cost~\cite{Birdsall1991,Hockney1988}. Fully kinetic models, such as continuum and particle-in-cell (PIC) codes, resolve ions and electrons in the 6D phase space and capture non-Maxwellian distributions and kinetic instabilities~\cite{Taitano2015,Markidis2010}. However, their immense computational demands confine the simulations to small spatial and temporal scales. In contrast, MHD models treat plasmas as flowing conductive fluids~\cite{Goedbloed2004}, allowing researchers to study large-scale behavior but ignoring the small-scale dynamics and fast time scales of particle motion. 

Hybrid models bridge the gap between detailed kinetic models and practical fluid approaches by coupling kinetic ions with fluid electrons~\cite{Kunz2014,Valentini2007}. Hybrid models enhance the computational efficiency of larger-scale simulations by compromising certain electron-scale physics~\cite{Told2015}. Gyrokinetic models simplify kinetics by averaging over ion/electron gyromotion, reducing the phase-space dimensionality for turbulence studies~\cite{Schekochihin2009,Groselj2017,Frieman1982}. However, this procedure eliminates waves at cyclotron frequencies (or higher), as well as processes such as cyclotron damping and stochastic heating. Therefore, existing approaches face critical gaps: fully kinetic models are computationally intractable for multi-scale systems, MHD lacks kinetic resolution, hybrid models neglect electron kinetic effects, and gyrokinetics fails at ion scales for weakly magnetized or unmagnetized plasmas~\cite{TenBarge2013}.

We present a hybrid gyrokinetic framework that couples fully kinetic ions with electrons modeled via a reduced gyrokinetic or drift-kinetic description to address these limitations~\cite{Lopes2022}. This approach retains ion kinetic fidelity while resolving essential electron dynamics (e.g., finite Larmor radius effects and wave-electron resonances) at a significantly lower computational cost than that of fully kinetic codes.

Previous work by Pommois~\cite{Karen2022} and Oliveira Lopes~\cite{Lopes2022} laid the foundation for this hybrid formulation. Pommois benchmarked both linear (FIDEL) and nonlinear (ssV) implementations against the gyrokinetic code GENE~\cite{Jenko2000}, demonstrating strong agreement with the theory. Oliveira Lopes extended this work by validating linear responses using FIDEL, including ion-acoustic and ion Bernstein waves.

We validate the model against standard benchmarks, including linear and nonlinear Landau damping setups, with results matching the theoretical predictions for damping rates~\cite{Yi2017,Filbet2001,Myers2017}. Verification through the recovery of ion-acoustic waves~\cite{Kunz2014,Valentini2007,Munoz2018}, ion Bernstein waves~\cite{Podesta2012,Schild2024}, and kinetic Alfven wave physics~\cite{Mandell2020,Pezzi2019,Dannert2004} confirm its ability to resolve multi-scale kinetic phenomena. Using high-order semi-Lagrangian schemes (PFC~\cite{Filbet2001}, FCV, FCV-Umeda~\cite{Umeda2009}, and SLMP5~\cite{Tanaka2017}), the code offers a practical tool to investigate unresolved questions in plasma astrophysics, such as instabilities, reconnection, and turbulent energy transfer in collisionless systems.

The following structure outlines the rest of the paper:  Section \ref{thoery}, describes the hybrid gyrokinetic
framework in detail, along with the necessary equations. Section \ref{num-implement} covers the numerical implementation
of the model in the ssV code, including the solver structure and the semi-Lagrangian schemes
used. Section \ref{num-tests} describes the theoretical setups used to assess the stability and accuracy of the
schemes and presents the results from standard test cases such as Landau damping.  Section \ref{num-verification} discusses the outcomes, showing how the simulations successfully capture the physics of ion-acoustic waves, ion Bernstein waves, and kinetic Alfven waves. Section \ref{conclusion} concludes the paper.

\section{Hybrid gyrokinetic model: Theory and equations}\label{thoery}
A consistent and physically grounded theoretical framework is essential for effectively investigating the kinetic processes and multi-scale interactions discussed in the previous section. In this study, we adopt a hybrid gyrokinetic model that captures the full kinetic physics of ions while retaining key electron dynamics through a reduced gyrokinetic or drift-kinetic approach. The following section, presents the theoretical foundation of the hybrid model. We begin with a Lagrangian formulation and a geometric framework that guides the phase-space reduction and sets the stage for deriving the key equations.
\subsection{Fully kinetic Vlasov equation for ions}
To derive the fully kinetic Vlasov equation within our framework, we use tools from symplectic geometry and Hamiltonian mechanics. This geometric approach not only provides a unified formalism for both fully kinetic and reduced models but also facilitates a clean transition to gyrokinetic coordinates. For readers less familiar with differential geometry and the symplectic structure of phase space, useful introductions can be found in Refs.~\cite{Morrison1998, Marsden1982, Littlejohn1983}. In brief, the method involves identifying a Lagrangian one-form that defines the dynamics in phase space, from which a non-canonical Poisson bracket and the corresponding Vlasov equation naturally follow. We begin with a tautological one-form~\cite{Wasserman2004} defined in a symplectic geometry framework, from which we identify the canonical symplectic form and perform the corresponding phase-space transformation. By writing the non-canonical Lagrangian one-form in terms of the new (transformed) coordinates and momenta, we can connect the canonical momentum and position variables to the physical velocity and the electromagnetic potentials. This approach reveals the symplectic structure underlying the fully kinetic ion dynamics and allows us to construct the corresponding Poisson tensor. With these pieces in place, the Poisson bracket arises naturally, capturing the Hamiltonian flow that governs the Vlasov dynamics. In the end, the fully kinetic description can be expressed through the following Poisson bracket:
\begin{equation}
\{f,g\}=\sum_{ij}\frac{\partial f}{\partial \Theta^{i}}\Pi^{ij}\frac{\partial g}{\partial \Theta^{j}}=\frac{1}{m}\left(\nabla f \cdot \frac{\partial g}{\partial v}-\frac{\partial f}{\partial v} \cdot \nabla g\right)+B\left(\frac{\partial f}{\partial v}\times\frac{\partial g}{\partial v}\right).
\end{equation}

Using Liouville’s theorem in this symplectic framework (noting that the phase-space distribution function is constant along Hamiltonian trajectories), we derive the Vlasov equation. After writing the time evolution of the distribution function in terms of the Hamiltonian flow and substituting the relations found above, the kinetic equation for the fully kinetic species takes the form:

\begin{equation}
\frac{\partial F_{}^{}}{\partial t}+v_{i}\frac{\partial F}{\partial x_{i}}+ \frac{q}{m}\left(\frac{\partial\phi}{\partial x_{i}}+\epsilon_{ijk}v_{j}\frac{B_{k}}{c}\right)\frac{\partial F}{\partial v_{i}}=0.\label{final_fully}
\end{equation}
with the understanding that here, implicitly, the repeated lower indices j,k are summed over informally.
It is worth noting that even though the result in Eq.\ref{final_fully} is not unexpected, the geometric approach provides a framework that is well-suited for a seamless gyrokinetic coordinate reduction. This is because we begin the transformation from the tautological one-form defined by the Lagrangian ~\cite{Lopes2022}, rather than starting from the final equations themselves.
\subsection{Gyro/Drift kinetic Vlasov equation for electrons}
For the electrons, we apply a gyrokinetic formulation by performing a coordinate transformation on the fully kinetic equations (as derived for the ions). This follows the standard approach introduced by Littlejohn ~\cite{Littlejohn1982,Littlejohn1983}  and others, and is designed to eliminate explicit dependence on the gyrophase angle $\theta$. By carrying out a sequence of carefully chosen transformations and perturbation expansions, we remove the complexities associated with the fast gyration motion and isolate the slower electron dynamics of interest. 

At first, this procedure mirrors the standard guiding-center transformation, preserving the general form of the original Lagrangian. We then introduce small corrections due to the background magnetic field in a systematic way, which ensures that the Lagrangian and its symplectic structure remain accurate to low order. In this way, the new gyrocenter coordinates capture the essential physics without straying far from the usual canonical structure found in the literature. Significant differences appear only at higher order in the expansion, reflecting the subtle effects of the background field and the removal of the gyroangle $\theta$. 

Once the gyrocenter coordinates and their Lagrangian are established, the Hamiltonian dynamics lead to a modified Poisson bracket. This new bracket is derived from the inverse of the transformed symplectic form and governs the evolution of gyrokinetic variables. Taking into account the modified gradient operators, effective fields, and the redefined magnetic field $B^*$, the gyrokinetic Poisson bracket is:

\begin{multline*}
\left\{ f,g\right\} _{gc}=\frac{e}{mc}\frac{1}{\epsilon_B}\left(\frac{\partial f}{\partial\theta}\frac{\partial g}{\partial\mu}-\frac{\partial f}{\partial\theta}\frac{\partial g}{\partial\mu}\right) \\
+\frac{B^{*}}{mB_{\parallel}^{*}}\cdot\left(\nabla^{*}f\frac{\partial g}{\partial V_{\parallel}}-\nabla^{*}g\frac{\partial f}{\partial V_{\parallel}}\right)-\epsilon_B\frac{c\hat{b}}{eB_{\parallel}^{*}}\cdot\left(\nabla^{*}f\times\nabla^{*}g\right) \label{gk_Poisson_bracket},
\end{multline*}

where the modified gradient is given by 

$\nabla^{*}=\nabla-\boldsymbol{R}^{*}\partial_{\theta_{}}$, $\boldsymbol{B}^{*}=\boldsymbol{B}+\frac{m}{e}cv_{\parallel}\nabla\times\hat{\boldsymbol{b}}-\frac{mc^{2}}{e^{2}}\mu\nabla\times \boldsymbol{R}^{*}$. 

The field perturbations are handled by applying a series of contact transformations, following the comprehensive gyrokinetic reduction used in this model~\cite{Lopes2022}. It is important to clarify some assumptions in our model. The perturbation theory requires a certain ordering: in our case, we use an expansion parameter defined by $(k_\perp \rho_{th}), e \phi_1 / T_i = \epsilon_\perp \epsilon_\delta$, which characterizes the strength of the perturbation (here in terms of the perpendicular wavenumber $k_\perp$ and the ion thermal gyroradius $\rho_{th}$). For a standard gyrokinetic derivation we set $\epsilon_\perp = 1$. We assume an ordering for the background magnetic field such that $\epsilon_B = \epsilon_\delta^2$. Similarly, the gradient of the perturbed magnetic potential is ordered as $\nabla A_1 / A_1 = \epsilon_\delta^2$. 

Applying the same procedure to derive the Vlasov equation for the gyrokinetic electrons, we obtain: 

\begin{equation*}
   \frac{\partial F}{\partial t}+\frac{\textbf{B}^{*}}{m{B}_{\parallel}^{*}}\boldsymbol{\cdot}\left(mv_{gy, \parallel}+\frac{e}{c}\left\langle {A}_{1\parallel}\right\rangle \right)\nabla_{gy}F \\
   \end{equation*}
   \begin{equation*}
   +\frac{c\hat{\textbf{b}}}{e{B}^{*}_{\parallel}}\times\left(\mu_{gy}\nabla_{gy}B(\textbf{X}_{gy})-\varepsilon_{\delta}e\nabla\left\langle \phi_{1}\right\rangle +\varepsilon_{\delta}\frac{e}{c}v_{gy, \parallel}\left\langle \nabla {A}_{1\parallel}\right\rangle \right) \boldsymbol{\cdot} \nabla_{gy}F \\
   \end{equation*}
   \begin{equation}
   -\frac{\textbf{B}^{*}}{m{B}_{\parallel}^{*}} \boldsymbol{\cdot} \left(\mu_{gy}\nabla_{gy}B(\textbf{X}_{gy})-\varepsilon_{\delta}e\nabla\left\langle \phi_{1}\right\rangle +\varepsilon_{\delta}\frac{e}{c}v_{gy, \parallel}\left\langle \nabla {A}_{1\parallel}\right\rangle \right)\partial_{v_{gy, \parallel}}F=0,
\end{equation}
In this equation, the second term (first line) represents the contribution from the canonical momentum, while the terms in the second line correspond to the $\nabla B$ drift, the $E \times B$ drift, and the magnetic flutter effects. The final term represents the total parallel acceleration.

\subsection{Poisson equation}
To derive the field equations of the model, we consider the system to reside on a heterogeneous manifold ~\cite{Wasserman2004,Fan2021}. This manifold is "heterogeneous" because it combines the phase space of the fully kinetic ions with the reduced phase space of the gyrokinetic electrons, in addition to incorporating the electromagnetic fields. In practice, the ions are described in the full phase space $\Omega = (x, v)$, while the electrons are described in a reduced gyrocenter phase space $\Omega_{gy} = (X_{gy}, v_{\parallel gy}, \mu_{gy}, \theta)$ (after averaging out the fast gyromotion). 

In this framework, the electromagnetic field acts as a bridge between the two descriptions (fully kinetic ions and gyrokinetic electrons) since it couples to both species. We assume the field Lagrangian is defined on the same manifold as the fully kinetic particles. This choice lets us couple the fields consistently to the ion dynamics and also include their influence on the gyrokinetic electrons via the proper coordinate transformations. 

Next, we derive the field equations using a variational principle. In particular, we consider a small variation of the action $S$ with respect to a field $\chi(\Omega)$, which leads to the stationary-action condition:

\begin{equation}
    \frac{\delta\mathcal{S}[\chi(\Omega)]}{\delta\chi(\Omega)}\circ\hat{\chi}(\Omega)=0.
\end{equation}
Here, $S$ is the action of the system, $\chi(\Omega)$ is the field being varied, and $\hat{\chi}(\Omega)$ is an arbitrary test variation. (For details on the functional derivative in the context of a heterogeneous manifold, see ~\cite{Wasserman2004,Fan2021}.) 

Because the ions and electrons live in different phase spaces and the fields are defined in physical space $x$, we must convert quantities defined in gyrokinetic coordinates back into physical coordinates to perform the variation in a single, common coordinate system. For example:
\begin{equation*}
\frac{\delta\phi_{1}(\mathbf{X}_{gy}+\rho)}{\delta\phi_{1}(\mathbf{x})}\circ\hat{\chi}(\mathbf{x})=\frac{\delta}{\delta\phi_{1}(\mathbf{x})}\left(\phi_{1}(\mathbf{X}_{gy}+\rho)\right)\circ\hat{\chi}(\mathbf{x}).
\end{equation*}

Here, $\phi_{1}(\mathbf{X}_{gy}+\rho)$ represents the scalar potential evaluated at the position $\mathbf{X}_{gy}+\rho$, where $\rho$ is the gyroradius vector, and we express its variation in terms of the physical coordinate $\mathbf{x}$.

Starting from a single Lagrangian that includes the gyrokinetic electrons, the fully kinetic ions, and the electromagnetic fields, we can derive an action integral for the entire system. Varying this action yields a Poisson equation for the scalar potential that intrinsically includes the electrons’ gyrokinetic effects. In the derivation, we integrate over both the electron and ion phase spaces to relate their densities to the electromagnetic potentials. After writing out the potentials explicitly and accounting for the parallel electron flow, the resulting field equation couples the scalar potential with the vector potential. The final form of this Poisson equation, which includes the thermal gyroradius and the Debye length and their interplay with the electron parallel flow and density, is:

\begin{equation}\label{eq:poissoneq}
\frac{1}{4\pi}\nabla_{\perp}^{2}\phi_{1}\left(4\pi\frac{\rho_{th}^{2}}{\lambda_{D}^{2}} - 1\right) + u_{e \parallel} \frac{\rho_{th}^{2}}{\lambda_{D}^{2}} \nabla_{\perp}^2 A_{1 \parallel}(\mathbf{x})  = \sum_i q_i n_i(\mathbf{x}) + e n_e(\mathbf{x}).
\end{equation}

This equation highlights how the potentials and charge densities interplay, and it shows how the gyrokinetic corrections modify the standard Poisson equation. The factor $\big(4\pi, \rho_{th}^2/\lambda_D^2 - 1\big)$ acts as an effective dielectric coefficient that alters the Laplacian of the scalar potential, and the coupling to the parallel vector potential $A_{1\parallel}$ is mediated by the electron parallel flow $u_{e\parallel}$.

\subsection{Parallel Ampere equation}

If we instead vary the action with respect to the vector potential (the magnetic potential), we obtain the homogeneous part of the electromagnetic field equations. Including all relevant terms in the Lagrangian and accounting for the electron dynamics as well as a long-wavelength (low-$k$) approximation, we derive a modified parallel Ampère’s law. This parallel Ampère equation includes effects such as density fluctuations, magnetic field perturbations, and electron gyro-averaging. The final form of the parallel Ampère equation is:

\begin{equation}\label{eq:Ampereeq}
\frac{c}{4\pi}\nabla_{\perp}^{2} A_{1\parallel}\left(1+\frac{\beta_e}{2}\right) + u_{e \parallel}\frac{\rho_e^2}{\lambda_D^2}\nabla_{\perp}^2\phi_1(\textbf{x})=\frac{e^2}{m_e c} n_e A_{1 \parallel}(\textbf{x})- I_e+\sum_i I_{i \parallel}
\end{equation}
In the above, $n_{i}=\int d\Omega F_{i}(x,v)$ is the ion number density, and we define $I_{i\parallel} = q_i n_i u_{i\parallel}$ for each ion species (so that $\sum_i q_i \int F_i(x,v) u_{i\parallel} d\Omega = \sum_i I_{i\parallel}$). Similarly, $n_e = \int d\Omega F_e(x)$ is the electron density, and $I_e = e u_{e\parallel} n_e$ is the electron parallel current. The first term on the right-hand side of Eq.\ref{eq:Ampereeq} comes from the velocity shift introduced by the gyrokinetic coordinate transform ~\cite{Lopes2022}; this term corresponds to the magnetization current. In particular, for electrons it shows that using a reduced (gyrokinetic) coordinate description eliminates the usual diamagnetic current ~\cite{FrankKamenetskii1972}. Finally, note that since we have restricted our attention to perturbations perpendicular to the background magnetic field, we defer the derivation of the perpendicular Ampere equation to future work.

\section{Numerical implementation into the code ssV}\label{num-implement}
Numerical implementation plays a critical role in accurately solving plasma model equations. Semi-Lagrangian schemes have been widely adopted in plasma simulation codes due to their stability and ability to handle large time steps without stringent CFL constraints~\cite{Filbet2001,Sonnendrucker1998}. 

In this section, we describe the numerical implementation of the hybrid gyrokinetic model within the super simple Vlasov (ssV) code, written in FORTRAN90, numerically implements the simulation model discussed in Section \ref{thoery}. This code has been designed to efficiently simulate weakly magnetized, collisionless plasmas using fully kinetic ions and drift-kinetic electrons. The code operates in slab geometry and applies periodic boundary conditions in the $X$ space. The base version of the code was derived from the MUPHY I code in collaboration with the Theoretical Physics Department (Ruhr University Bochum).
\subsection{Geometry of the system}

The super simple Vlasov (ssV) code uses a slab geometry for both the electrostatic and electromagnetic branches, where the magnetic field ($B$) is oriented in the $z$-direction, denoted as $\hat{b} = \hat{z}$. This implies that the parallel gradient is defined as $\nabla_{\parallel} = \partial / \partial z $, the perpendicular gradient is defined as $\nabla_{\perp} = \left( \partial / \partial x, \partial / \partial y \right)$, and the parallel velocity is $v_{\parallel} = v_z$.
\subsection{Normalized equations}
While some codes implement equations without any normalization and are therefore not strictly necessary, they are commonly used to help guide physics intuition. In this section, we outline the rules we follow for normalizing the ssV equations. The quantity $L$ is normalized by defining a reference quantity $L_R$ and dimensionless quantity $\hat{L}$ such that $L = L_R \hat{L}$. The reference quantities ($L_R$) are functions of reference mass, charge, temperature, magnetic field, and particle density. The Time and length were normalized using a reference cyclotron frequency ($\Omega_{cR}$) and radius ($\rho_R$) as $t = \Omega_{cR}^{-1} \hat{t}$ and $d = \rho_R \hat{d}$, where $\Omega_{cR} = q_R B_R / m_R c $ and $\rho_R = m_R c_R / q_R B_R $. where, $c_R$ is the reference thermal velocity, which is calculated as $c_R = \sqrt{T_R / m_R} $. The velocity was normalized using the thermal velocity of each species $\sigma$ as $v = c_R \hat{v} \hat{v}_{T\sigma}$, where $\hat v_{T\sigma}  = \sqrt{ 2 T_\sigma / m_\sigma} $ is the species $\sigma$ thermal velocity.
\subsection{Parallelization and boundary exchange}

The super simple Vlasov (ssV) code employs a parallelization strategy that combines domain decomposition with ghost-cell communication to efficiently solve high-dimensional Vlasov-Maxwell systems. The simulation domain was partitioned into subdomains, each processed independently on separate processor cores. Data synchronization was achieved through optimized boundary exchanges to ensure scalable performance while maintaining accuracy.

\begin{figure}[h]
    \centering
    \subfigure[Illustration of sub domains and ghost cells]{\label{fig:subgrid}
    \includegraphics[width=0.48\textwidth]{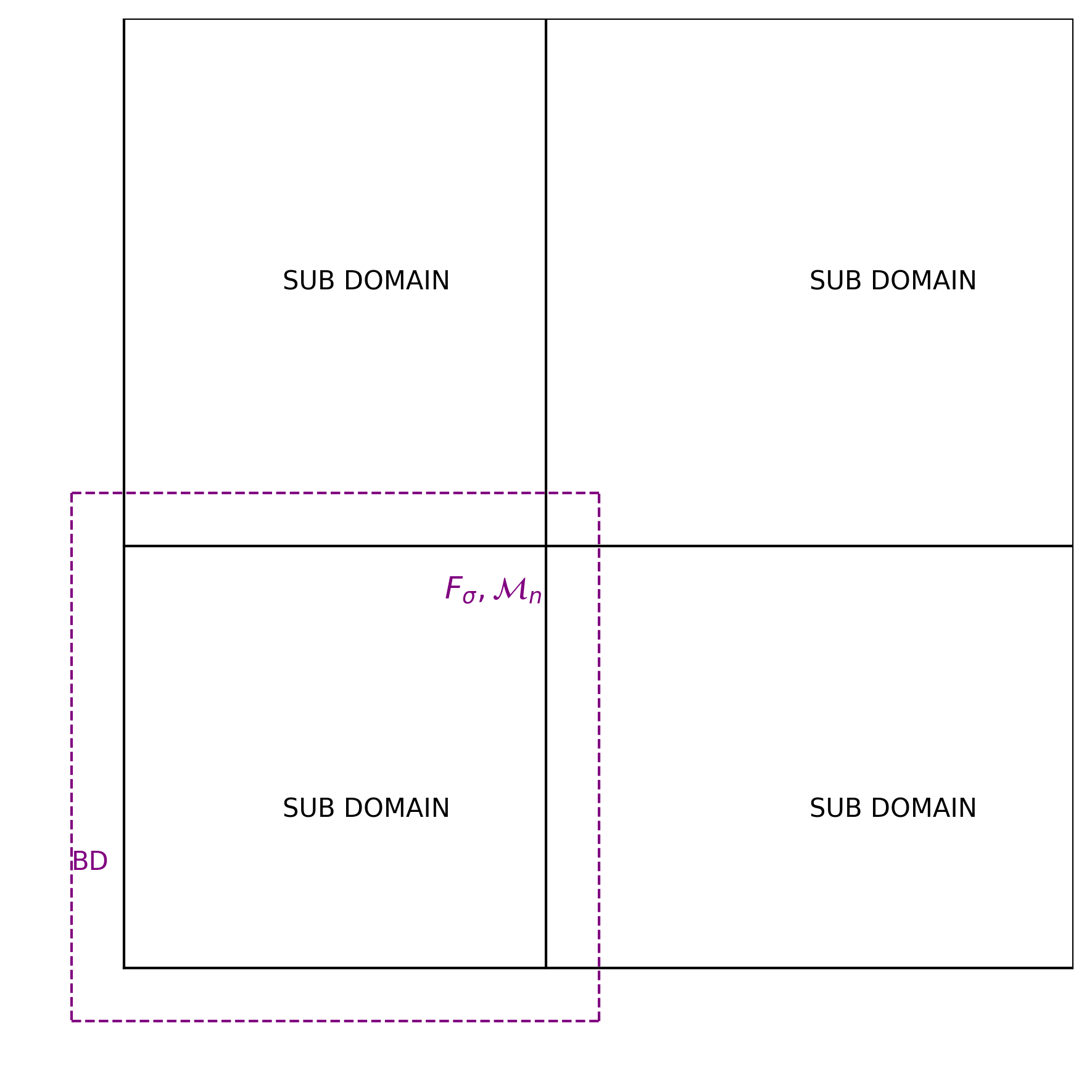} 
    }
    \subfigure[Illustration of kinetic boundary exchange]{\label{fig:kineticboundary}
    \includegraphics[width=0.48\textwidth]{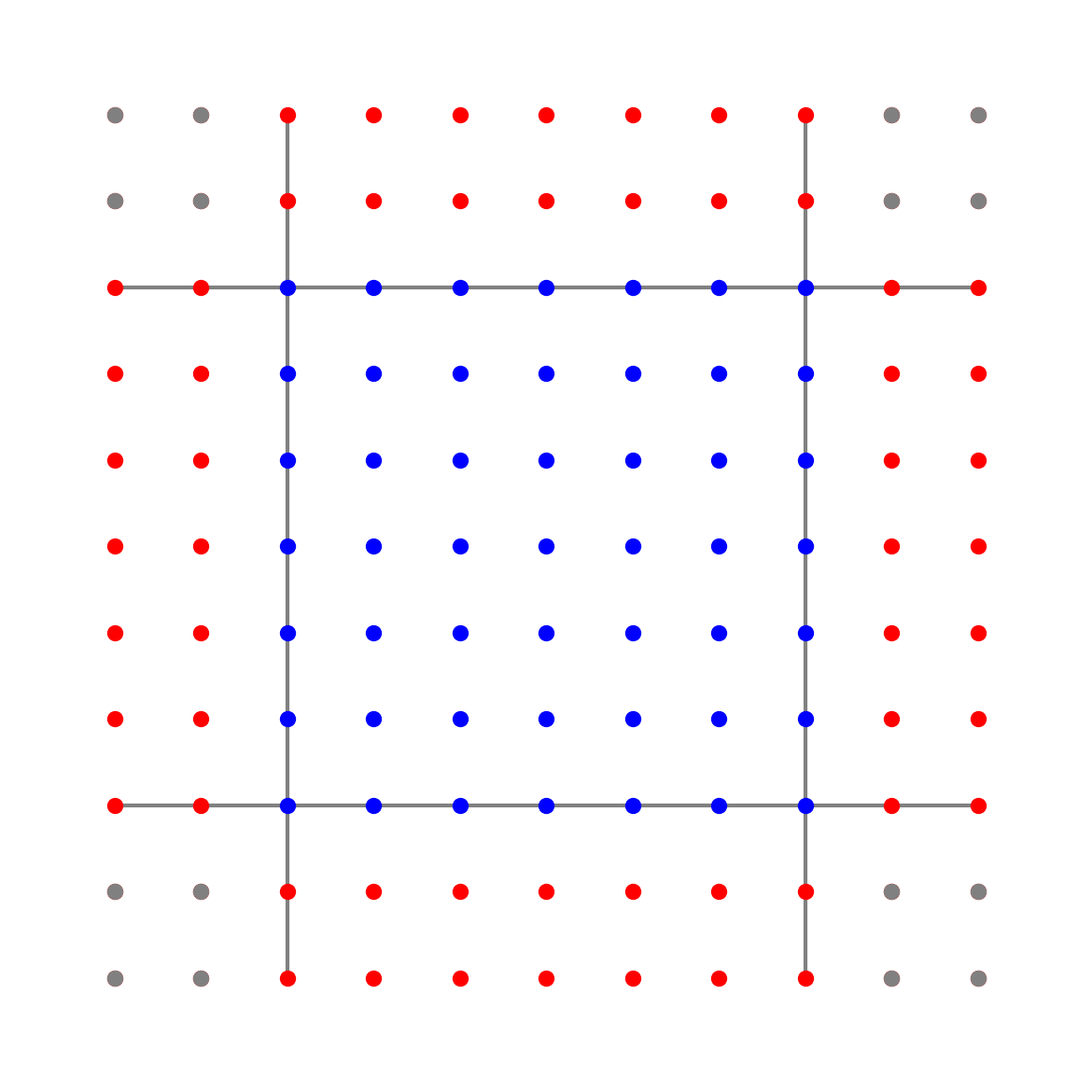}} 
    \caption{Examples of the simulation domain in ssV. a) The solid black lines divide the domain into subdomains, while the dashed purple lines represent subdomains with BD additional ghost cells. For distribution functions and their moments, BD ghost cells (dashed purple lines) are added, and the electromagnetic fields are computed in Fourier space. b) illustrates the kinetic boundary exchange mechanism where the gray corner cells are excluded and only red cells are exchanged, ensuring consistency across domain partitions.}
    \label{fig:subdomain}
\end{figure}

As shown in Figure \ref{fig:subgrid}, the ssV code accelerates plasma simulations by decomposing the computational domain into subdomains assigned to individual processor cores. This parallel execution uses modern multicore systems to achieve the necessary phase space resolution while maintaining affordable run times. A scheme-dependent number of ghost cells (labeled BD in Figure \ref{fig:subgrid}) containing distribution function values from neighboring subdomains is maintained in each process to enable local computation and ensure consistency of the numerical schemes.

The ssV code employs localized numerical schemes, where the cell updates depend only on the immediate neighbors. This minimizes inter-core communication and enhances scalability, which is critical for high-dimensional systems. Partial differential equations (PDEs) are solved using the Domain Decomposition Method (DDM)~\cite{Quarteroni1999}, where the subdomains independently compute local solutions.

To maintain consistency in the charge and current densities before solving the coupled Poisson and Ampere equations, the ssV code uses a kinetic boundary exchange protocol (Figure \ref{fig:kineticboundary}). This exchange method ensures accurate synchronization of the boundary data without the need for additional corner-based exchanges.

Unlike the other components of the Vlasov-Maxwell system, the electromagnetic field solver in ssV (which presently assumes periodic boundary conditions) does not utilize domain decomposition. Instead, it is computed entirely in Fourier space, avoiding the need for inter-core communication during field updates. This approach enhances the numerical stability and eliminates the parallelization overhead for the electromagnetic field solver, thereby ensuring efficiency in large-scale simulations.

\subsection{Splitting algorithm and advection forms}
\begin{figure}[h]
    \centering
    \includegraphics[width=0.8\textwidth]{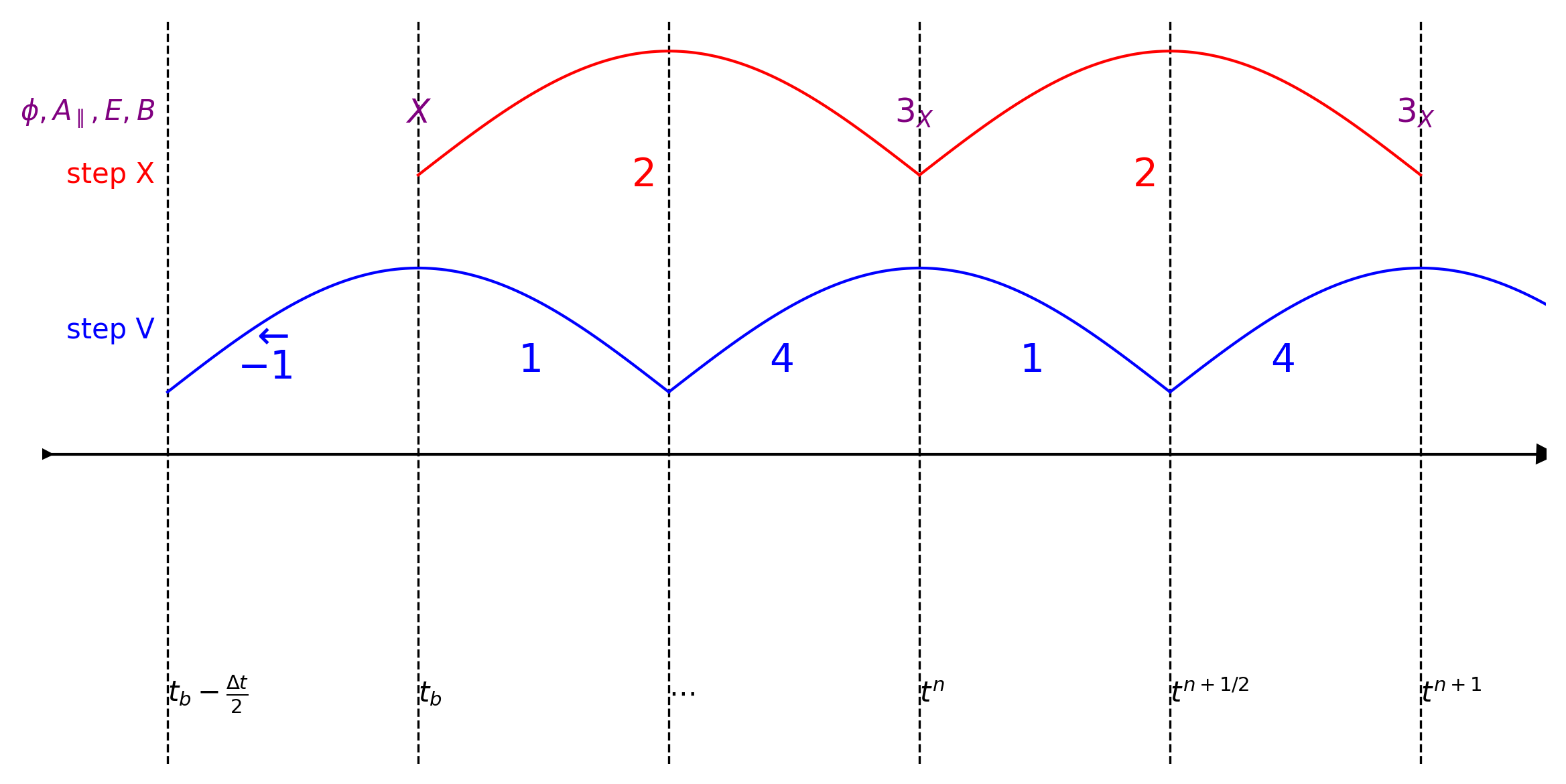} 
    \caption{illustrates the splitting algorithm. Before starting the time cycle, the distribution function was shifted back in time by half a time step in the velocity space (step -1). The time evolution proceeds in four steps: the first and fourth steps evolve the function in velocity space ($V$), the second step evolves it in the position space ($X$), and the third step computes the fields.
}
    \label{fig:stepxv}
\end{figure}

The super simple Vlasov (ssV) code employs a splitting algorithm to decompose the high-dimensional Vlasov equation into manageable components. This method separates the evolution of the distribution function \( F_\sigma \) into sequential steps in position space (\( X \)-space) and velocity space (\( V \)-space). First, the \( X \)-space equation is as follows:
\begin{equation}
    \frac{\partial F_\sigma}{\partial t} + \dot{\mathbf{X}} \cdot \nabla_X F_\sigma = 0
\end{equation}
and the \( V \)-space advection
\begin{equation}
    \frac{\partial F_\sigma}{\partial t} + \dot{\mathbf{V}} \cdot \nabla_V F_\sigma = 0
\end{equation}
separately. The advection operations in the velocity space for the ions do not commute. To minimize error and reduce computational effort, the ssV code applies a cascade interpolation method combined with back-substitution~\cite{Schmitz2006a,Rieke2015,Trost2016}. The discretization followed the Boris push method~\cite{Boris1970}. The ssV code employs the Strang splitting algorithm illustrated in Figure  \ref{fig:stepxv} to preserve second-order temporal accuracy. This approach evolves \( V \)-space for \( \Delta t/2 \), \( X \)-space for \( \Delta t \), and \( V \)-space again for \( \Delta t/2 \).
The split equations are reduced to one-dimensional advection problems of the form
\begin{equation}
    \frac{\partial F_\sigma}{\partial t} + U(R, t) \frac{\partial F_\sigma}{\partial R} = 0,
\end{equation}
where \( U(R, t) \) represents species- and direction-dependent advection velocities. For kinetic ions, \( X \)-space advection uses \( U_K = v_i \), whereas \( V \)-space acceleration follows
\begin{equation}
    U_K = \frac{q_i}{m_i} \left( \mathbf{E} + \frac{\mathbf{v} \times \mathbf{B}_0}{c} \right) \nabla_v.
\end{equation}
For drift-kinetic electrons, \(X\)-space advection is implemented in a standard manner, where non-commutative behavior in the perpendicular direction is inherent. This is justified because the electron $E\times B$ drift evolves on timescales much slower than the ion gyrofrequency, allowing the semi-Lagrangian method to capture the dynamics accurately when a sufficiently small time step is used. However, if the time step is not adequately constrained, numerical errors can accumulate over time.
\(X\)-space advection uses
\begin{equation}
    U_{DK} =v_\parallel +\frac{q_e}{c m_e} A_\parallel +\frac{c}{B_0} \hat{\mathbf{b}} \times\left( \mathbf{E}_\perp  + \frac{1}{c}v_\parallel \nabla A_\parallel\right) 
\end{equation}
and in \( V \)-space acceleration follows
\begin{equation}
    U_{DK} = \frac{q_e}{m_e}\left( E_\parallel -\frac{1}{c} v_\parallel \frac{\partial A_\parallel}{\partial z}\right)\nabla_{v_\parallel}
\end{equation}
  
\subsection{Semi-Lagrangian method for electrons}

The semi-Lagrangian method integrates Eulerian grid structures with Lagrangian trajectory calculations to solve the DK electron advection equations in the ssV code~\cite{Staniforth1991,Sonnendrucker1998,Crouseilles2010}. This method is specifically applied to solving the electron equations and is formulated as:

\begin{equation}
    \frac{\partial F_\sigma}{\partial t} + U(R,t)\frac{\partial F_\sigma}{\partial R} = 0.
\end{equation}

This approach combines a hybrid grid system, characteristic trajectory tracing, and adaptive interpolation to achieve accurate solutions without stringent Courant-Friedrichs-Lewy (CFL) constraints. This method employs a hybrid grid system that merges fixed Eulerian grids with Lagrangian particle tracking. Characteristic tracing determines backward trajectories by using an iterative Newton method to ensure precision in the advection process. To minimize the numerical diffusion, the scheme interpolates the distribution function using third- or fifth-order Lagrange polynomials. The characteristic trajectory is calculated as follows. First, the grid points were initialized at time $t^{n+1}$. Backward trajectories are then traced to locate their origins at $t^n$. Displacement $\alpha$ is computed as follows:

\begin{equation}
    F(\alpha) = \alpha - \Delta t U(X^n) = 0.
\end{equation}

An iterative Newton method refines the solution:

\begin{equation}
    \alpha^k = \alpha^{k-1} - \frac{F(\alpha^{k-1})}{F'(\alpha^{k-1})}.
\end{equation}

The cell-averaged velocity $U$ and centered finite differences contribute to the numerical stability. The interpolation scheme updates the distribution function as:

\begin{equation}
    f_i^{n+1} = f^n(x - \alpha).
\end{equation}

The interpolation order ditermines the number of ghost cells required. A memory-optimized implementation ensures efficient computation using a single time-step update for the electrons. This approach enhances computational efficiency while preserving second-order accuracy in electron dynamics.
\subsection{Positive flux conservative (PFC) scheme}

The Positive flux conservative (PFC) scheme~\cite{Filbet2001} enables divergence-free ion advection in the ssV code through flux-controlled trajectory integration. This method preserves the integral of the distribution function while preventing non-physical negative values. The PFC scheme achieves these properties through three core mechanisms: cell-averaged conservation, flux limitation enabled by slope limiters, and the accurate reconstruction of interface values.

Cell-averaged conservation ensures the integrity of \( f_\sigma \) by maintaining volume integrals:

\begin{equation}
    f^n_i = \frac{1}{\Delta x}\int_{x_i}^{x_{i+1}} f(x,t^n)dx.
\end{equation}

Limited flux computation enforces positivity using adaptive flux terms:

\begin{equation}
    \Phi_i = -\mathrm{sgn}(U)\alpha_i\left[f_j + \frac{\varepsilon_j^+}{6}(1-\alpha_i)(2-\alpha_i)(f_{j-\mathrm{sgn}(U)} - f_j) + \frac{\varepsilon_j^-}{6}(1-\alpha_i)(1+\alpha_i)(f_j - f_{j+\mathrm{sgn}(U)})\right].
\end{equation}

Slope limiters control interpolation gradients to prevent excessive oscillations:

\begin{align*}
    \varepsilon_j^+ &= \min\left(1, \frac{2f_j}{f_{j-\mathrm{sgn}(U)} - f_j}\right) \quad \text{if } (f_{j-\mathrm{sgn}(U)} - f_j) > 0, \\
    \varepsilon_j^- &= \min\left(1, \frac{2f_j}{f_{j+\mathrm{sgn}(U)} - f_j}\right) \quad \text{if } (f_{j+\mathrm{sgn}(U)} - f_j) > 0.
\end{align*}

The ssV code applies ion advection splitting to separate phase-space evolution into X-space and V-space operations for all implemented schemes, including PFC. In X-space, direct displacement calculations are performed using:

\begin{equation}
    \alpha = \frac{v\Delta t}{\Delta x}.
\end{equation}

For V-space advection, in which operations do not commute, the ssV code utilizes a cascade interpolation method combined with back-substitution in time. The discretization follows the Boris push method, ensuring accurate velocity-space evolution. The PFC scheme which uses third-order polynomial interpolation maintains accuracy during cascade back-substitution for non-commuting V-space operations.

The PFC method ensures the conservation of the average value. By preventing numerical oscillations, it remains well-suited for applications requiring positivity preservation. The method keeps the global error controlled when the distribution function has limited total variation. It offers both high-order accuracy and strong numerical stability, making it suitable for flux-conservative plasma simulations.
\subsection{Flux-conservative fifth-order (FCV) scheme}

The flux-conservative fifth-order (FCV) scheme extends the principles of the positive flux conservative (PFC) method by incorporating fifth-order polynomial interpolation while omitting the positivity constraints. This approach enhances accuracy but may introduce non-physical values in regions with steep gradients. The scheme preserves the integral of the distribution function as follows:

\begin{equation}
    f^n_i = \frac{1}{\Delta x}\int_{x_i}^{x_{i+1}} f(x,t^n)dx.
\end{equation}

To achieve a higher accuracy, the scheme employs enhanced interpolation using fifth-order Lagrangian polynomials based on six grid points:

\begin{equation}
    P(x) = \sum_{j=0}^5 f_j \prod_{\substack{k=0\\k\neq j}}^5 \frac{x - x_k}{x_j - x_k}.
\end{equation}

Flux calculation is performed without limiters, relying on a fifth-order expansion:

\begin{equation}
    \Phi_i = -\mathrm{sgn}(U)\, \alpha_i \sum_{m=0}^{5} c_m f_{j + \delta m},
\end{equation}
where:

    \( \alpha_i = |v \Delta t / \Delta x| \) is the normalized displacement.
     \( j \) is the index of the cell containing the foot of the characteristic traced from \( x_{i+1/2} \) at time \( t^{n+1} \).
    \( \delta = -1 \) for \( U > 0 \) (left-biased stencil), and \( \delta = +1 \) for \( U < 0 \) (right-biased stencil).
    \( f_{j + \delta m} \) are the distribution function values at the stencil points.
     \( c_m \) are the Lagrange interpolation coefficients corresponding to the fifth-order stencil.

The FCV scheme requires two additional cells per dimension compared with the PFC scheme to accommodate the extended interpolation stencil.

The FCV scheme differs from PFC in three critical aspects. First, it enhances the accuracy by reducing numerical diffusion, particularly for smooth solutions, owing to its fifth-order polynomial terms. Second, the absence of positivity enforcement allows the distribution function \( f_\sigma \) to take non-physical values in regions with sharp gradients. To address this, external limiters such as the Umeda limiter~\cite{Umeda2009} can be implemented to restore positivity. Finally, the computational cost is notably higher than that of the PFC scheme. This method provides a high-fidelity approach to flux conservation, particularly for smooth distributions; however, it requires additional corrections in cases where positivity preservation is essential.

\subsection{Semi-Lagrangian monotonicity- and positivity-preserving (SLMPP) schemes}

The SLMPP framework~\cite{Tanaka2017} enables high-order Vlasov simulations through constrained semi-Lagrangian advection. This method solves the advection equation as follows:

\begin{equation}
    \frac{\partial f}{\partial t} + c\frac{\partial f}{\partial x} = 0.
\end{equation}
We define the cell-averaged value of $f(x, t)$ over cell $i$ as:
\begin{equation}
    f_i^n = \frac{1}{\Delta x} \int_{x_{i-1/2}}^{x_{i+1/2}} f(x, t^n) \, dx
\end{equation}
The updated value using the Semi-Lagrangian method is:
\begin{equation}
    f_i^{n+1} = \frac{1}{\Delta x} \int_{x_{i-1/2}}^{x_{i+1/2}} f(x - c \Delta t, t^n) \, dx
\end{equation}

This can be expressed using numerical fluxes:
\begin{equation}
    f_i^{n+1} = f_i^n - \left( \Phi_{i+1/2} - \Phi_{i-1/2} \right)
\end{equation}
where
\begin{equation}
    \Phi_{i+1/2} = \frac{1}{\Delta x} \int_{x_{i+1/2} - c \Delta t}^{x_{i+1/2}} f(x, t^n) \, dx
\end{equation}
we compute the fifth-order interpolated flux using the stencil:

\begin{equation}
\Phi_{i+1/2}^{\text{int}} = \frac{1}{60} \left(
2f_{i-3} - 13f_{i-2} + 47f_{i-1} + 27f_i - 3f_{i+1}
\right)
\end{equation}

Then, we apply the MP limiter to ensure monotonicity of the flux:

\begin{equation}
\Phi_{i+1/2}^{\text{MP}} = \text{MP}\left( \Phi_{i+1/2}^{\text{int}}, f_{i-3}, f_{i-2}, f_{i-1}, f_i, f_{i+1} \right)
\end{equation}

The limiter ensures that the final flux satisfies:

\begin{equation}
\min(f_i, f_{i+1}) \leq \Phi_{i+1/2}^{\text{MP}} \leq \max(f_i, f_{i+1})
\end{equation}

The MP operator ensures that the resulting flux lies within a monotonic range defined by its neighboring stencil values. Specifically, it enforces the following:

\begin{equation}
    \min(f_i, f_{i+1}) \leq \Phi_{i+1/2}^{\text{MP}} \leq \max(f_i, f_{i+1})
\end{equation}

This guarantees that interpolation does not introduce a new extrema into the solution, preserving monotonicity in the transported quantity.

In practice, we apply:
\begin{equation}
    f_{i+1/2}^{\text{MP}} = \max\left( \min(f_i, f_{i+1}), \min\left( \max(f_i, f_{i+1}), f_{i+1/2}^{\text{int},5} \right) \right)
\end{equation}

Optionally, we can define a tighter interval using the parameter $\alpha$ (typically $\alpha = 4$) to restrict the range:
\begin{align}
    f_{\min}^{\alpha} &= f_i - \alpha (f_{i+1} - f_i) \\
    f_{\max}^{\alpha} &= f_i + \alpha (f_{i+1} - f_i) \\
    f_{i+1/2}^{\text{MP}} &= \max \left( f_{\min}^{\alpha}, \min\left( f_{\max}^{\alpha}, f_{i+1/2}^{\text{int},5} \right) \right)
\end{align}

This formulation preserves monotonicity while retaining high-order accuracy in smooth regions.
To ensure the non-negativity of $f$, we define a convex combination between the MP value and first-order upwind value:
\begin{equation}
    \hat{f}_{i+1/2} = \theta_{i+1/2} f_{i+1/2}^{\text{MP}} + (1 - \theta_{i+1/2}) f_i
\end{equation}
with
\begin{equation}
    \theta_{i+1/2} = \min\left(1, \frac{f_i^n}{f_i^n - \hat{U}_{i+1/2}} \right)
\end{equation}
ensuring positivity of the update.

The updated solution is then:
\begin{equation}
    f_i^{n+1} = f_i^n - \nu \left( \hat{f}_{i+1/2} - \hat{f}_{i-1/2} \right), \quad \nu = \frac{c \Delta t}{\Delta x}
\end{equation}
\begin{equation}
    f_i^{n+1} = f_i^n - \left( \hat{\Phi}_{i+1/2} - \hat{\Phi}_{i-1/2} \right)
\end{equation}
where
\begin{equation}
    \hat{\Phi}_{i\pm1/2} = \text{PP}\left( \Phi_{i\pm1/2}^{\text{MP}}, f_i, f_{i\pm1} \right)
\end{equation}
The SLMPP scheme incorporates a splitting approach to solve multi-dimensional problems using directional operator splitting. For stability and accuracy, a Courant-Friedrichs-Lewy (CFL) condition of \(\nu < 1/2\) was used for positivity preservation. Despite its high accuracy, its computational cost remains comparable to that of the PFC scheme~\cite{Tanaka2017}, making it an efficient choice for high-order simulations.

The SLMPP5 framework achieves a fifth-order accuracy, significantly reducing the numerical diffusion compared to the PFC scheme. It maintains physically consistent distribution function bounds even under sharp gradients and is effective for modeling collisionless astrophysical plasmas. The constrained interpolation method facilitates multi-scale modeling of phase mixing and collisionless damping while preventing spurious oscillations. The ssV code specifically implements the SLMP5 (semi-Lagrangian monotonicity-preserving scheme), a fifth-order numerical method optimized for high-fidelity plasma simulations-while omitting the positivity-preserving limiter.

\section{Numerical tests (validation results)}\label{num-tests}
To validate the hybrid gyrokinetic model and its numerical implementation in the ssV code, we conducted a series of benchmark tests using standard plasma physics problems. These include comparisons to known theoretical predictions and previously published results. In this section, we present the outcomes of these tests, highlighting the trade-offs between numerical diffusion and oscillations, and evaluating the ability of each scheme to accurately reproduce fundamental plasma phenomena.
\subsection{Numerical oscillation and diffusion trade-offs}
\begin{figure}[H]
\centering
\subfigure[Linear advection for a step function]{\label{fig:step_function}
\includegraphics[width=0.47\textwidth]{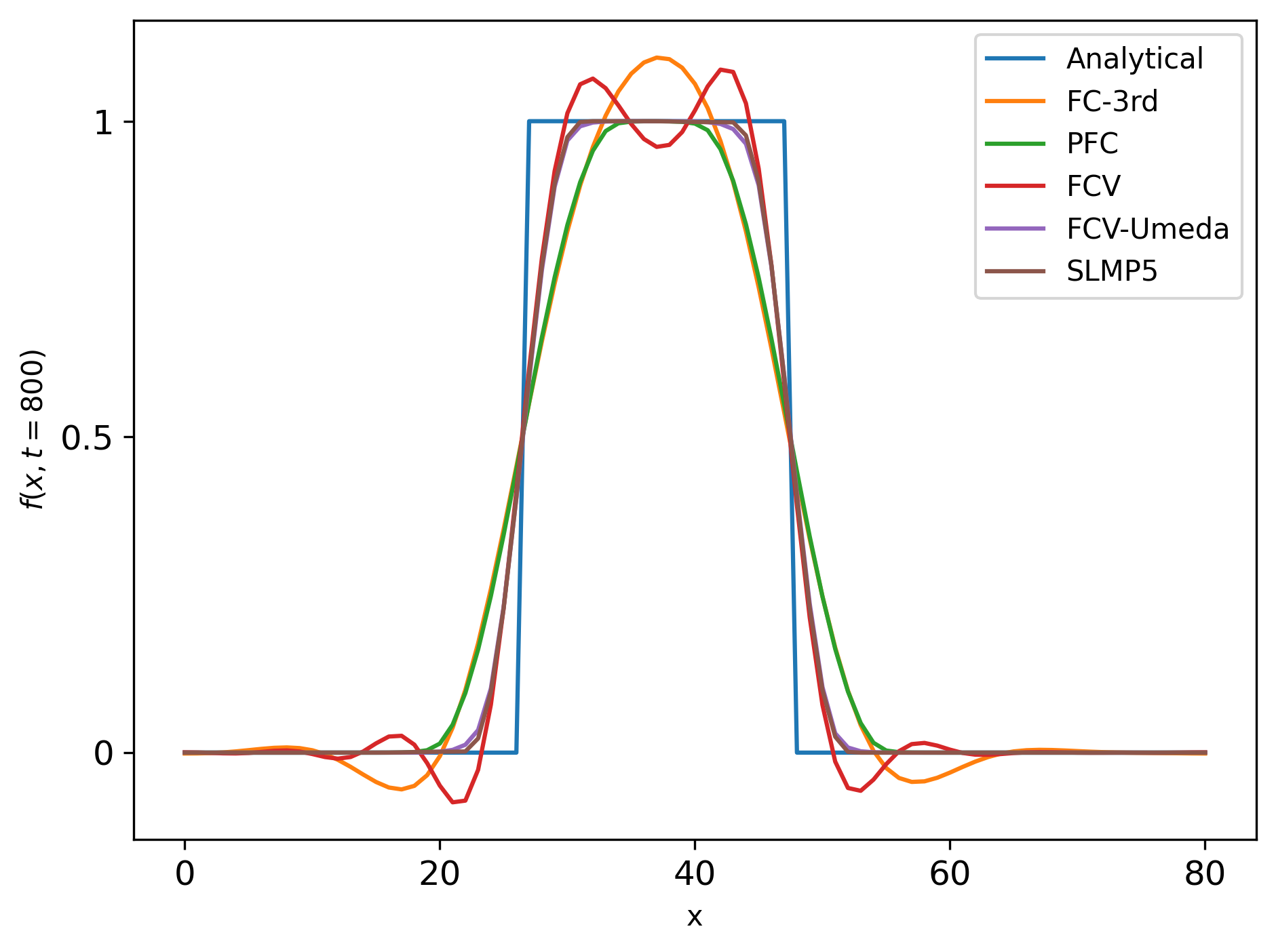}
}
\subfigure[Linear advection for two step function]{\label{fig:2step_function}
\includegraphics[width=0.47\textwidth]{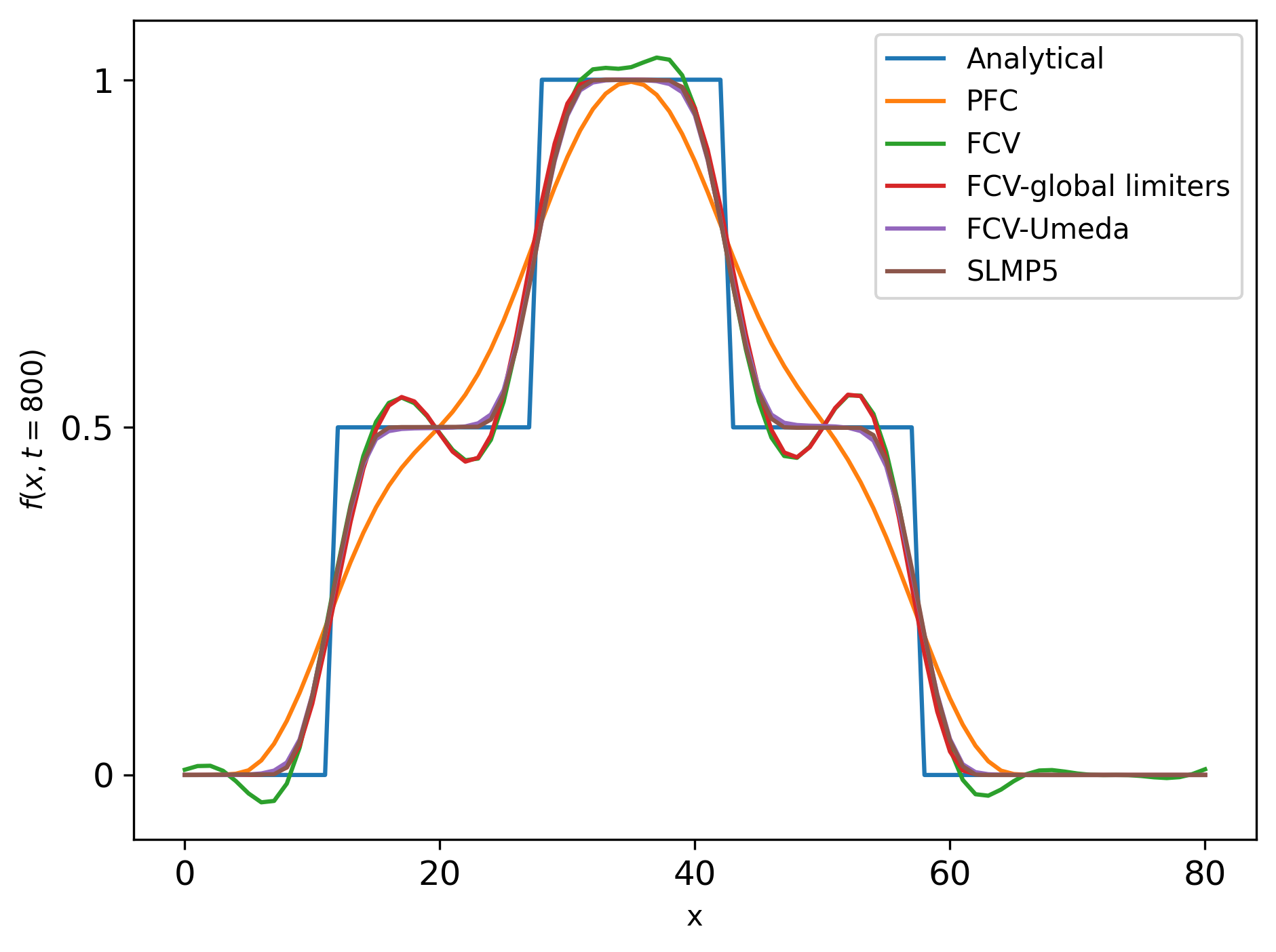}
}

\subfigure[Linear advection for sinusoidal function]{\label{fig:sin_function}
\includegraphics[width=0.48\textwidth]{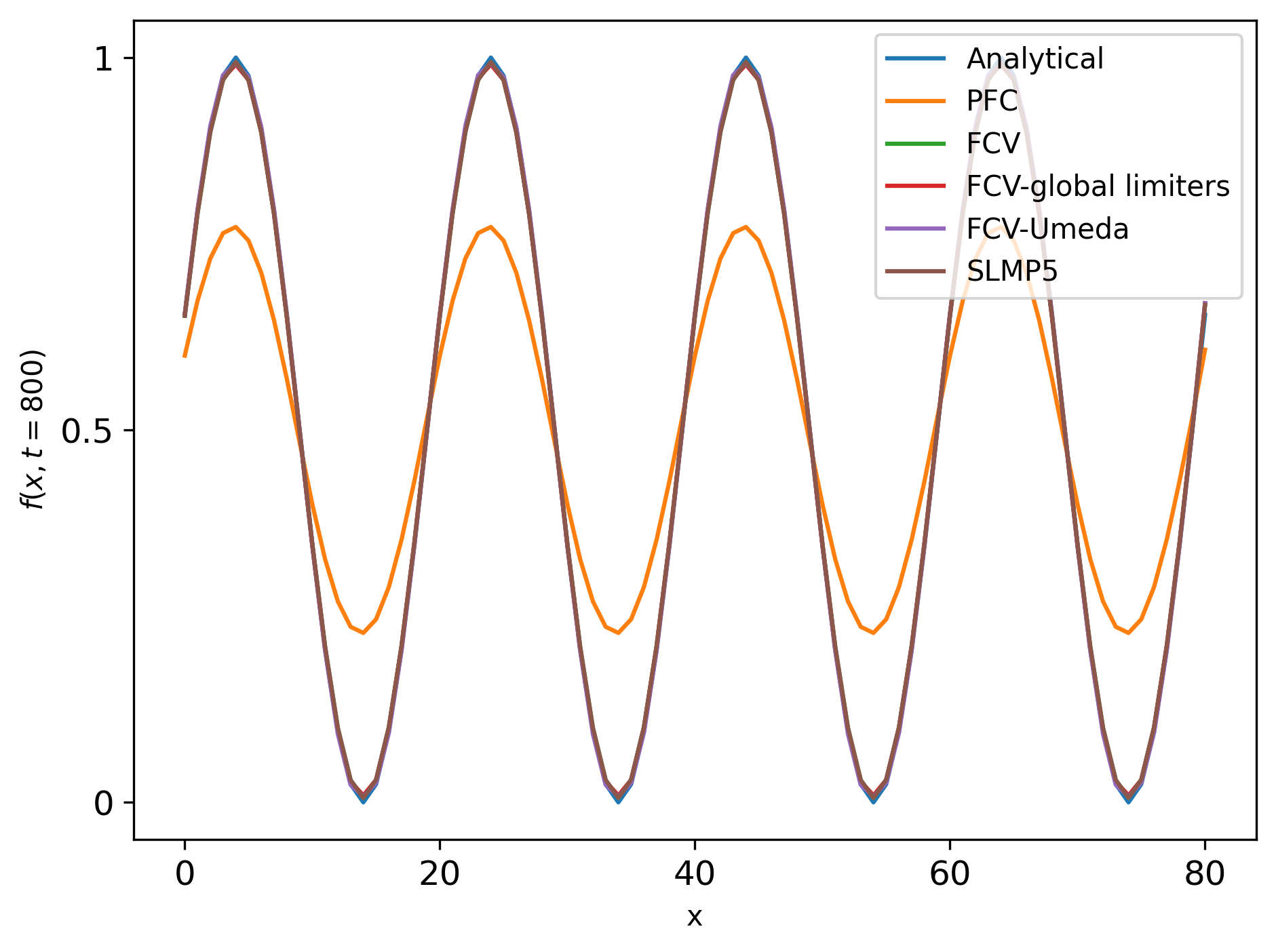}
}
\caption{The 1D linear advection equation is solved with the Flux conservative 3rd order scheme  (FC-3rd), positive flux conservative scheme (PFC), Flux conservative 5th order scheme (FCV), Flux conservative 5th order scheme with Umeda limiters (FCV-Umeda), Flux conservative 5th order scheme with global limiters (FCV-global limiters), semi-Lagrange monotonicity preserving 5th order scheme  (SLMP5) for different initial profiles.}
\label{fig:linear-advection-comparison_plots}
\end{figure}

Godunov's theorem~\cite{Godunov1959} establishes fundamental constraints for advection schemes, mandating oscillations in linear schemes above first-order accuracy~\cite{LeVeque2002}. To analyze the performance of different numerical schemes, we considered various test distributions, including step functions, double step functions, and sinusoidal profiles~\cite{Filbet2001,Umeda2009,Tanaka2017}, as shown in Figure~\ref{fig:linear-advection-comparison_plots}. These test cases allow for a systematic evaluation of numerical oscillations and diffusion across different methods.

Here is the Linear advection setup(for ions in X space):
\begin{equation}
\frac{\partial f_i}{\partial t} + v_x \frac{\partial f_i}{\partial x} = 0
\end{equation}
Semi-Lagrangian methods control these effects through constrained interpolation and limiters, with performance varying across schemes. Using a general third-order flux-conservative scheme, the step functions generate spurious minima and maxima, as shown in Figure~\ref{fig:step_function}. To address this, the positive flux conservative (PFC) scheme ensures positivity and reproduces the step function with some smoothing at the top corners owing to its inherent diffusion, as shown in Figure~\ref{fig:step_function}. However, when applied to sinusoidal profiles, the PFC scheme exhibits significant diffusion, as illustrated in Figure~\ref{fig:sin_function}.

To accurately recover sinusoidal profiles, the fifth-order flux conservative (FCV) scheme was used, showing excellent performance in preserving their shape and accuracy as shown in Figure~\ref{fig:step_function}. However, when applied to step functions, the FCV scheme introduces oscillations owing to its reduced dissipation, as shown in Figure~\ref{fig:step_function}. To control these oscillations, limiters were introduced, improving performance on step functions and yielding better results compared to PFC (Figure~\ref{fig:step_function}). Yet, when tested with two step functions, the limiters only suppressed global extrema, leaving oscillations at local extrema, as seen in Figure~\ref{fig:2step_function}.

The introduction of Umeda limiters to the FCV scheme yielded improved results for all the profiles. The step functions, sinusoidal profiles, and two step functions were accurately resolved with controlled oscillations, as shown in Figures~\ref{fig:linear-advection-comparison_plots}. Finally, the semi-Lagrangian monotonicity preserving scheme (SLMP5) outperformed both the PFC and FCV schemes with Umeda limiters. It delivered superior results for the step functions, sinusoidal profiles, and two step functions, as shown in Figure \ref{fig:linear-advection-comparison_plots}.
These findings highlight the effectiveness of advanced semi-Lagrangian schemes, particularly SLMP5, in addressing the limitations of traditional flux-conservative methods while achieving high accuracy across the diverse profile types discussed in Figure~\ref{fig:linear-advection-comparison_plots}.

\begin{table}[h]
    \centering
    \caption{Scheme Comparison for Advection Artifacts}
    \begin{tabular}{lccc}
        \hline
        Property & PFC & FCV & SLMP5 \\
        \hline
        Order & 3rd & 5th & 5th\\
        Diffusion & High & Medium & Low \\
        
        Oscillations & Low & High (without limiter) & Controlled \\
        
        Limiters & Positivity & External limiters (Umeda) & MP \\
        Stability & Unconditional& CFL-dependent & CFL-constrained \\
        & (Theoretically) &(limiters require small $\Delta t$)&(requires $\nu<0.5$ for positivity)\\
        \hline
    \end{tabular}
    \label{tab:scheme_compare}
\end{table}

\subsubsection{Scheme comparison with different distribution profiles discussed in Figure~\ref{fig:linear-advection-comparison_plots}}

The Positive flux conservative (PFC) scheme preserves positivity for step and two-step functions but introduces excessive diffusion, which leads to strong smoothing in step functions and significant amplitude loss in sinusoidal functions. The Flux conservative fifth order (FCV) scheme effectively captures sharp discontinuities but suffers from Gibbs oscillations, requiring external limiters to control these artifacts. The semi-Lagrangian monotonicity-preserving fifth-order scheme (SLMP5) resolves step, two-step, and sinusoidal functions with greater accuracy than the other methods. Despite its superior performance, SLMP5 has a computational cost comparable to that of PFC while offering improved accuracy~\cite{Tanaka2017}.

\subsection{Linear Landau damping in 1D}  
This study validates the PFC (Positive flux Conservative), FCV (Flux conservative 5th Order), FCV-Umeda (Flux conservative 5th Order scheme with Umeda limiter), and SLMP5 (semi-Lagrangian monotonicity preservation 5th Order) schemes using a 1D linear Landau damping test case~\cite{Yi2017,Filbet2001,Myers2017}. The Vlasov-Poisson system governs the evolution of the particle distribution function in the phase space and computes the self-consistent electric field via the Poisson equation.

The initial distribution function perturbs a Maxwellian equilibrium:

\begin{equation}
f(0, x, v) = \frac{1}{\sqrt{2\pi}} e^{-v^2/2} \left(1 + \alpha \cos(kx)\right),
\end{equation}
with the parameters $\alpha = 0.01$, $k = 0.5$, spatial domain $x \in (0, L)$ ($L = 4\pi$), and periodic $x$-boundary conditions. The numerical configurations included $N_x = 32$ spatial cells, $N_v = 32$ velocity cells ($v_{\text{max}} = 4.5$), and $\Delta t = 0.01$. A time-splitting algorithm advances solutions by alternating between $x$-advection (velocity-driven) and $v$-advection (electric field-driven).

\begin{figure}[H]
\centering
\subfigure[Electric energy evolution in log scale]{
\includegraphics[width=0.47\textwidth]{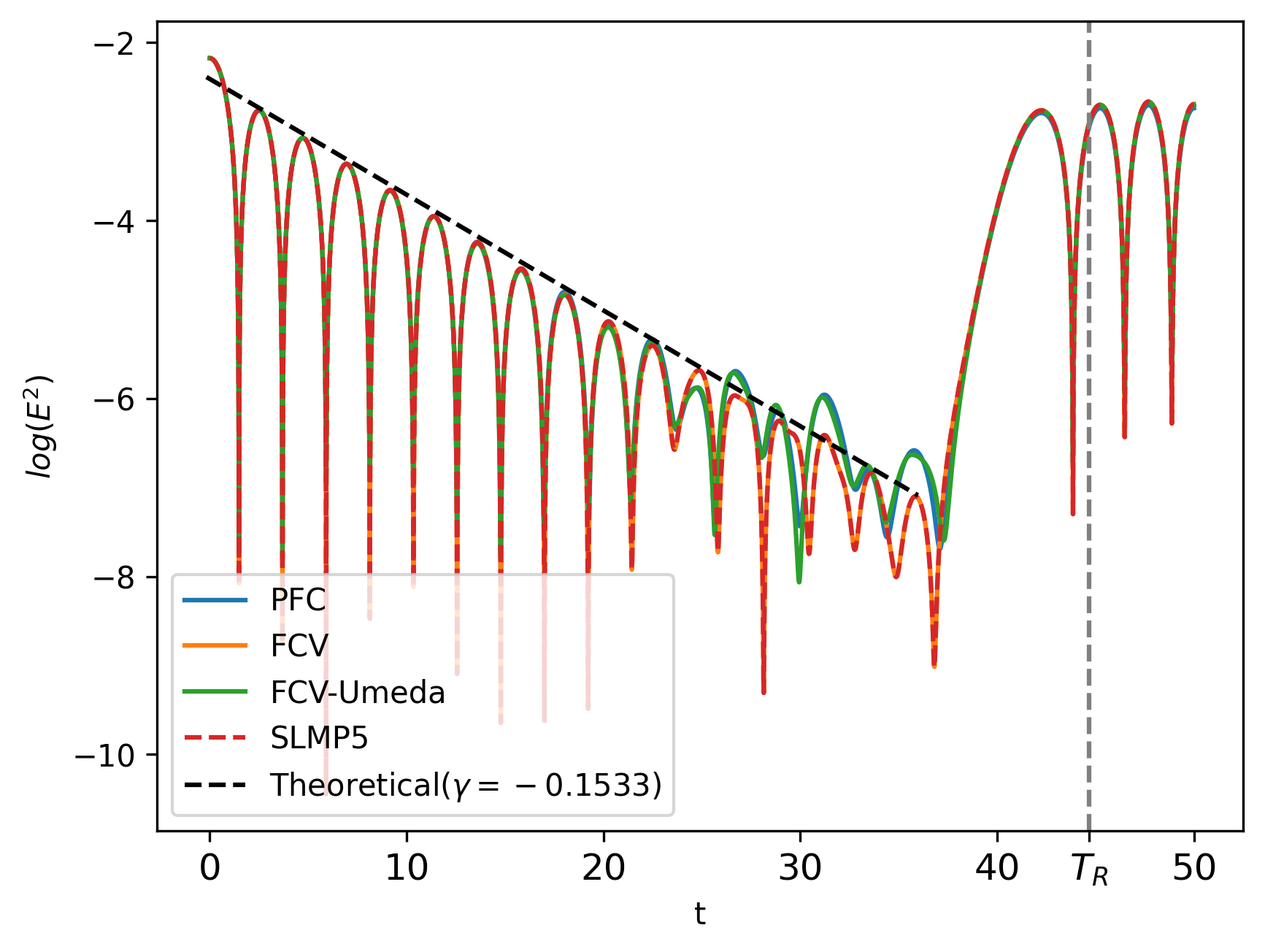}
}
\subfigure[Total energy error versus time]{
\includegraphics[width=0.47\textwidth]{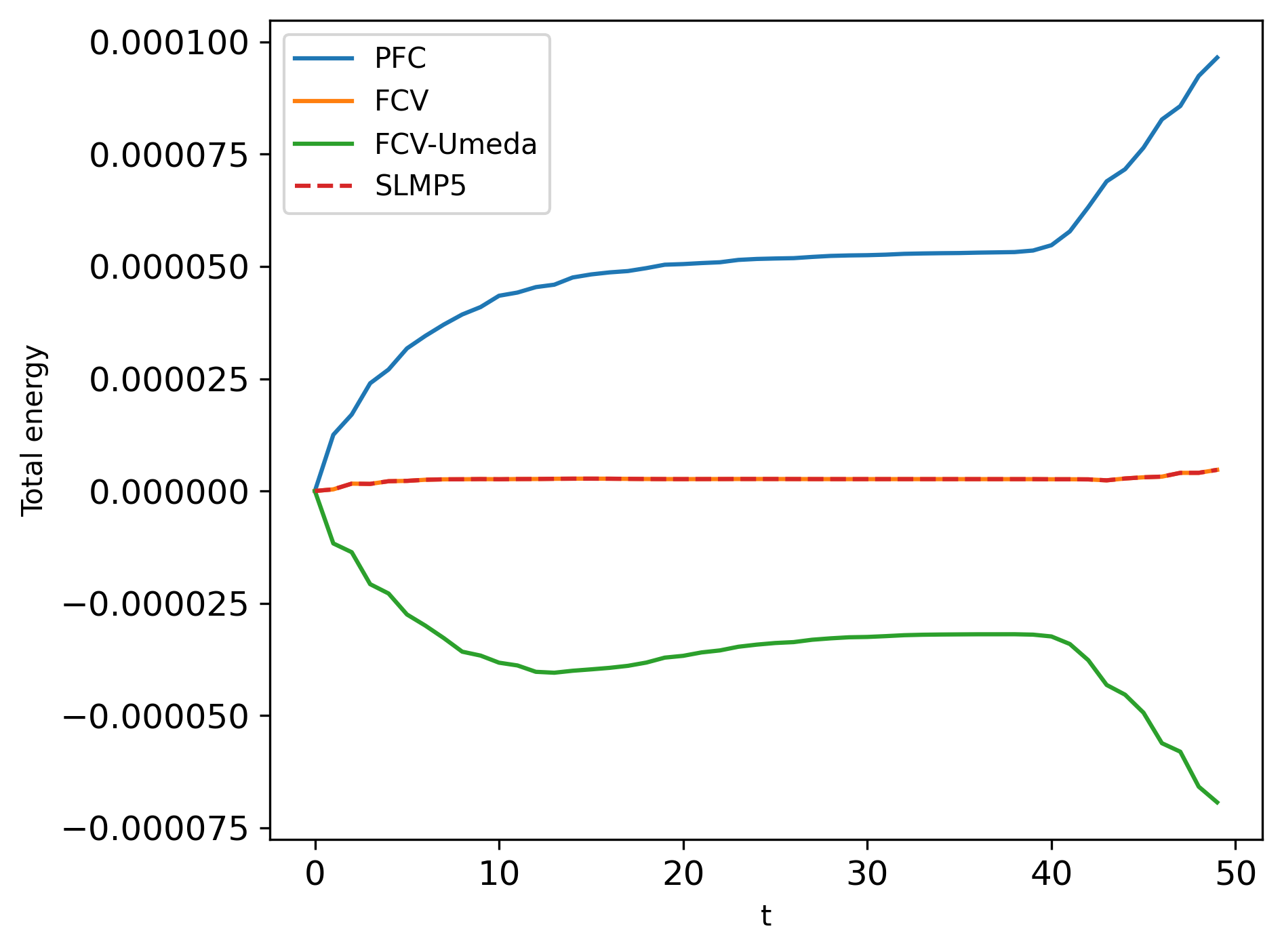}
}
\caption{Comparison of numerical schemes for all four schemes: (a) Evolution of electric energy $E^2$ in log scale over time with theoretical damping (the black line) with a slope of 0.1533 and the vertical dashed line at $T_R=44.68$ is the recurrence time (b) Total energy error versus time.}
\label{fig:comparison_plots}
\end{figure}

The electric energy $\sum |E_i(t)|^2$ decays exponentially, consistent with the linear Landau theory, as shown in Figure~\ref{fig:comparison_plots} and all the schemes achieve a damping rate ($\gamma = 0.153$) and oscillation frequency ($\omega = 1.415$) with $N_v = 32$, closely matching the theoretical values ($\gamma_{\text{theory}} = 0.1533$, $\omega_{\text{theory}} = 1.4156$). The accuracy declines for the PFC and FCV-Umeda schemes as the simulation nears the recurrence time $T_R =44.68= 2\pi/(k|\delta v|)$, while FCV and SLMP5 exhibit superior long-term stability.

All four schemes performed well and preserved positivity, as the distribution function remained non-negative. The PFC and FCV-Umeda schemes exhibited errors in energy conservation of approximately $10^-4$. The SLMP5 and FCV schemes achieve a good balance between accuracy and conservation, replicating the theoretical damping rates with minimal energy conservation errors $\leq 10^-5$. Combined with its non-oscillatory behavior, this suggests that the SLMP5 method is the preferred choice for Vlasov-Poisson systems.
 
\subsection{Linear Landau damping in 2D}  
Next, we extended the validation of all four schemes to a 2D linear Landau damping test case~\cite{Filbet2001,Myers2017}. The Vlasov-Poisson system governs the evolution of the particle distribution function $f(t, x, y, v_x, v_y)$ in a four-dimensional phase space, coupled with the Poisson equation for self-consistent electric field computation.

The initial distribution function perturbs a 2D Maxwellian equilibrium:

\begin{equation}
f_0(x, y, v_x, v_y) = \frac{1}{2\pi} e^{-(v_x^2 + v_y^2)/2} \left(1 + \alpha \cos(k_x x) \cos(k_y y)\right),
\end{equation}

with perturbation amplitude $\alpha \ll 1$, wave numbers $k_x = 0.5$ and $k_y = 0.5$, and periodic spatial domains $x, y \in (0, L)$, where $L = 4\pi$. The numerical parameters included $64$ grid points per direction ($x, y, v_x, v_y$), $v_{\text{max}} = 4.5$, and $\Delta t = 0.01$. A time-splitting algorithm advances the solution by decoupling the advection in the physical and velocity space.

\begin{figure}[H]
\centering
\subfigure[Electric energy evolution in log scale for setup $(t, k_x=0, k_y=0.5)$]{
\includegraphics[width=0.47\textwidth]{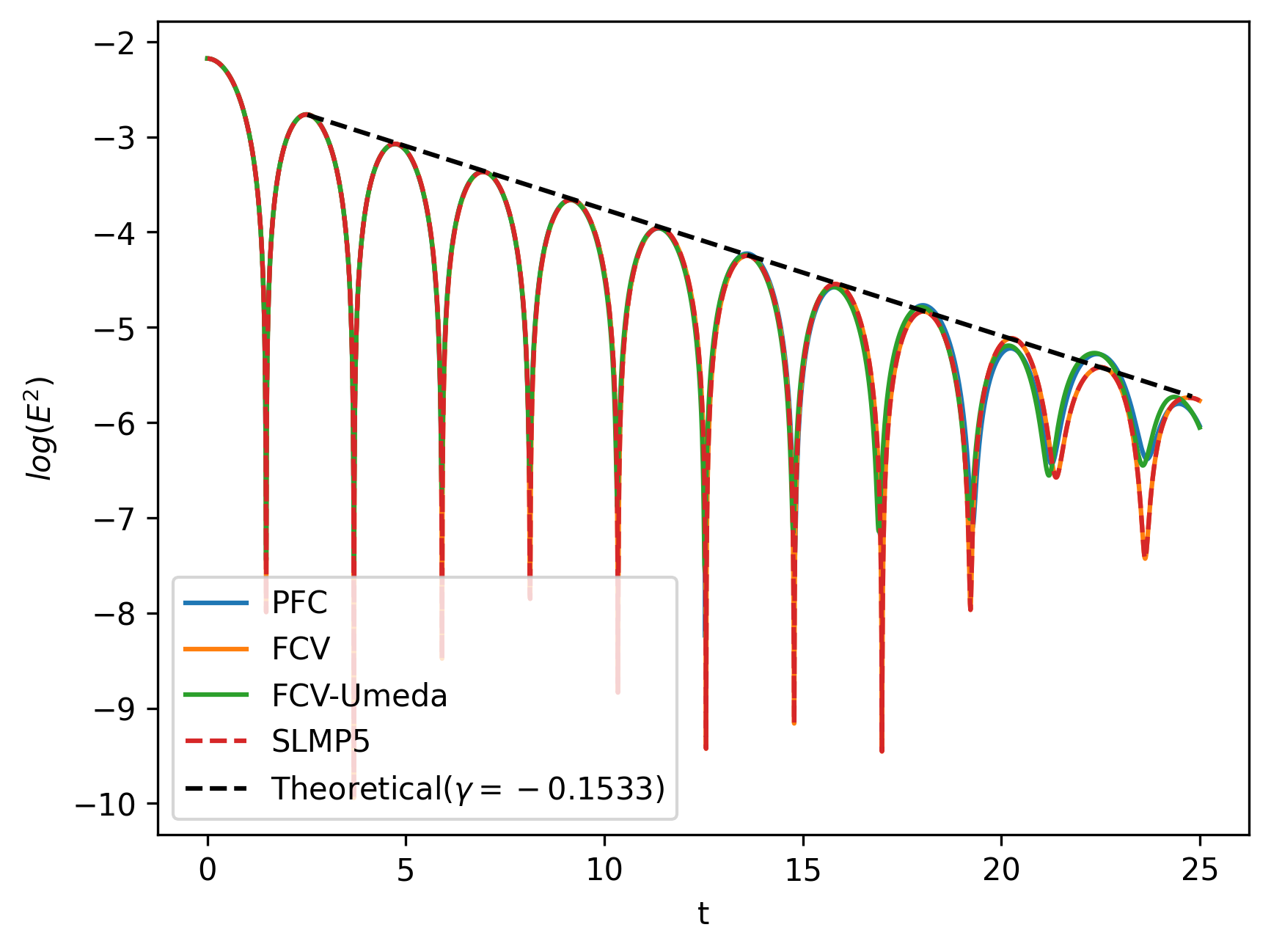}
}
\subfigure[Electric energy evolution in log scale for setup $(t, k_x=0.5, k_y=0.5)$]{
\includegraphics[width=0.47\textwidth]{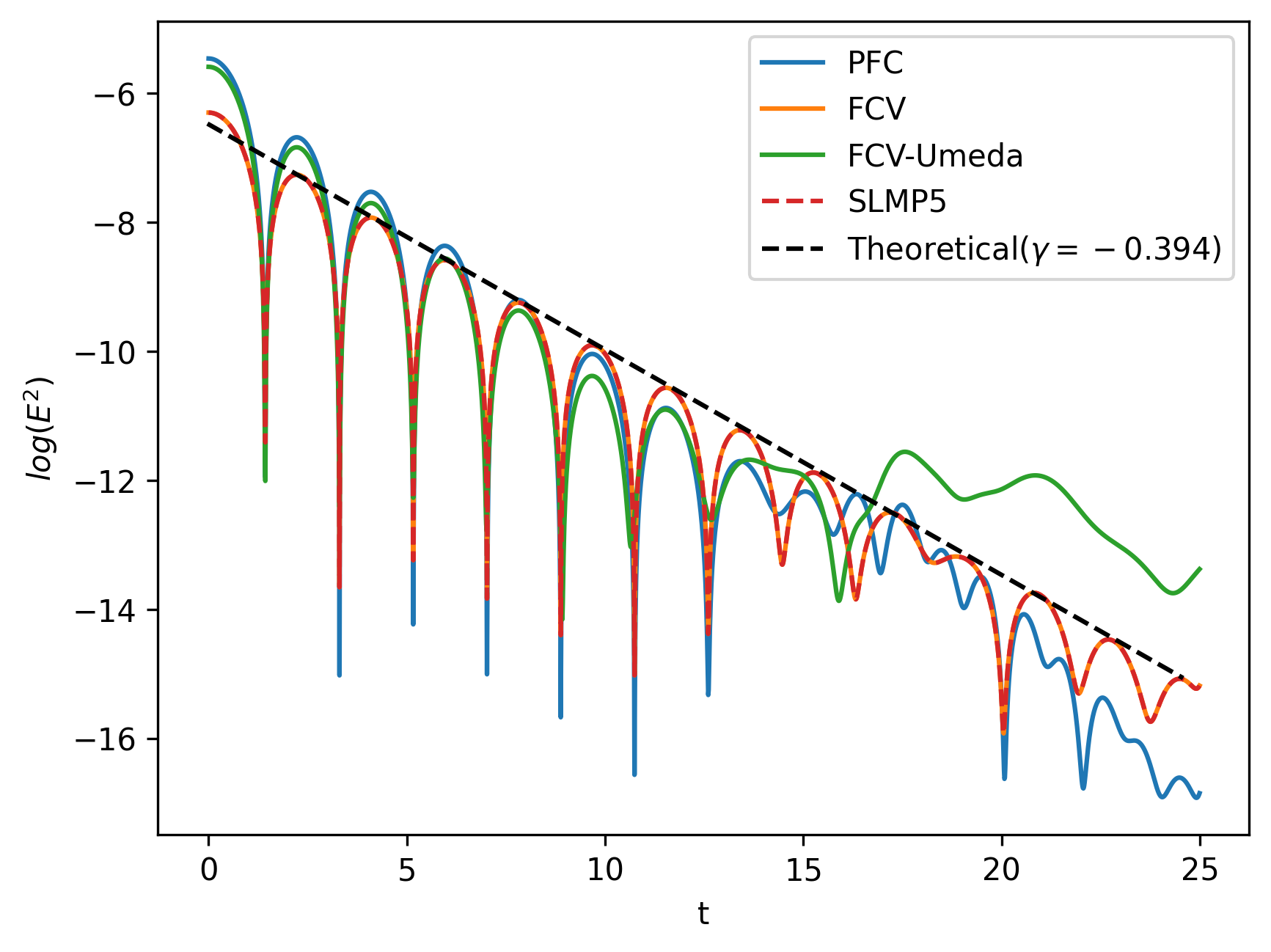}
}
\caption{Comparison of numerical schemes for the 2D case: (a) Evolution of electric energy $E^2$ for setup $(t, k_x=0, k_y=0.5)$ in log scale over time with
theoretical damping (the black dashed line) with a slope of 0.1533 (b)Evolution of electric energy $E^2$ for setup$(t, k_x=0.5, k_y=0.5)$ in log scale over time with
theoretical damping (the black dashed line) with a slope of 0.394  for PFC, FCV, FCV-Umeda and SLMP5 schemes.}
\label{fig:comparison_plots_2D}
\end{figure}

The electric energy $\sum |E_{x,y}(t)|^2$ decays exponentially, which is consistent with linear Landau theory, as shown in Figure~\ref{fig:comparison_plots_2D}. All four schemes demonstrated good agreement with theoretical predictions. 

The Fourier modes $E_x(t, k_x=0, k_y=0.5)$ and $E_x(t, k_x=0.5, k_y=0.5)$ obtained by the SLMP5 scheme decay exponentially with the damping rates ($\gamma = 0.1533$, $\gamma = 0.394$) and oscillation frequencies ($\omega = 1.4156$, $\omega = 1.6973$), respectively. The PFC and FCV-Umeda schemes do not properly replicate the theoretical damping slope, as shown in Figure~\ref{fig:comparison_plots_2D}. However, the SLMP5 and FCV schemes balance accuracy and conservation, replicating theoretical damping rates with minimal deviation.

\subsection{Strong Landau damping in 1D}  

Next, we evaluated all four schemes for strong Landau damping~\cite{Yi2017,Filbet2001} in 1D, a benchmark test emphasizing nonlinear effects in the Vlasov-Poisson system. The initial distribution function introduces a large-amplitude perturbation to the Maxwellian equilibrium:

\begin{equation}
f(0, x, v) = \frac{1}{\sqrt{2\pi}} e^{-v^2/2} \left(1 + \alpha \cos(kx)\right),
\end{equation}
\begin{figure}[H]
\centering
\subfigure[Electric energy evolution in log scale for a $32 \times 64$ grid]{
\includegraphics[width=0.47\textwidth]{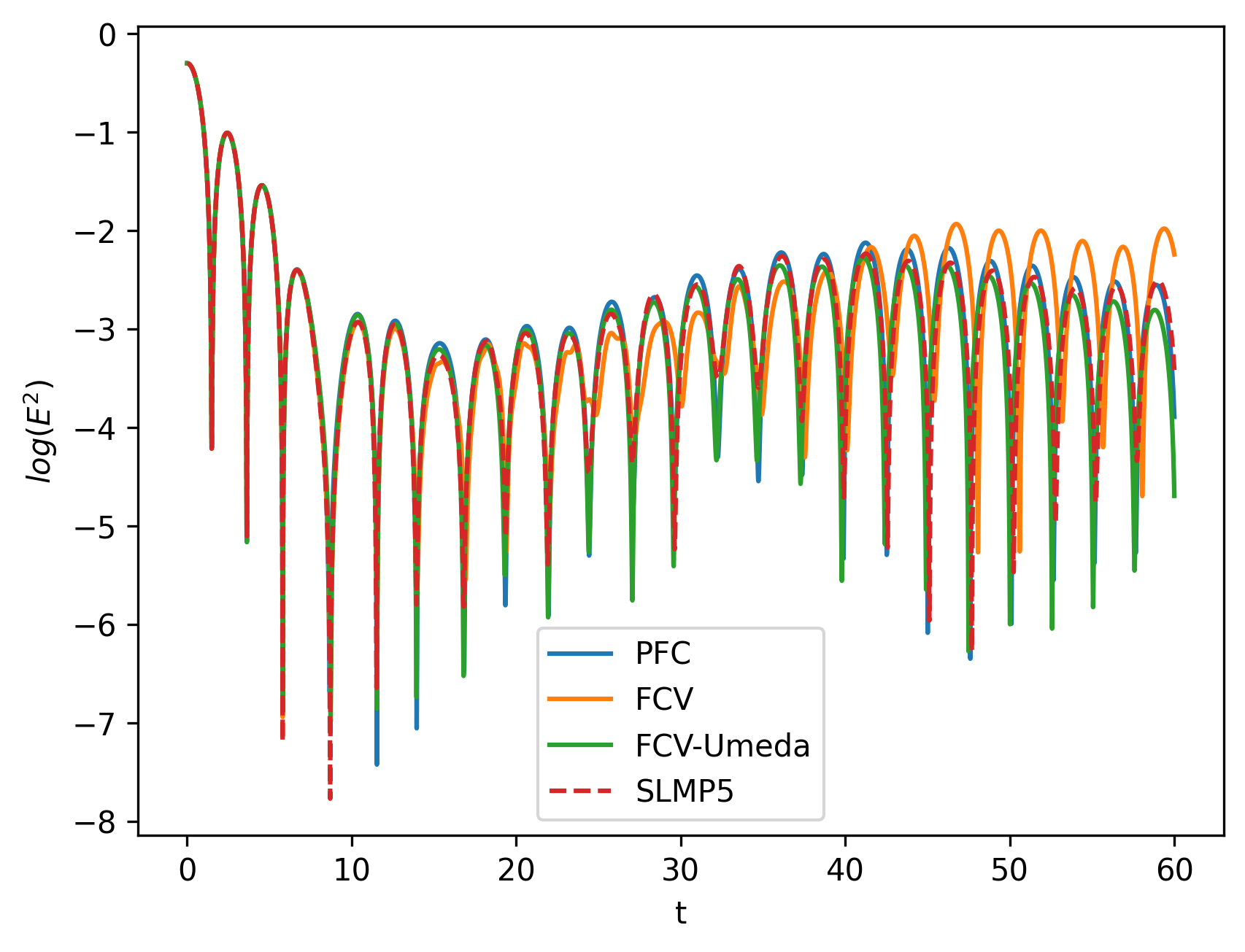}
}
\subfigure[Electric energy evolution in log scale for a $32 \times 128$ grid]{
\includegraphics[width=0.47\textwidth]{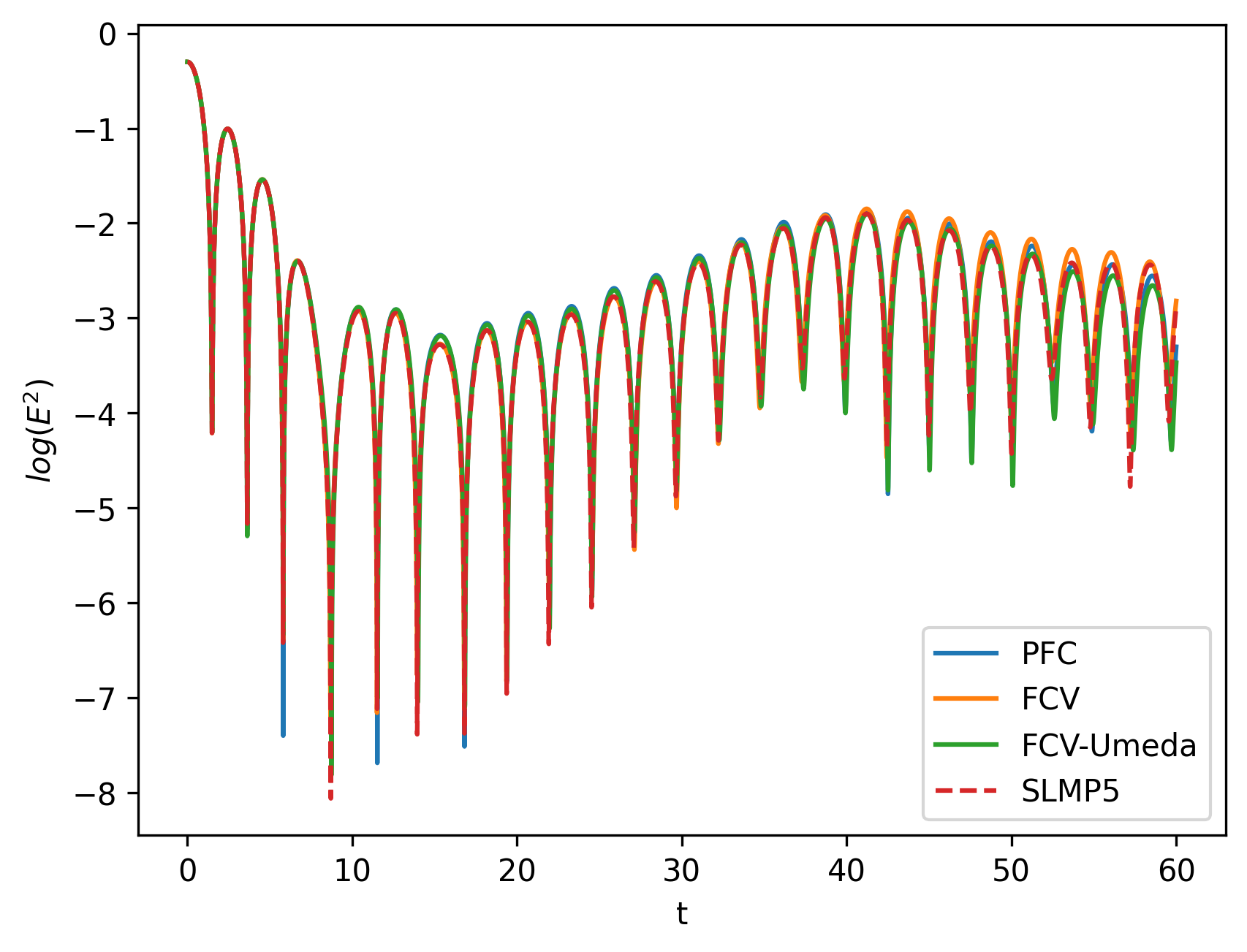}
}
\caption{Comparison of numerical schemes for the 1D strong Landau damping case: (a) Evolution of electric energy $E^2$ over time for a $32 \times 64$ grid (b) Evolution of electric energy $E^2$ over time for $32 \times 128$ grid points for the PFC, FCV, FCV-Umeda, and SLMP5 schemes.}
\label{fig:comparison_plots_1D}
\end{figure}
with perturbation amplitude $\alpha \gg 0.01$ (nonlinear regime, for this setup $\alpha =0.5$ ), wave number $k = 0.5$, and a periodic domain $x \in (0, L)$, where $L = 4\pi$. Numerical parameters include $N_x \times N_V = 32 \times 64 $ and $32 \times 128$ cells and $\Delta t = 0.01$. A time-splitting algorithm decouples $x$-advection (velocity-dependent) and $v$-advection (electric field-driven).

\begin{figure}[H]
\centering
\subfigure[$\int f(x,v,t=0)dx$ for PFC]{\label{fig:intfxv-pfc-0}
\includegraphics[width=0.3\textwidth]{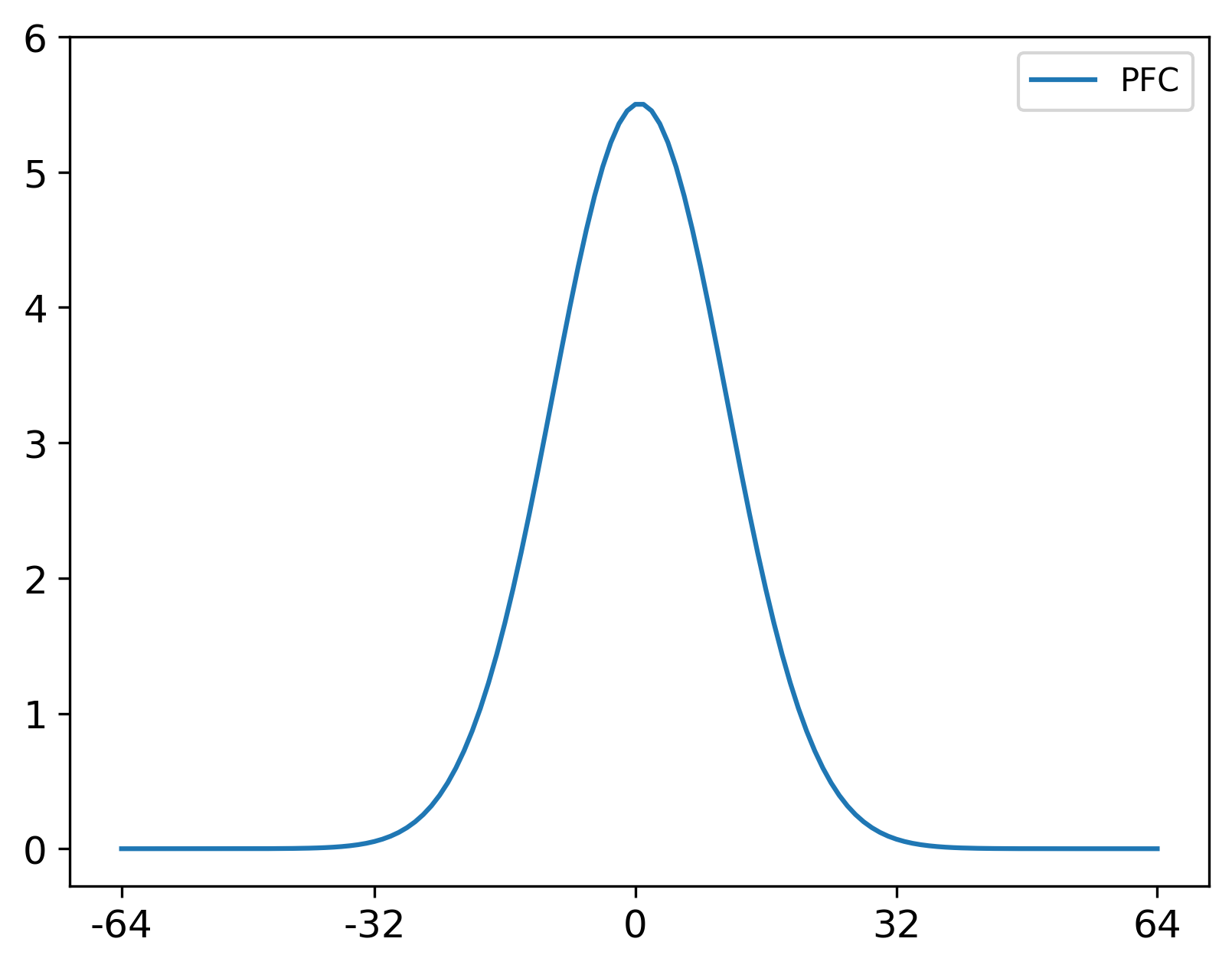}
}
\subfigure[$\int f(x,v,t=20)dx$ for PFC]{\label{fig:intfxv-pfc-20}
\includegraphics[width=0.3\textwidth]{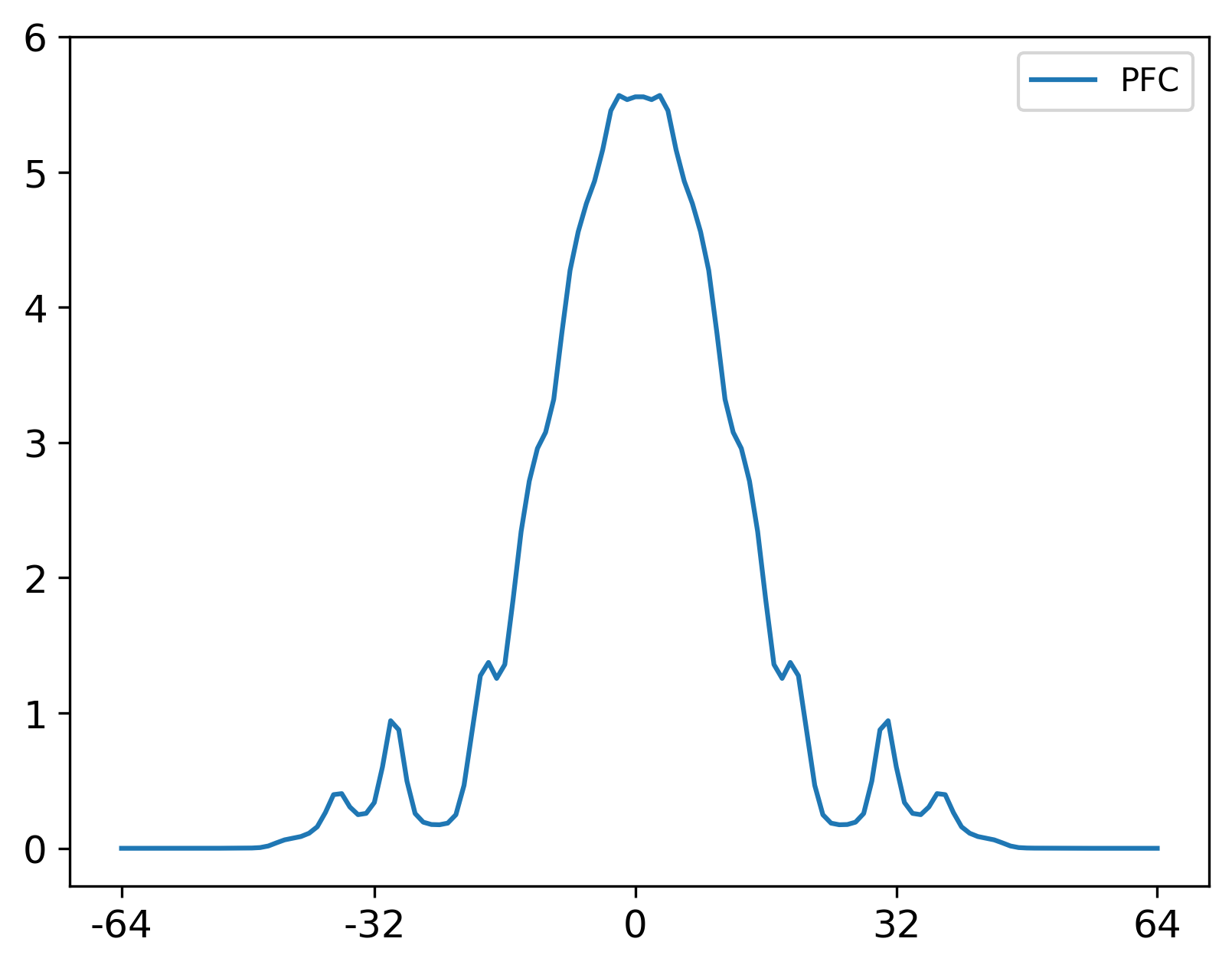}
}
\subfigure[$\int f(x,v,t=40)dx$for PFC]{\label{fig:intfxv-pfc-40}
\includegraphics[width=0.3\textwidth]{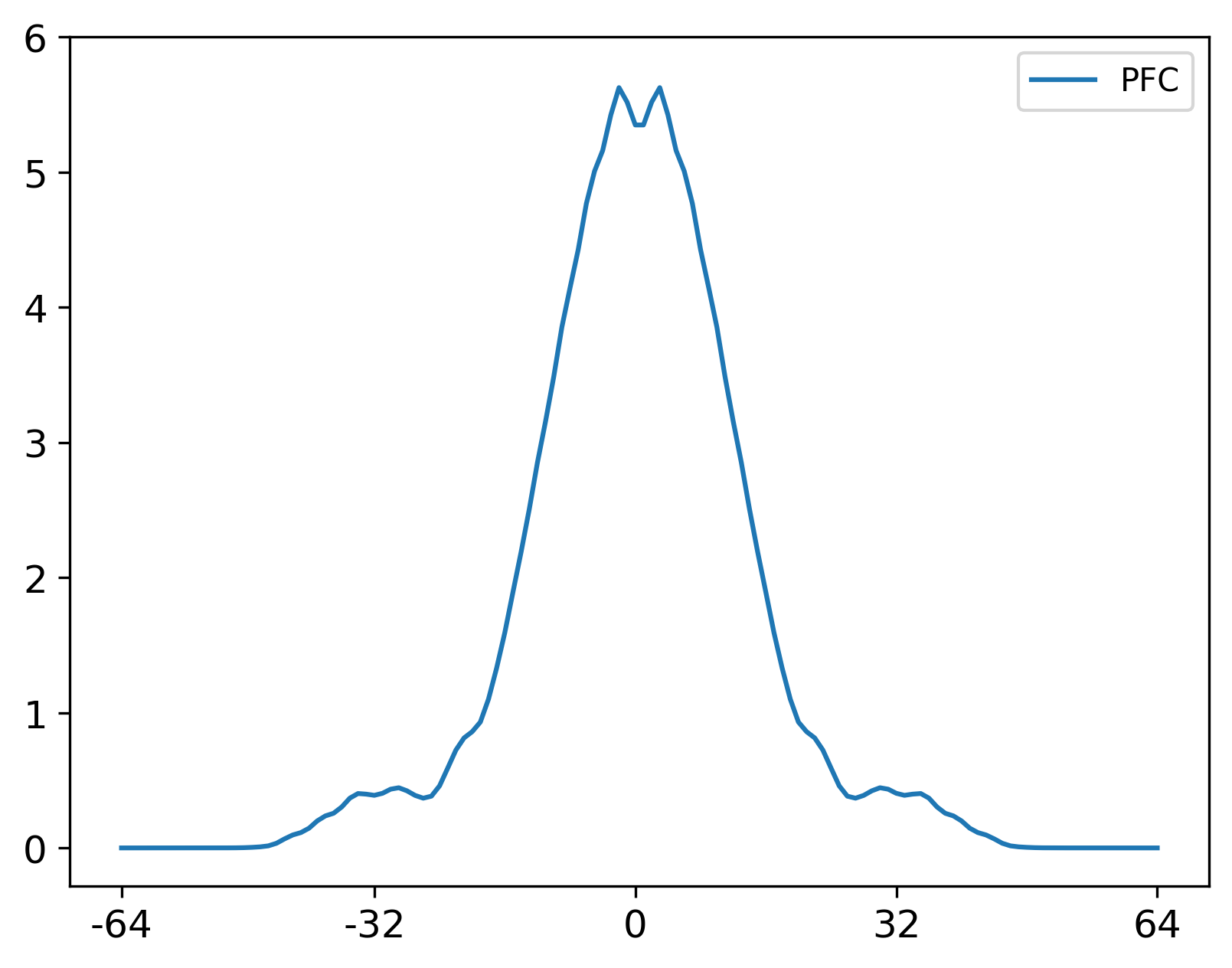}
}
\subfigure[$\int f(x,v,t=0)dx$ for FCV]{\label{fig:intfxv-fcv-0}
\includegraphics[width=0.3\textwidth]{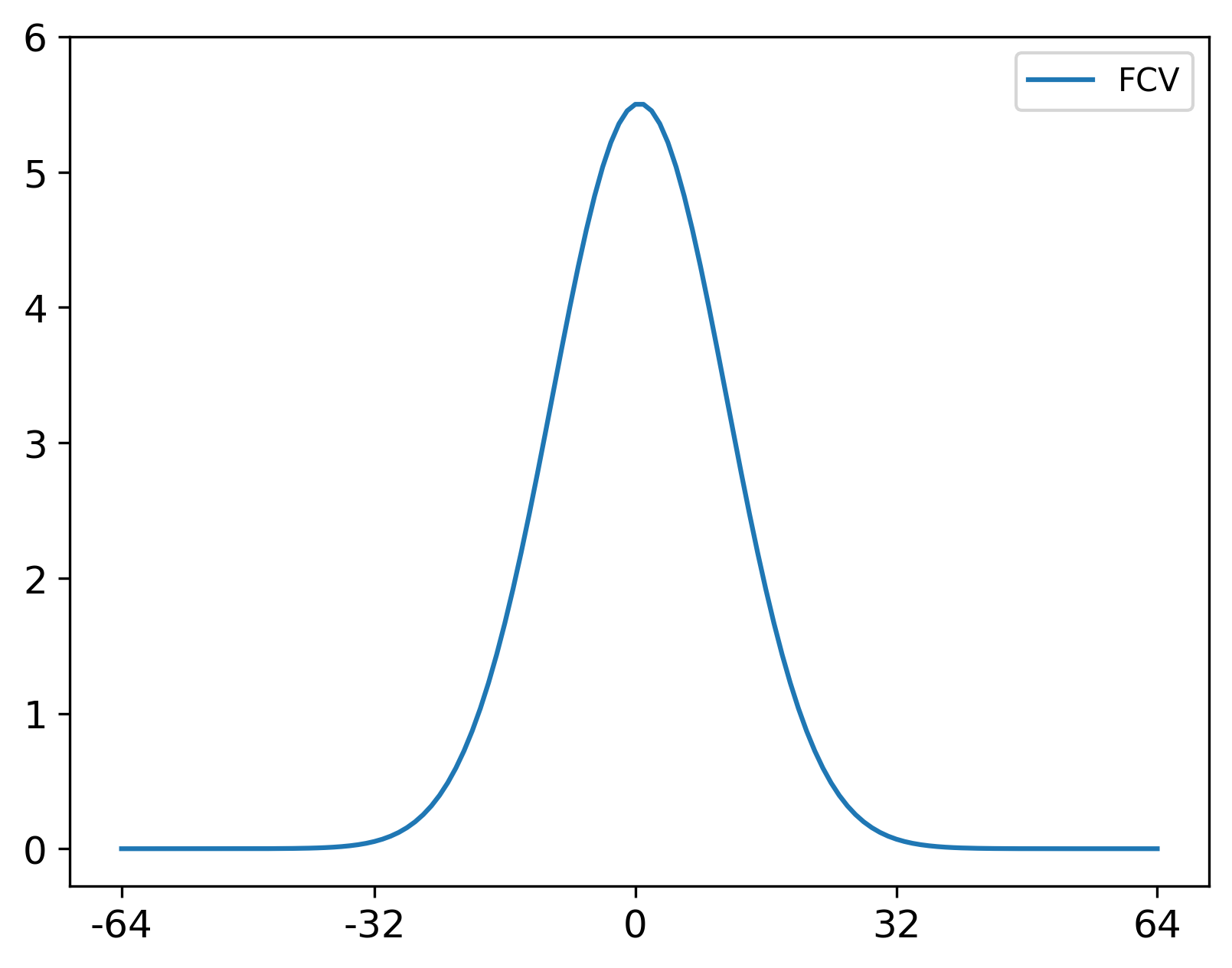}
}
\subfigure[$\int f(x,v,t=20)dx$ for FCV]{\label{fig:intfxv-fcv-20}
\includegraphics[width=0.3\textwidth]{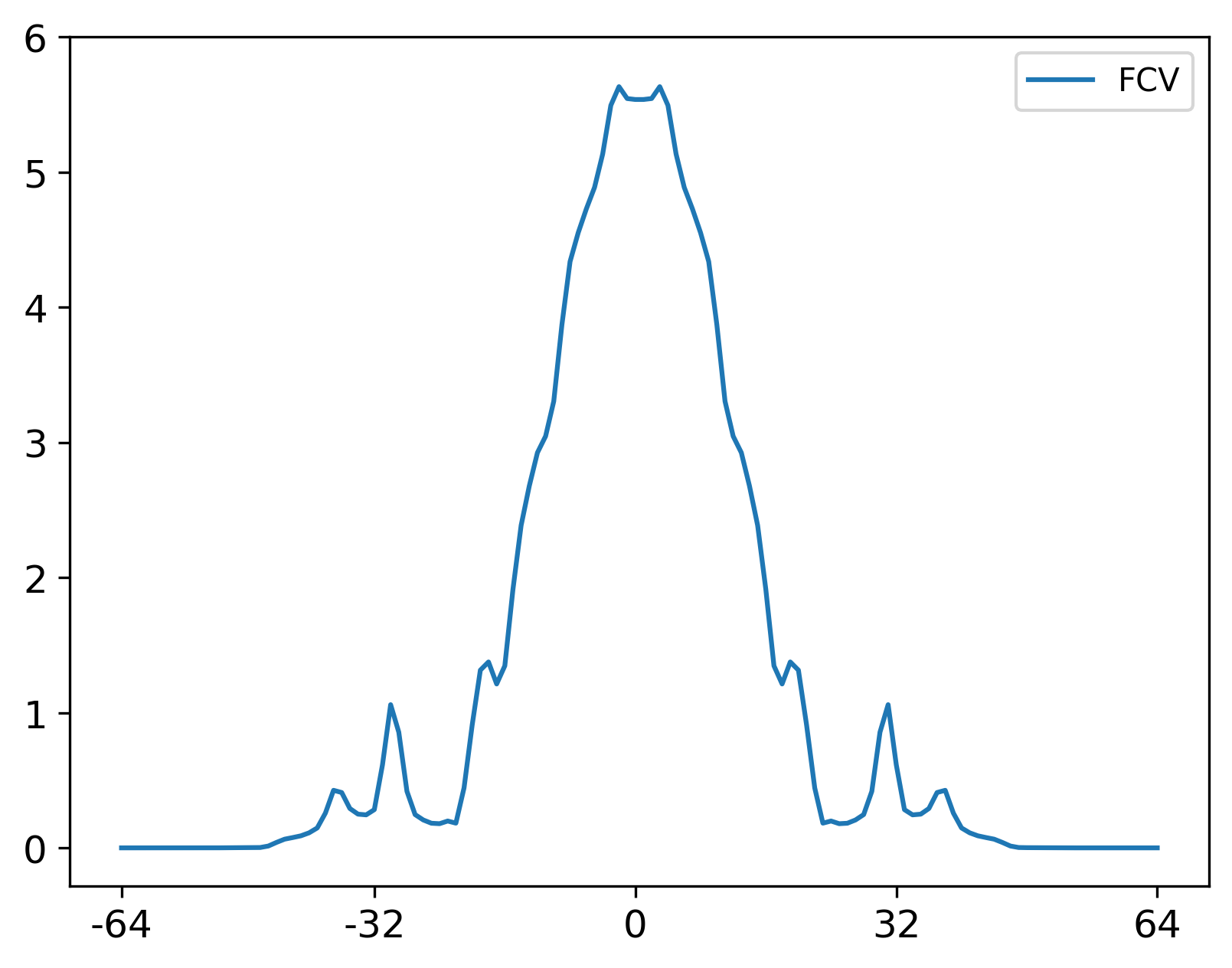}
}
\subfigure[$\int f(x,v,t=40)dx$ for FCV]{\label{fig:intfxv-fcv-40}
\includegraphics[width=0.3\textwidth]{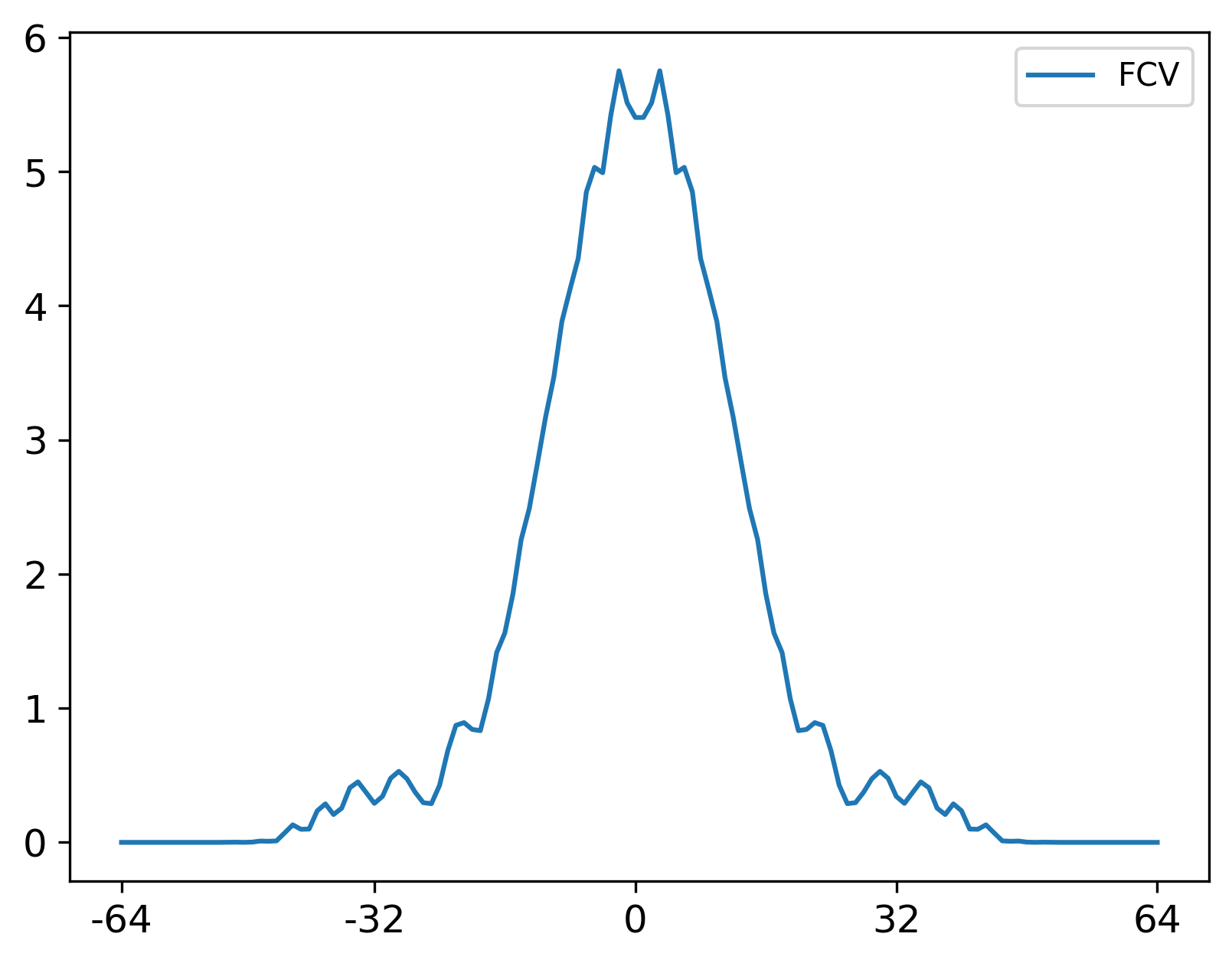}
}
\subfigure[$\int f(x,v,t=0)dx$ for FCV-Umeda]{\label{fig:intfxv-umeda-0}
\includegraphics[width=0.3\textwidth]{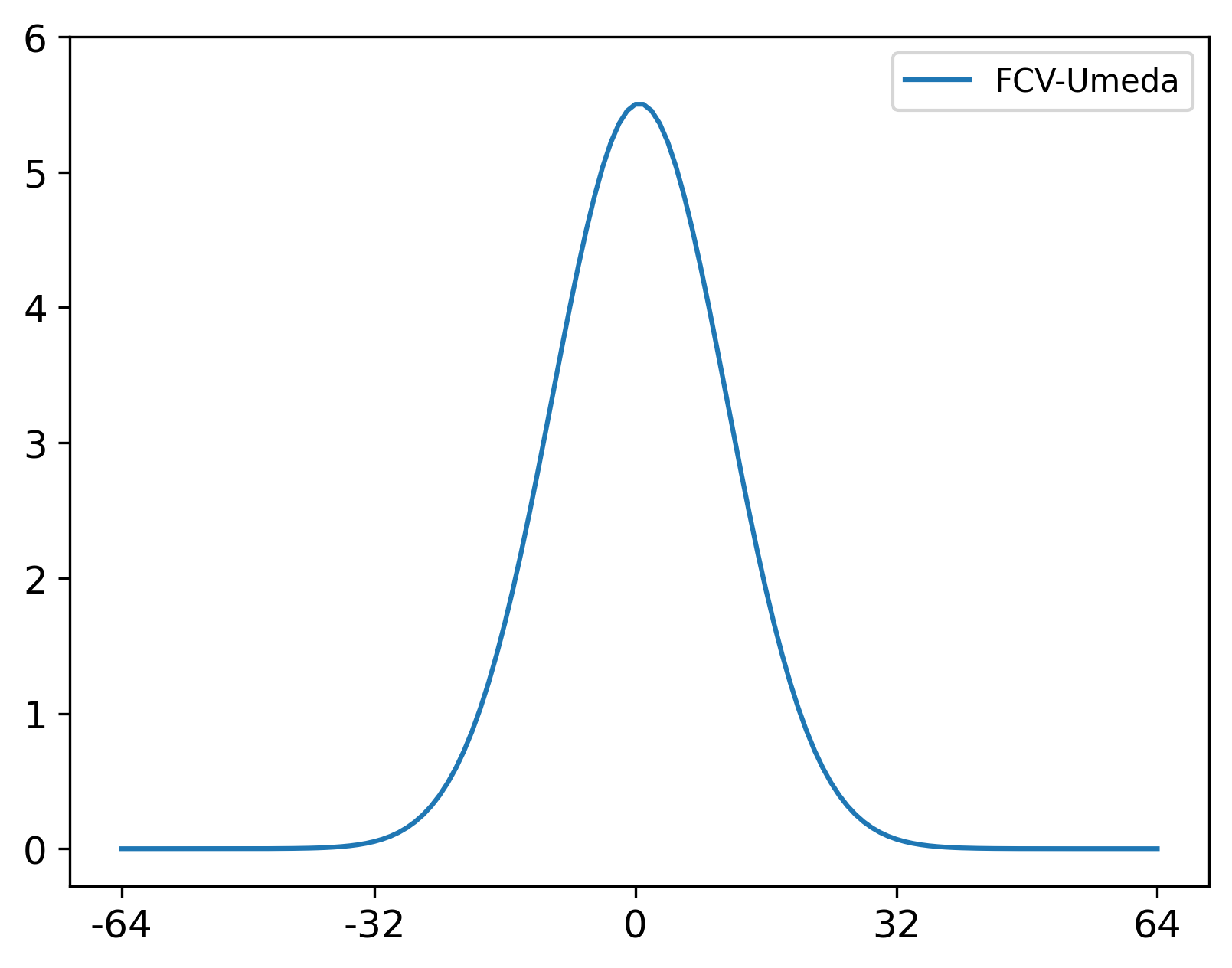}
}
\subfigure[$\int f(x,v,t=20)dx$ for FCV-Umeda]{\label{fig:intfxv-umeda-20}
\includegraphics[width=0.3\textwidth]{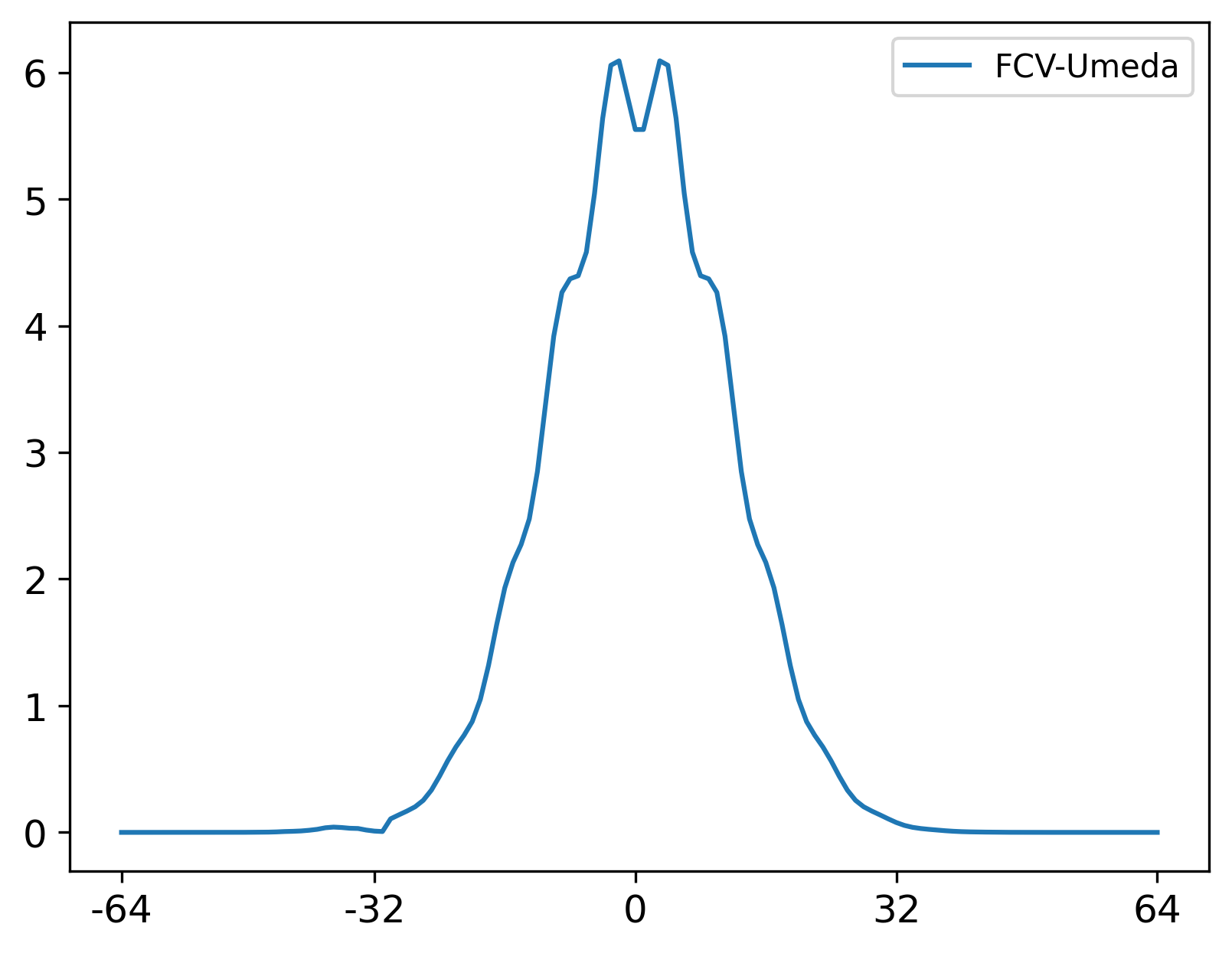}
}
\subfigure[$\int f(x,v,t=40)dx$ for FCV-Umeda]{\label{fig:intfxv-umeda-40}
\includegraphics[width=0.3\textwidth]{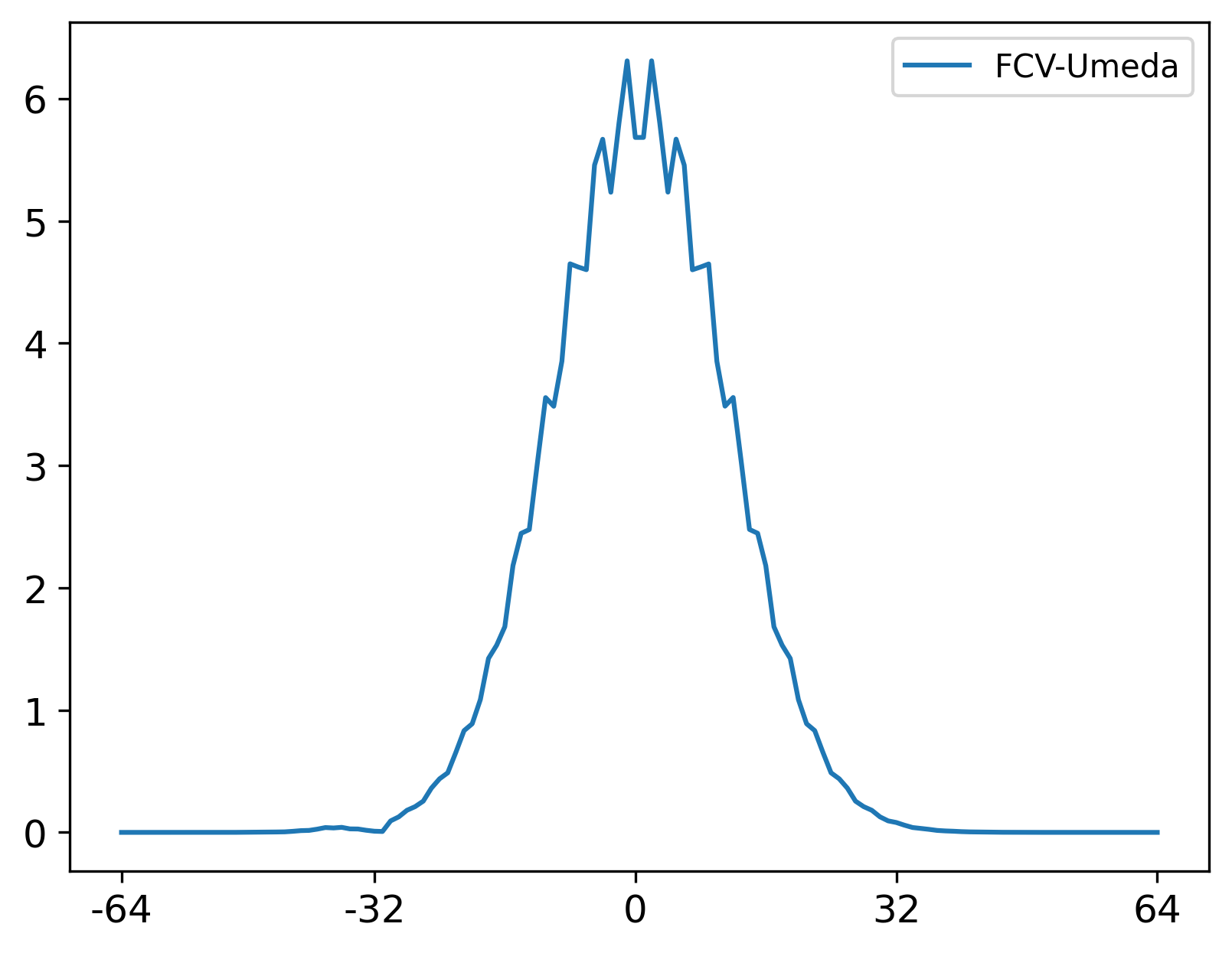}
}
\subfigure[$\int f(x,v,t=0)dx$ for SLMP5]{\label{fig:intfxv-slmp-0}
\includegraphics[width=0.3\textwidth]{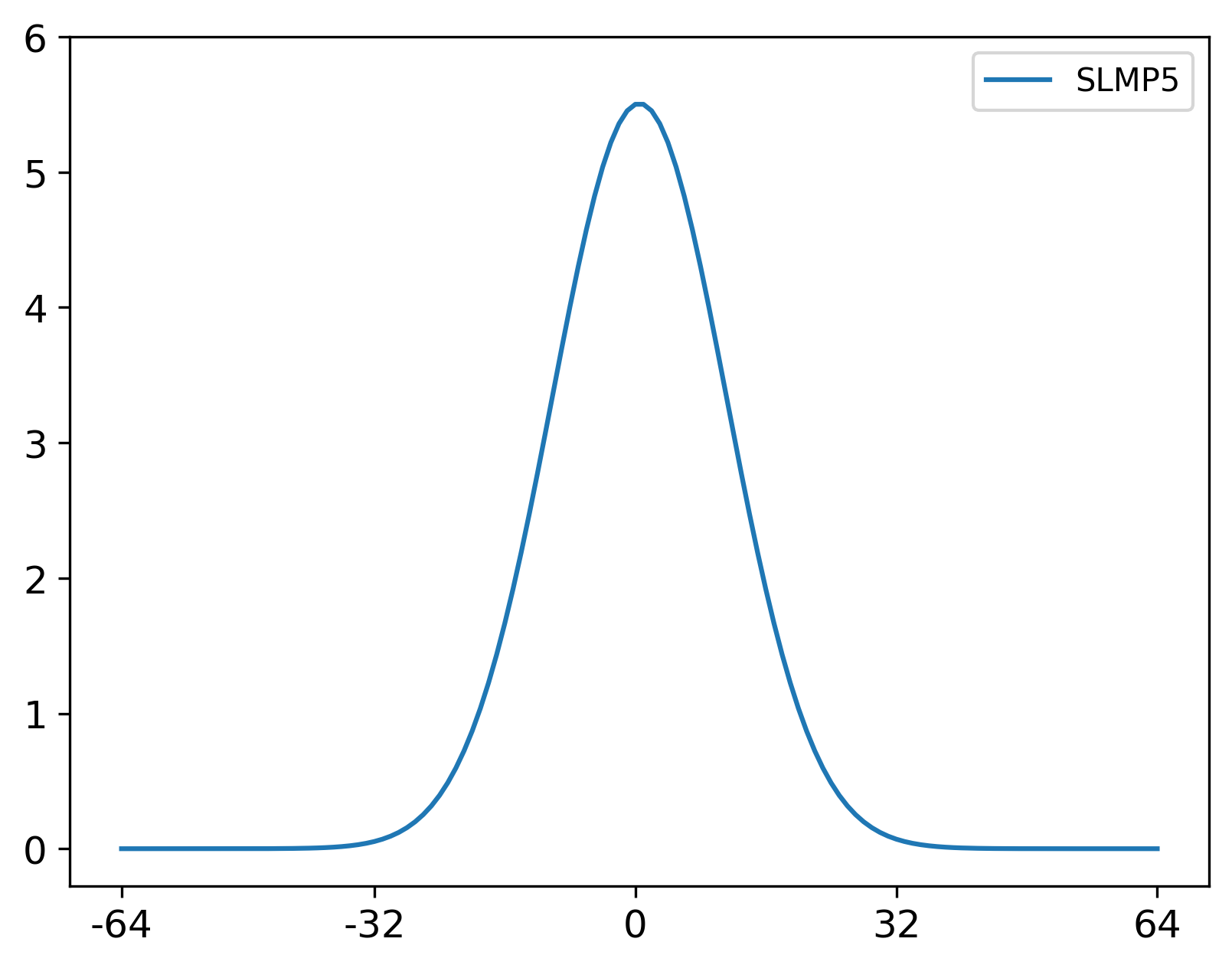}
}
\subfigure[$\int f(x,v,t=20)dx$ for SLMP5]{\label{fig:intfxv-slmp-20}
\includegraphics[width=0.3\textwidth]{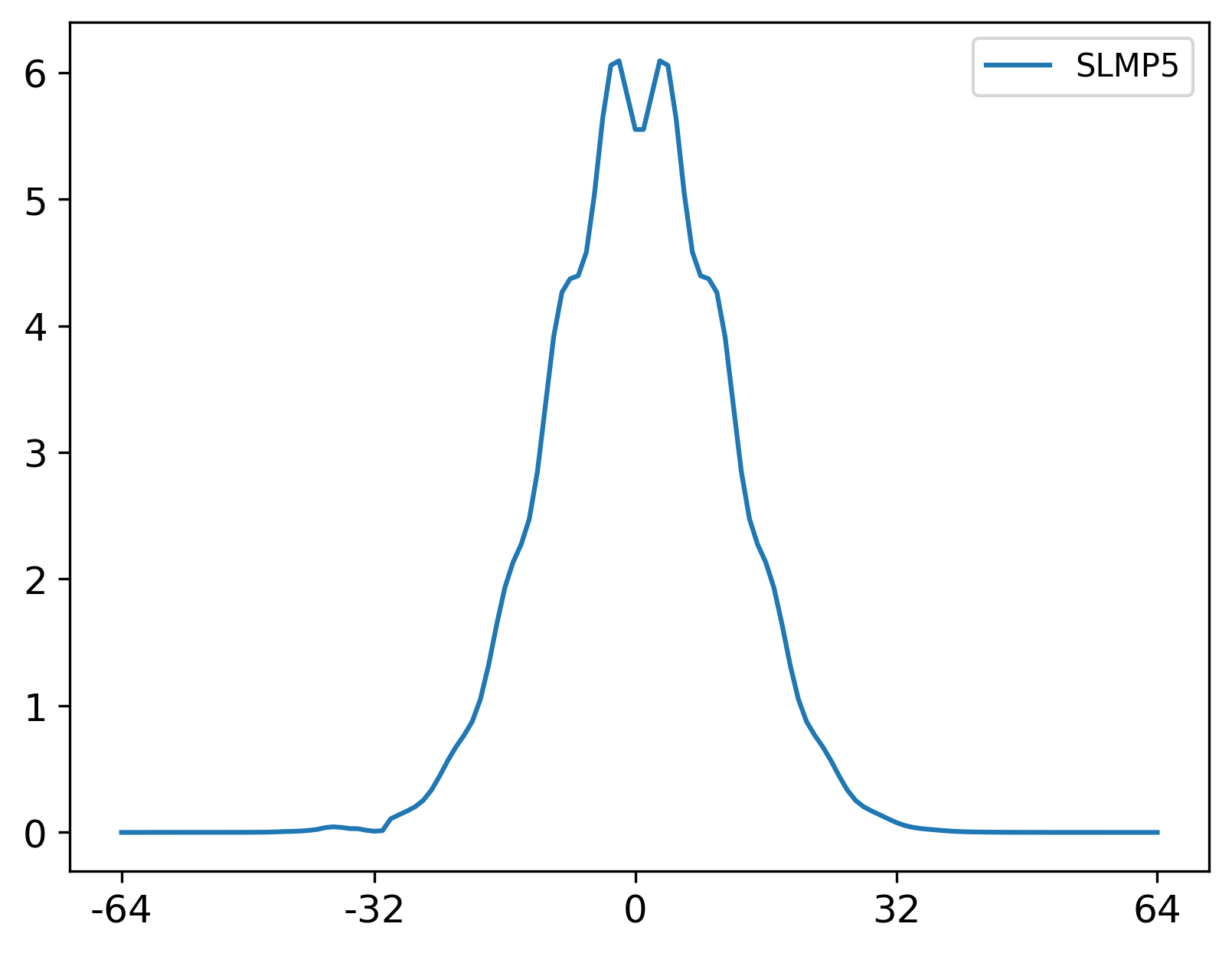}
}
\subfigure[$\int f(x,v,t=40)dx$ for SLMP5]{\label{fig:intfxv-slmp-40}
\includegraphics[width=0.3\textwidth]{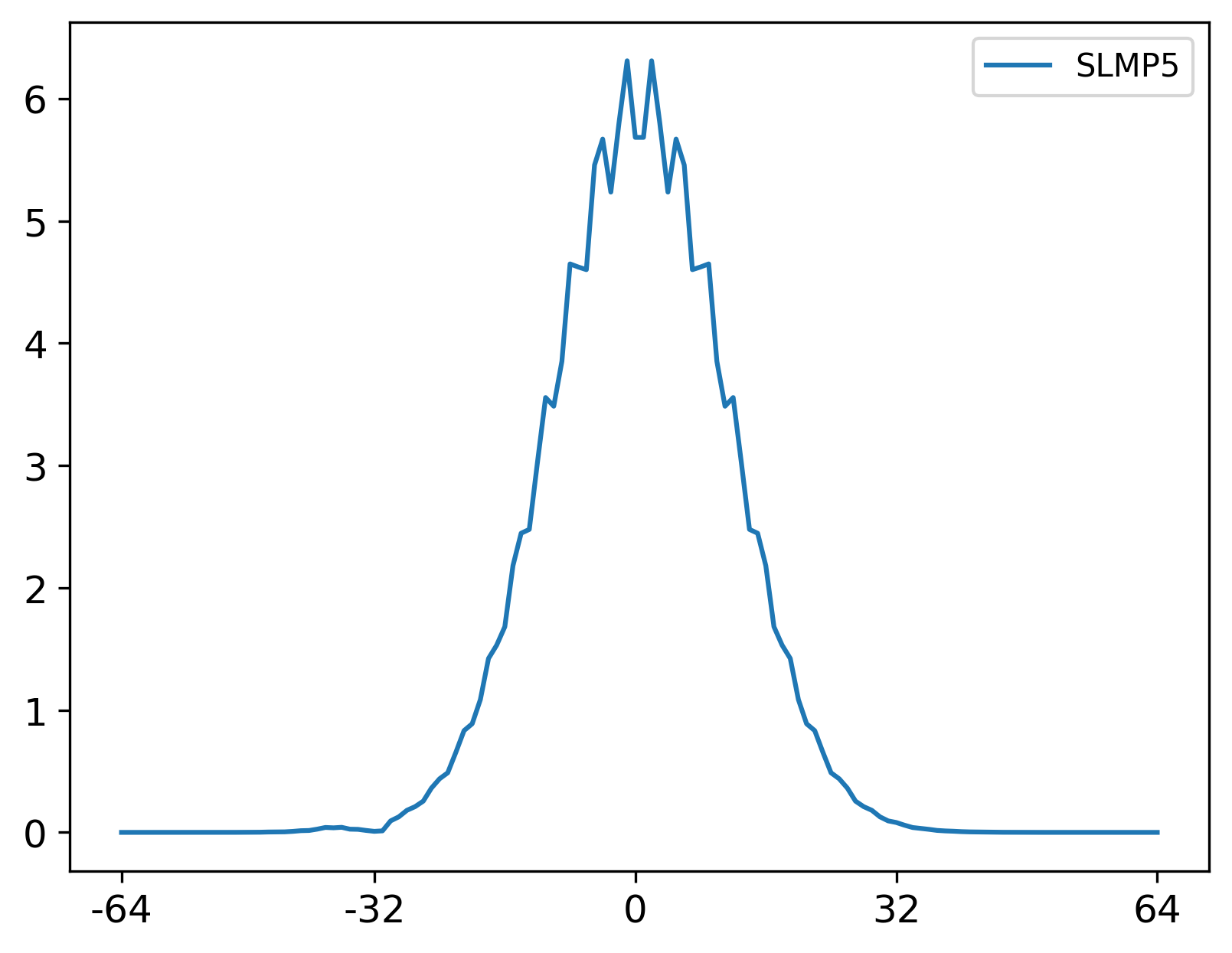}
}
\caption{Time development of the spatially integrated distribution function for the PFC, FCV, FCV-Umeda and SLMP5 schemes for strong Landau damping with grid ($N_x \times N_v = 32 \times 128$)}
\label{fig:integralf}
\end{figure}

\begin{figure}[H]
\centering
\subfigure[$f(x,v,t=0)$ for SLMP5]{\label{fig:fxv-slmp-0}
\includegraphics[height=0.42\textwidth]{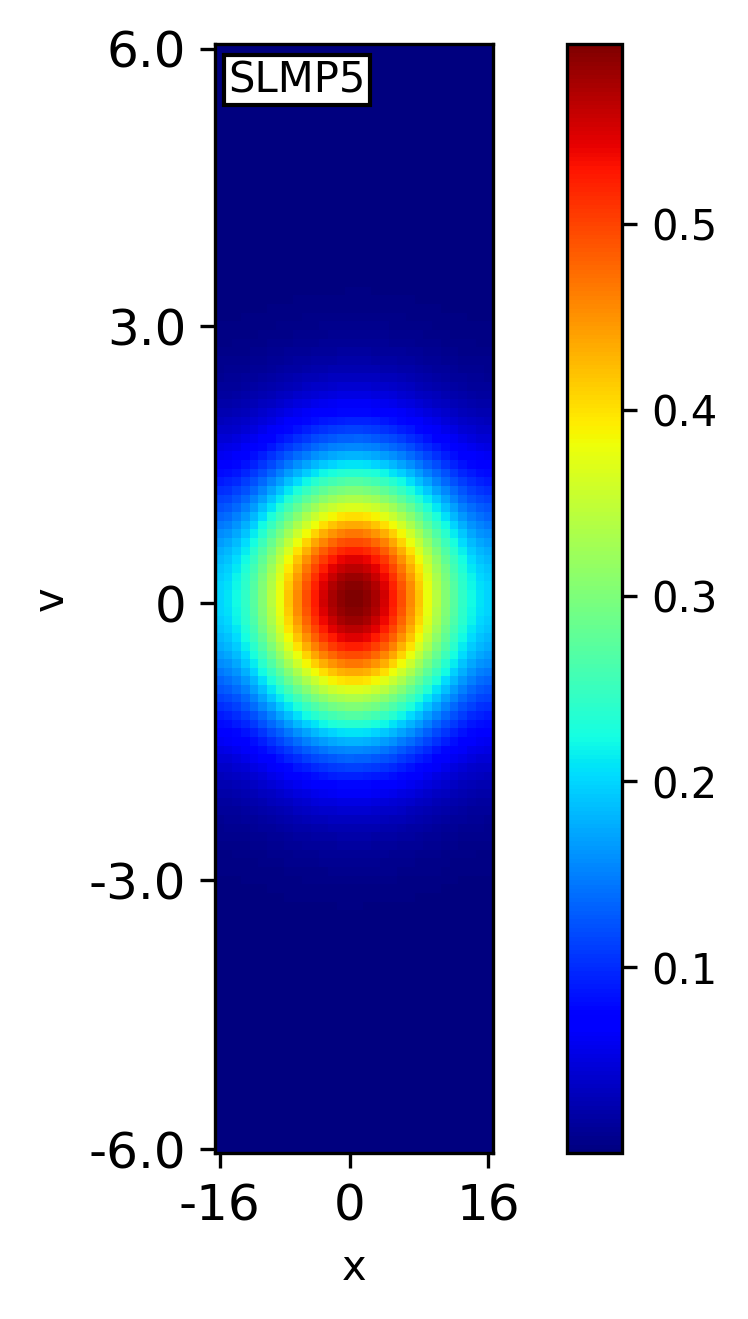}
}
\subfigure[$f(x,v,t=20)$ for SLMP5]{\label{fig:fxv-slmp-20}
\includegraphics[height=0.42\textwidth]{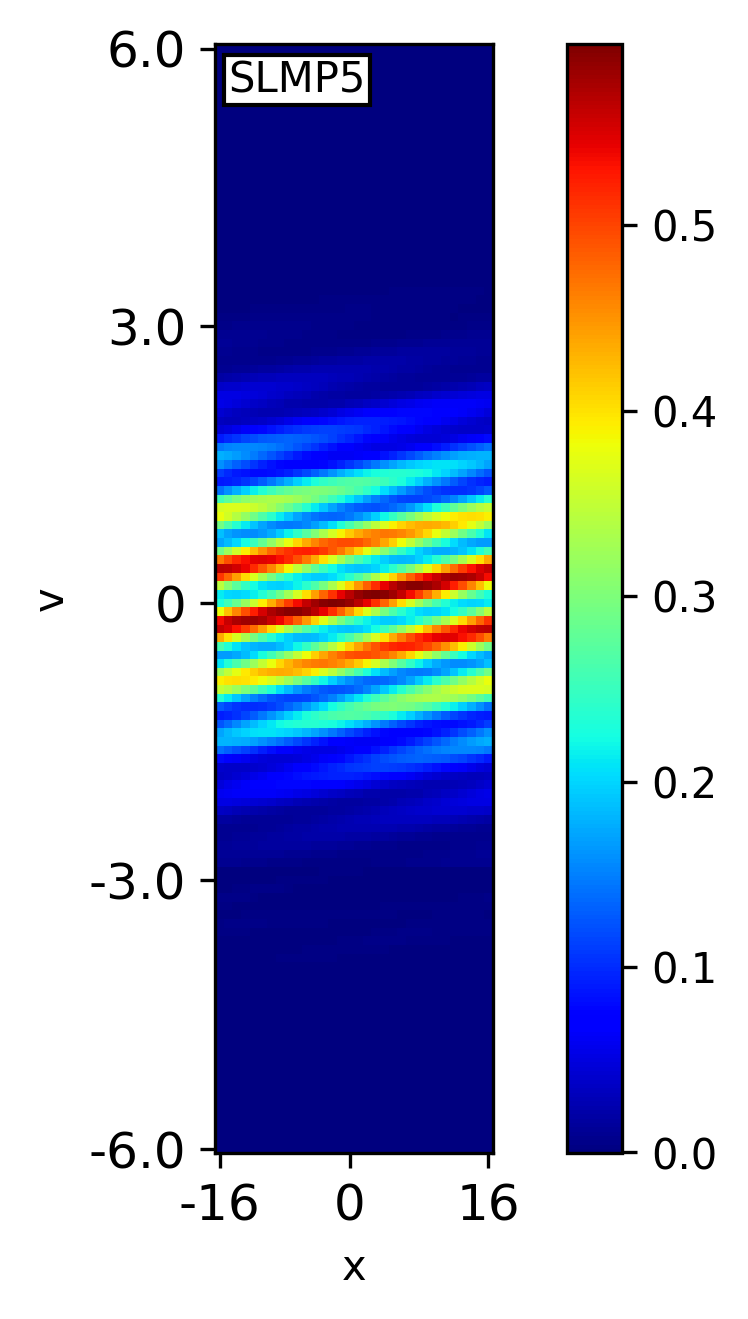}
}
\subfigure[$f(x,v,t=40)$ for SLMP5]{\label{fig:fxv-slmp-40}
\includegraphics[height=0.42\textwidth]{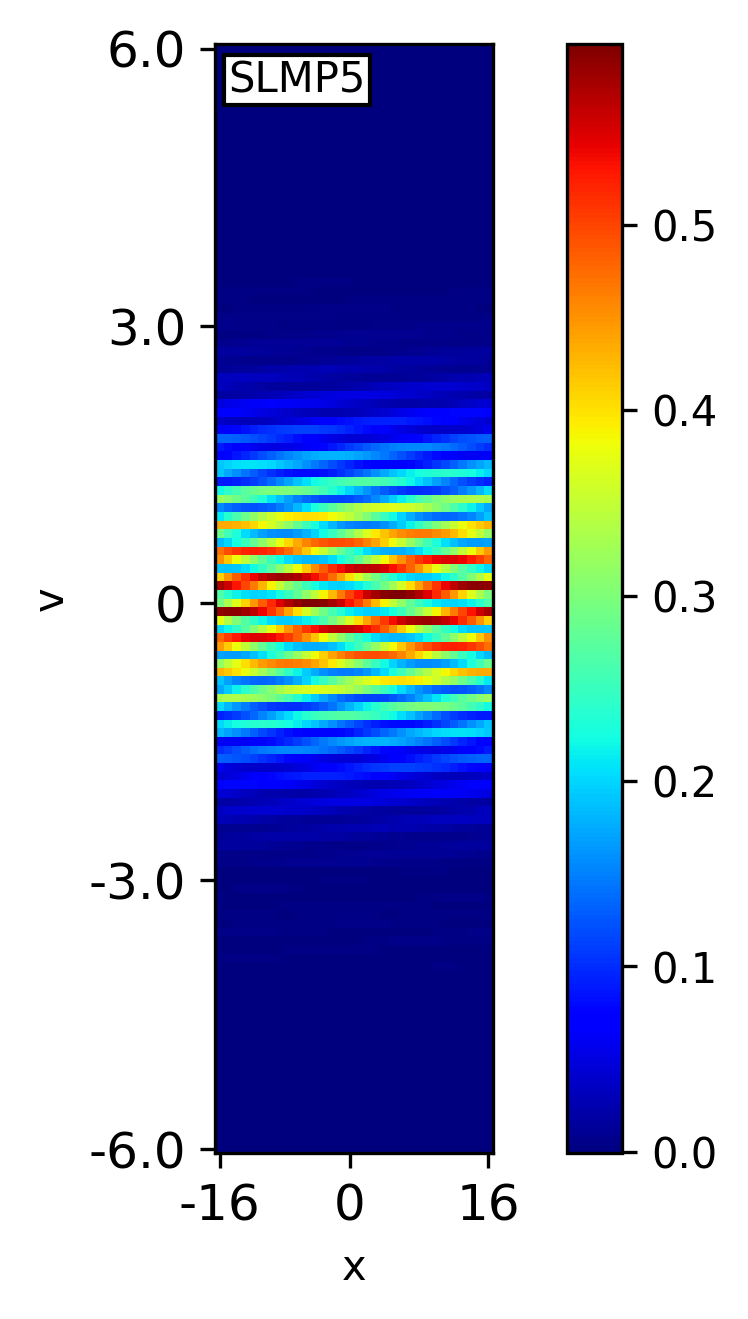}
}
\caption{Phase space plots of contour of the distribution function of strong Landau damping with grid ($N_x \times N_v = 32 \times 128$) at t=0,20,40.}
\label{fig:phasespace}
\end{figure}

Figure~\ref{fig:comparison_plots_1D} shows that the electric energy $\sum |E_i(t)|^2$ exhibits an initial exponential decay followed by periodic oscillations, which is characteristic of strong Landau damping. The SLMP5 scheme achieves superior accuracy on coarse grids ($N_x \times N_v = 32 \times 64$), outperforming the FCV, and FCV-Umeda schemes requiring finer resolution ($32 \times 128$) for comparable results.
Figure~\ref{fig:integralf} shows the time evolution of the integrated distribution function, and Figure~\ref{fig:phasespace} displays the corresponding phase space structures for the four numerical schemes: PFC, FCV, FCV-Umeda, and SLMP5. The PFC scheme exhibited stronger numerical diffusion and significantly damping fine-scale features in the velocity space. The FCV scheme preserves phase space filamentation more effectively but introduces non-physical oscillations, particularly in regions with steep gradients. The FCV-Umeda method effectively suppresses these oscillations, offering a more balanced solution. However, a higher resolution is required to maintain accuracy. Among the four schemes, the SLMP5 scheme provides a good compromise, delivering stable and accurate results at relatively lower resolutions.
\begin{figure}[H]
\centering
\subfigure[$f(x,v,t=0)$ for PFC]{\label{fig:fxv-pfc-0}
\includegraphics[height=0.42\textwidth]{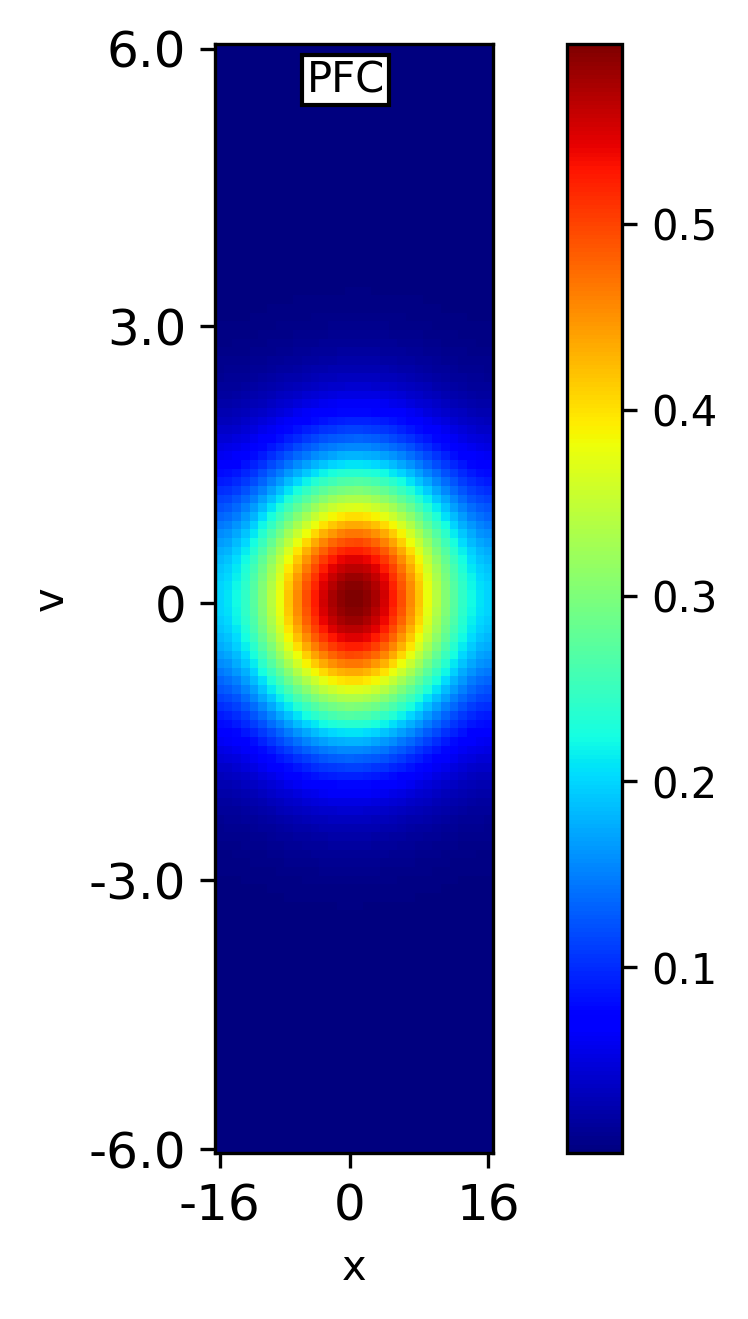}
}
\subfigure[$f(x,v,t=20)$ for PFC]{\label{fig:fxv-pfc-20}
\includegraphics[height=0.42\textwidth]{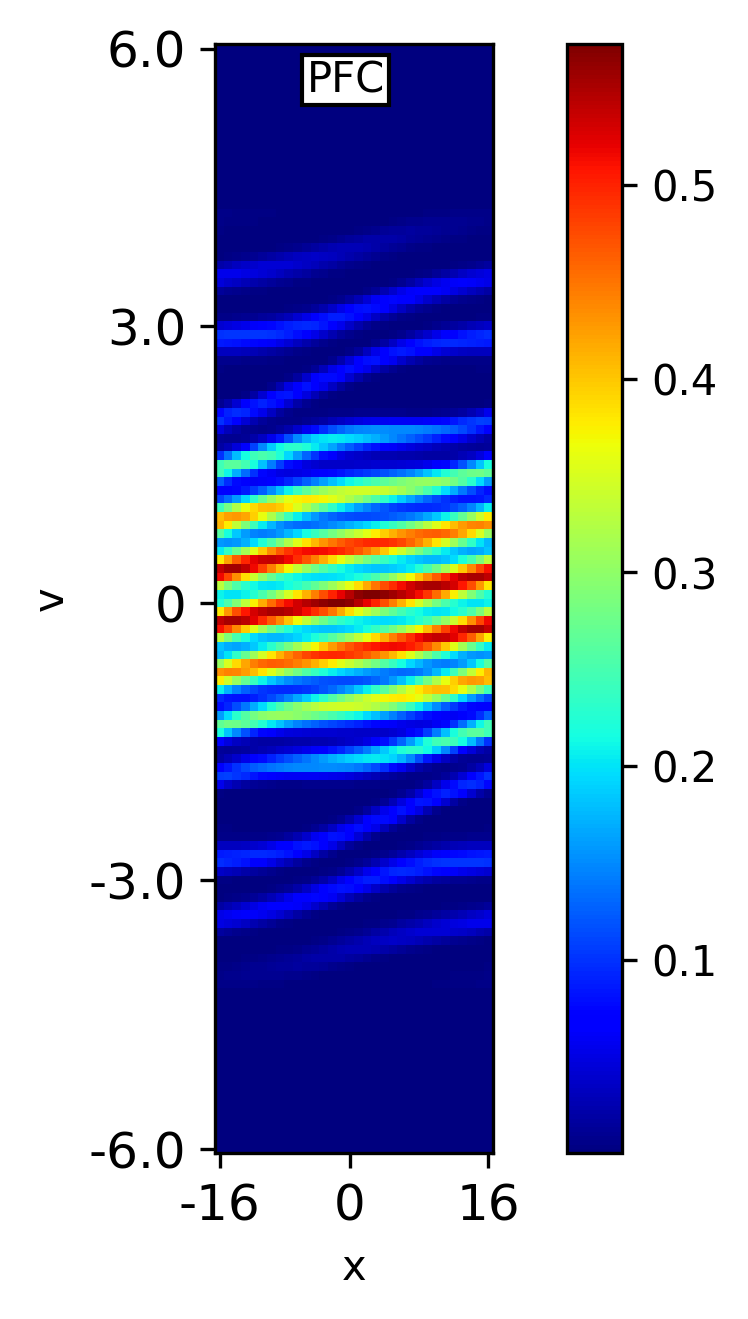}
}
\subfigure[$f(x,v,t=40)$ for PFC]{\label{fig:fxv-pfc-40}
\includegraphics[height=0.42\textwidth]{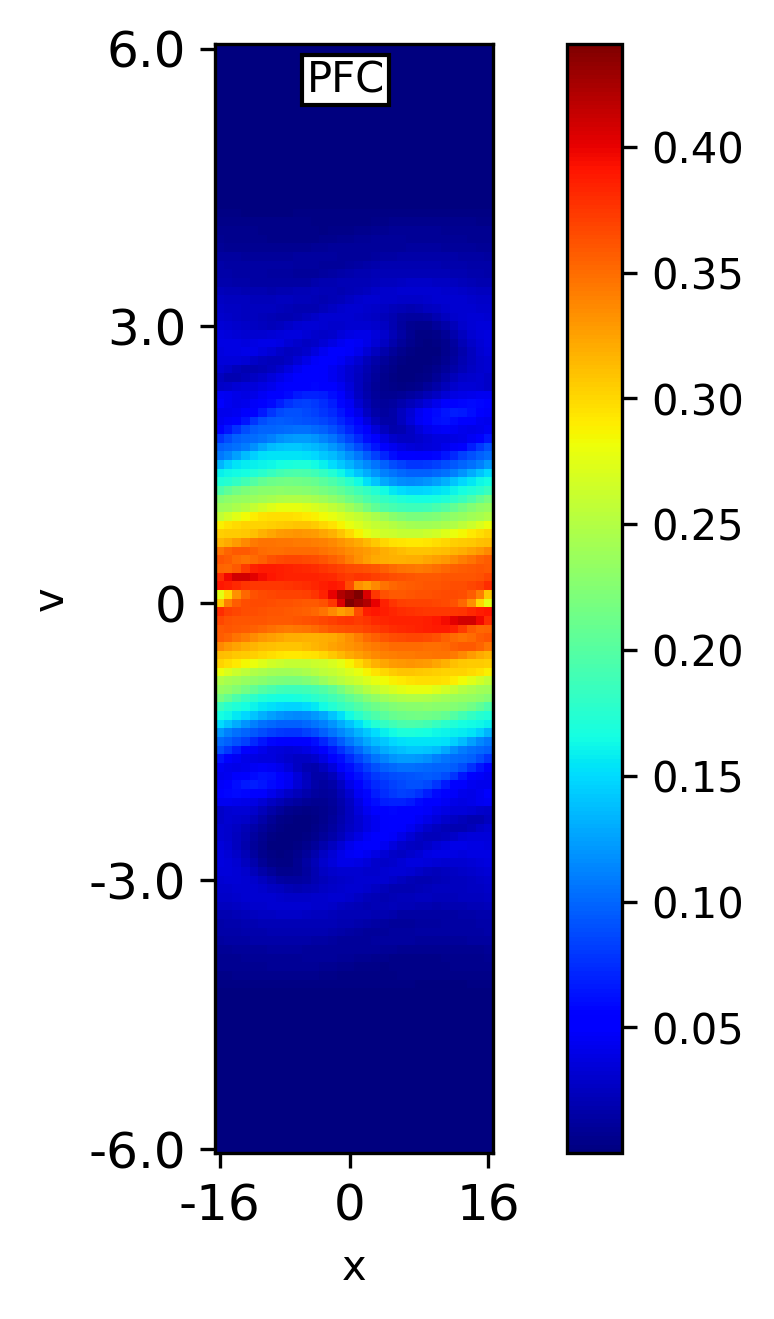}
}
\subfigure[$f(x,v,t=0)$ for FCV]{\label{fig:fxv-fcv-0}
\includegraphics[height=0.42\textwidth]{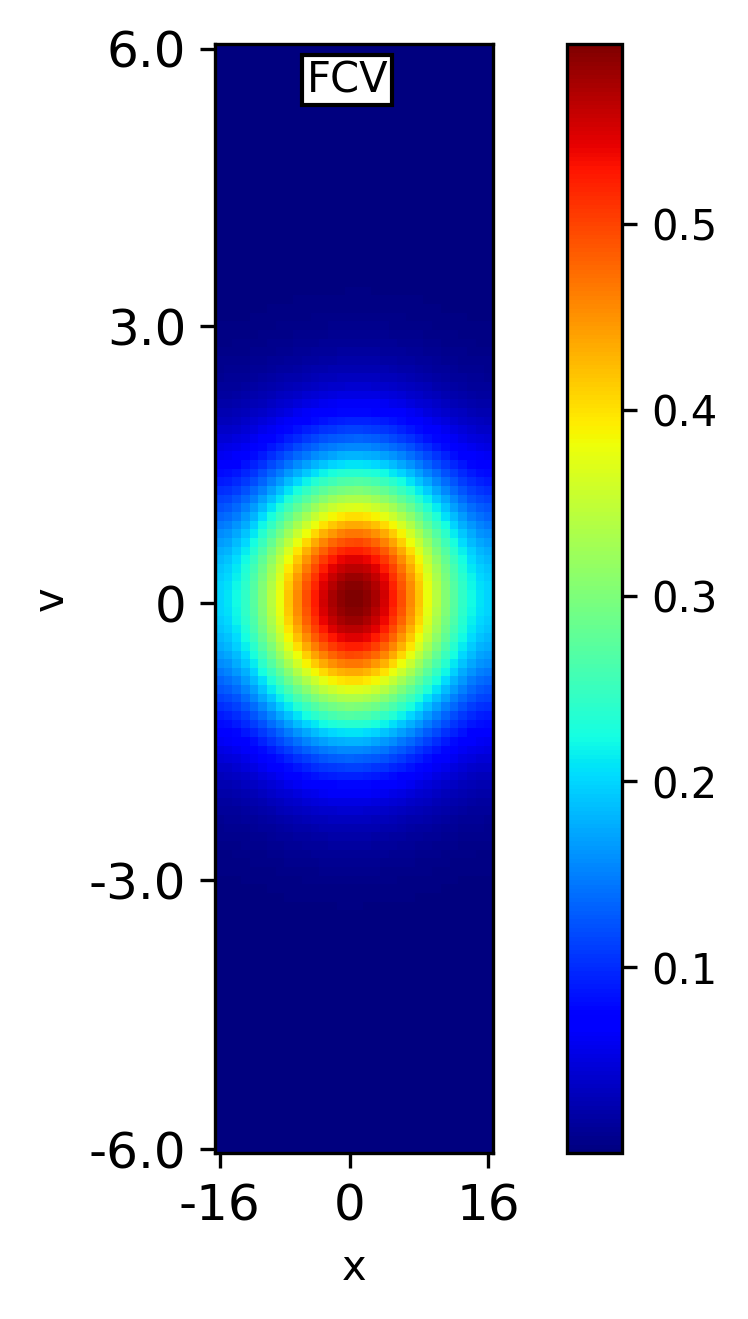}
}
\subfigure[$f(x,v,t=20)$ for FCV]{\label{fig:fxv-fcv-20}
\includegraphics[height=0.42\textwidth]{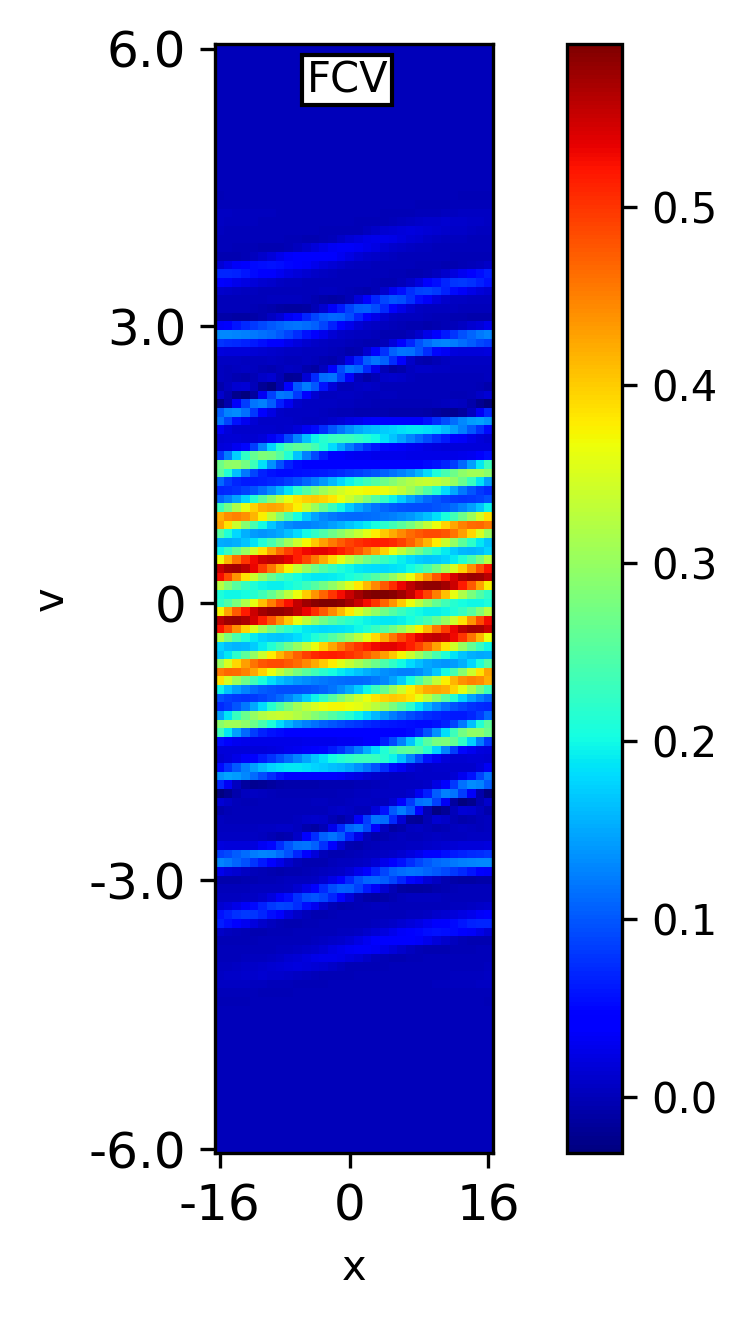}
}
\subfigure[$f(x,v,t=40)$ for FCV]{\label{fig:fxv-fcv-40}
\includegraphics[height=0.42\textwidth]{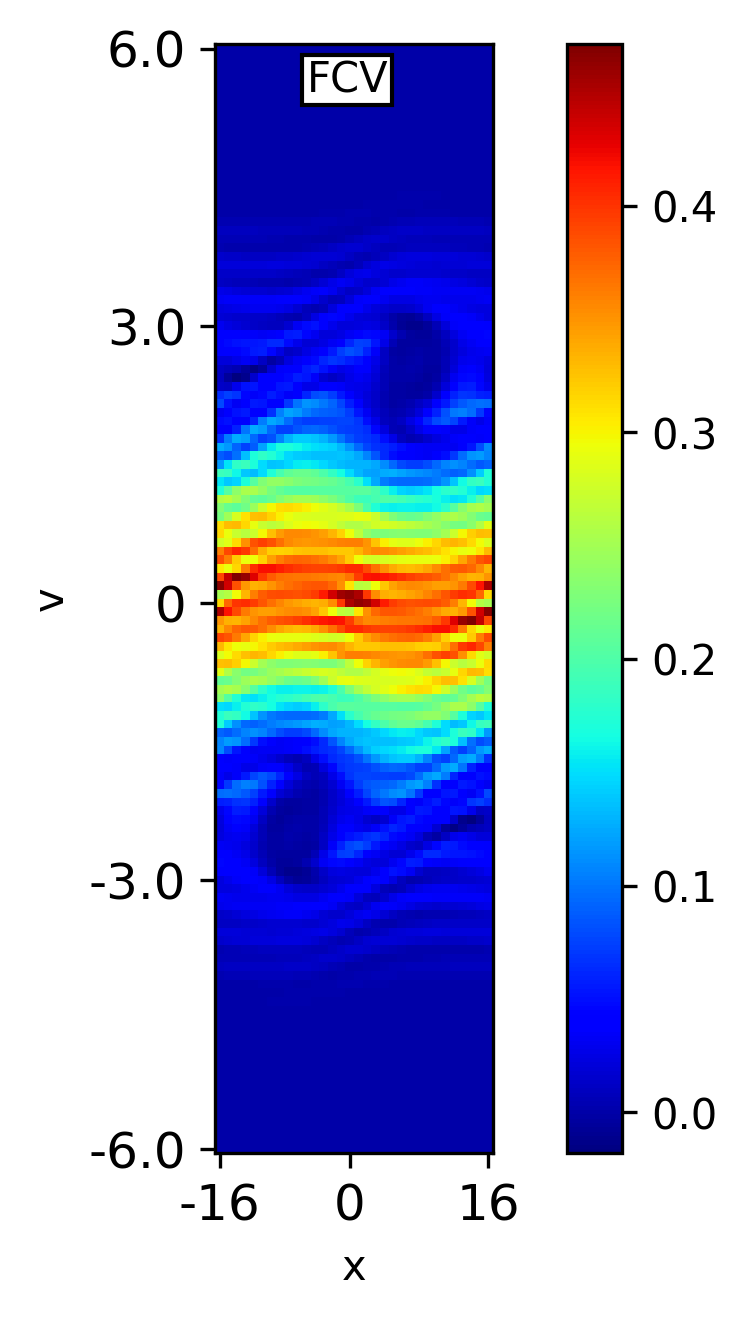}
}
\subfigure[$f(x,v,t=0)$ for FCV-Umeda]{\label{fig:fxv-umeda-0}
\includegraphics[height=0.42\textwidth]{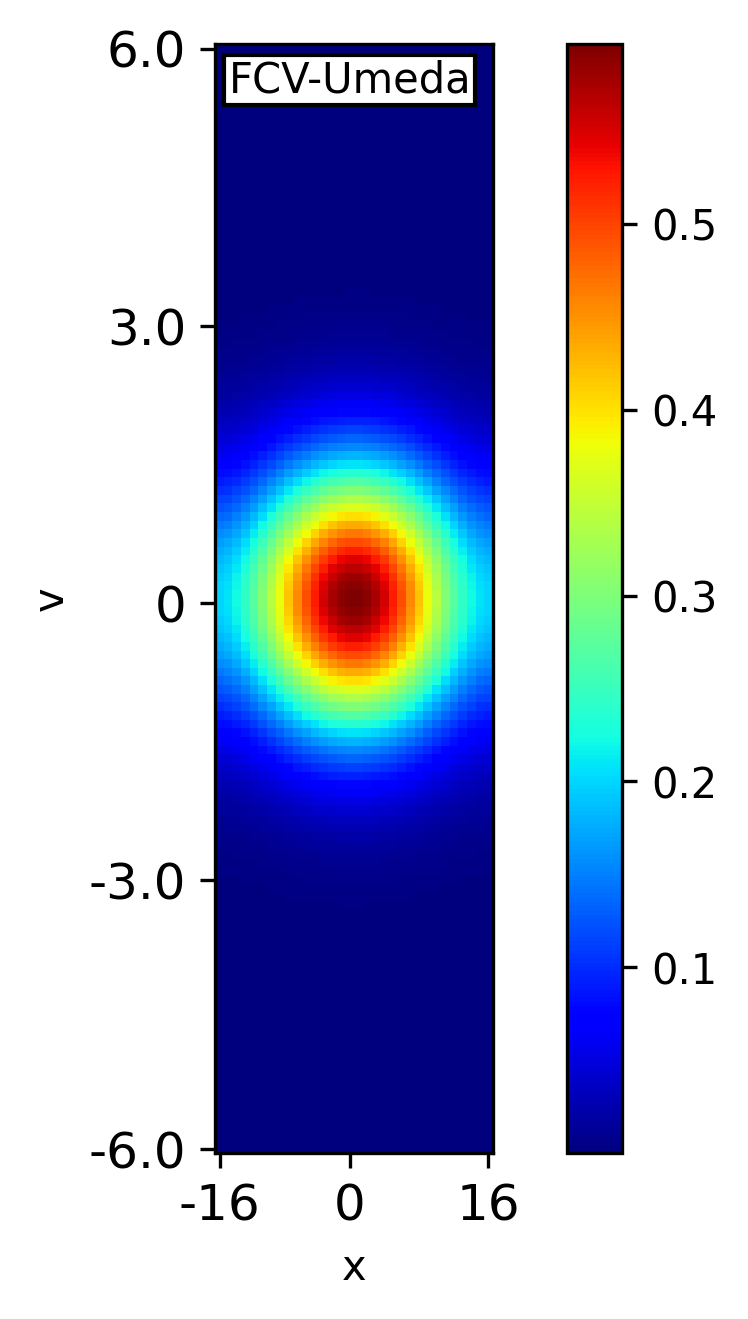}
}
\subfigure[$f(x,v,t=20)$ for FCV-Umeda]{\label{fig:fxv-umeda-20}
\includegraphics[height=0.42\textwidth]{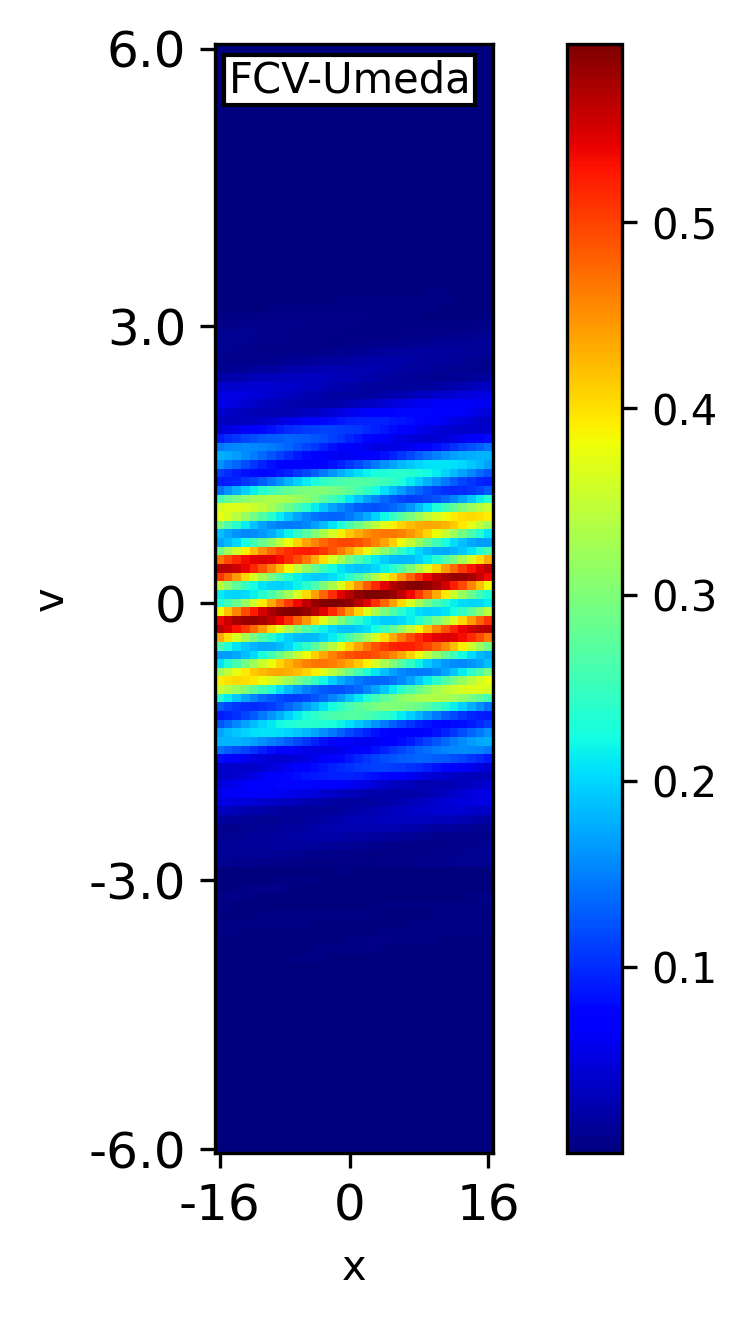}
}
\subfigure[$f(x,v,t=40)$ for FCV-Umeda]{\label{fig:fxv-umeda-40}
\includegraphics[height=0.42\textwidth]{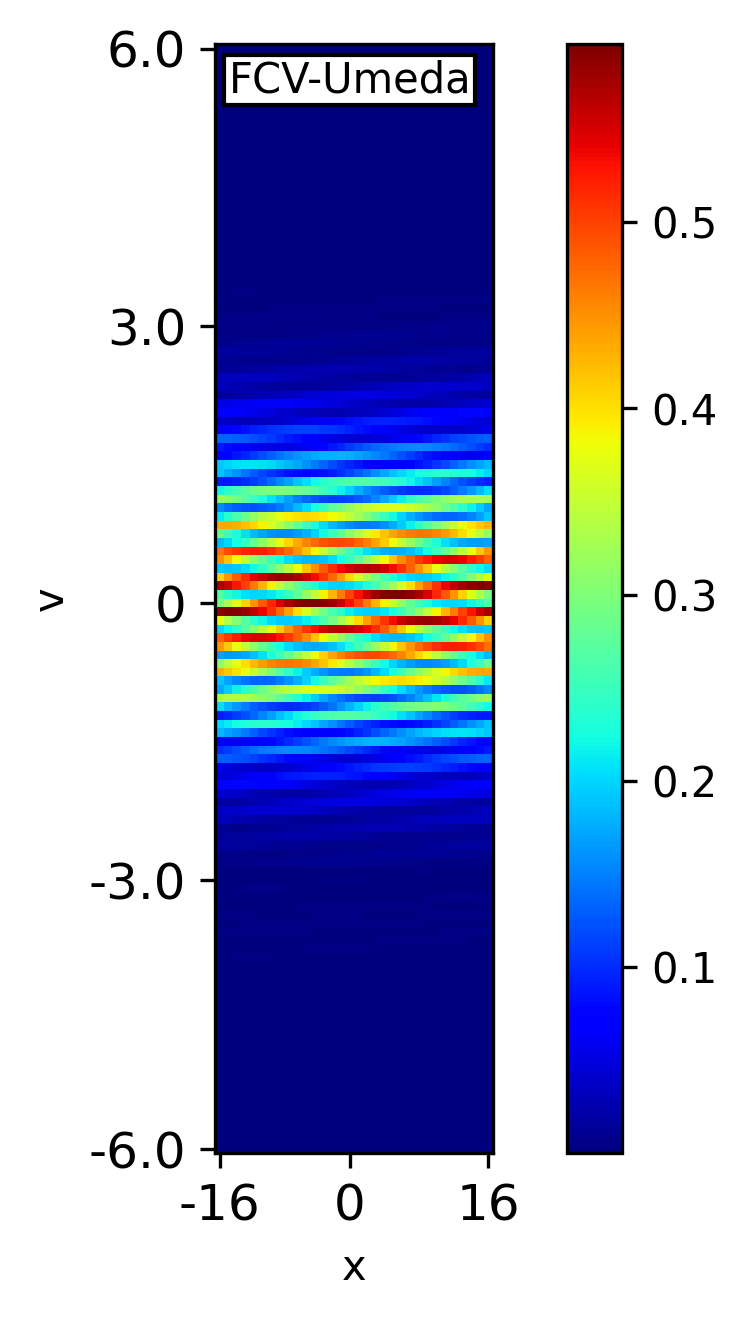}
}

\end{figure}

\section{Numerical experiments (verification results)}\label{num-verification}
Beyond Landau damping, we performed additional simulations to verify that the hybrid gyrokinetic model and the ssV code can capture the physics of specific plasma waves. In this section, we discuss tests for (i) ion-acoustic waves, (ii) ion Bernstein waves, and (iii) kinetic Alfven waves, including how the code handles the Ampere cancellation issue in the electromagnetic case.
\subsection{Ion-acoustic wave dynamics}  
Ion-acoustic waves are longitudinal plasma waves sustained by electron pressure and ion inertia, with phase velocities between the ion and electron thermal speeds~\cite{Kunz2014,Valentini2007,Munoz2018}. The Vlasov-Poisson system governs the dynamics:

Ion dynamics: Solved via the SLMP5 (semi-Lagrangian monotonicity preservation 5th order) scheme in $x$-$v$ phase space, ensuring conservation of ion distribution function $f_i(t, x, v)$.

Electron dynamics: Drift electrons were modeled using a semi-Lagrangian fifth-order scheme.

The dispersion relation for ion-acoustic waves in the long-wavelength limit~\cite{Lopes2022} is:

\begin{equation}
\omega^2 = k^2 C_s^2 \left(1 + \frac{3T_i}{T_e}\right),
\end{equation}

where $C_s = \sqrt{T_e/m_i}$ is the ion-sound speed, $T_i$ and $T_e$ are the ion and electron temperatures respectively, and $k$ is the wave number. The finite ion temperature corrections ($3T_i/T_e$) became significant when $T_i/T_e \geq 0.1$. For cold ions ($T_i/T_e \ll 1$), the relation is simplified to $\omega = k C_s$.

Collisionless damping arises from resonant energy transfer between the waves and ions. The damping rate $\gamma$ is derived from the imaginary part of the dielectric function as follows:

\begin{equation}
\gamma = -\frac{\text{Im}(\epsilon)}{\partial \text{Re}(\epsilon)/\partial \omega},
\end{equation}

where $\epsilon$ is the plasma dielectric function.

The numerical verification setup consists of the domain 1D physical space ($x \in [0, L]$, $L = 4\pi$) and 1D velocity space ($v \in [-v_{\text{max}}, v_{\text{max}}],v_{\text{max}}=4.5$), with periodic $x$-boundary conditions.

Initial Conditions: Perturbed Maxwellian ion distribution with small-amplitude cosine modes:

\begin{equation}
f_i(0, x, v) = \frac{1}{\sqrt{2\pi }} e^{-v^2/2} \left(1 + \alpha \sum_n \cos(k_n x)\right),
\end{equation}

where $v_{t,i} = \sqrt{2{T_i/m_i}}$ and $\alpha \ll 1$.

Parameters: $N_x = 64$, $N_v = 128$, $\Delta t = 0.01$, $T_e \gg T_i$ (weak damping) to $T_e \sim T_i$ (strong damping).

\begin{figure}[h]
\centering
\subfigure[Electric energy evolution in log scale for different $\tau = 0.1,0.2,0.33,0.6 $]{
\includegraphics[width=0.47\textwidth]{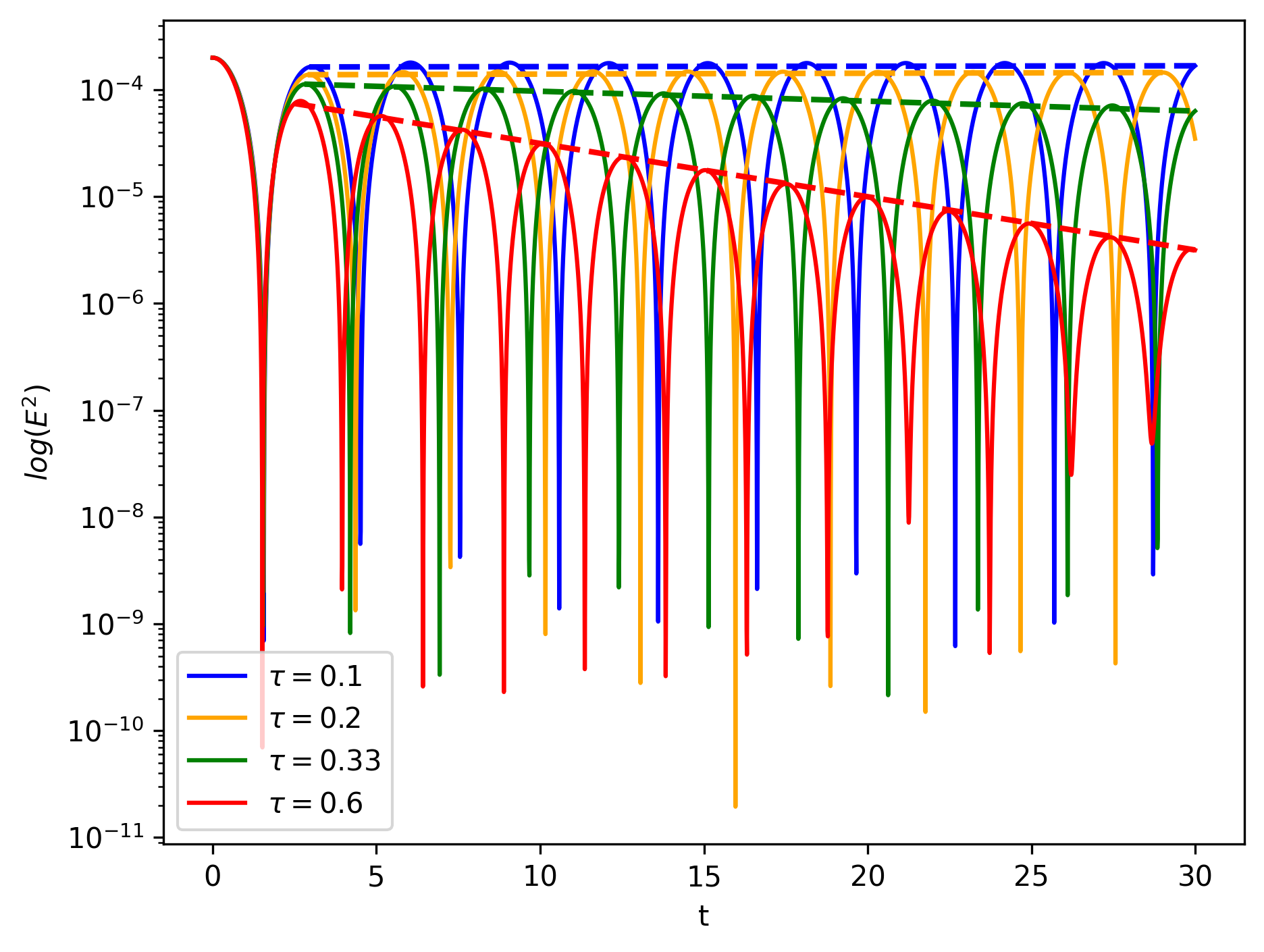}
}
\subfigure[Damping rate verification for different $\tau = 0.1,0.2,0.33,0.6 $ ]{
\includegraphics[width=0.48\textwidth]{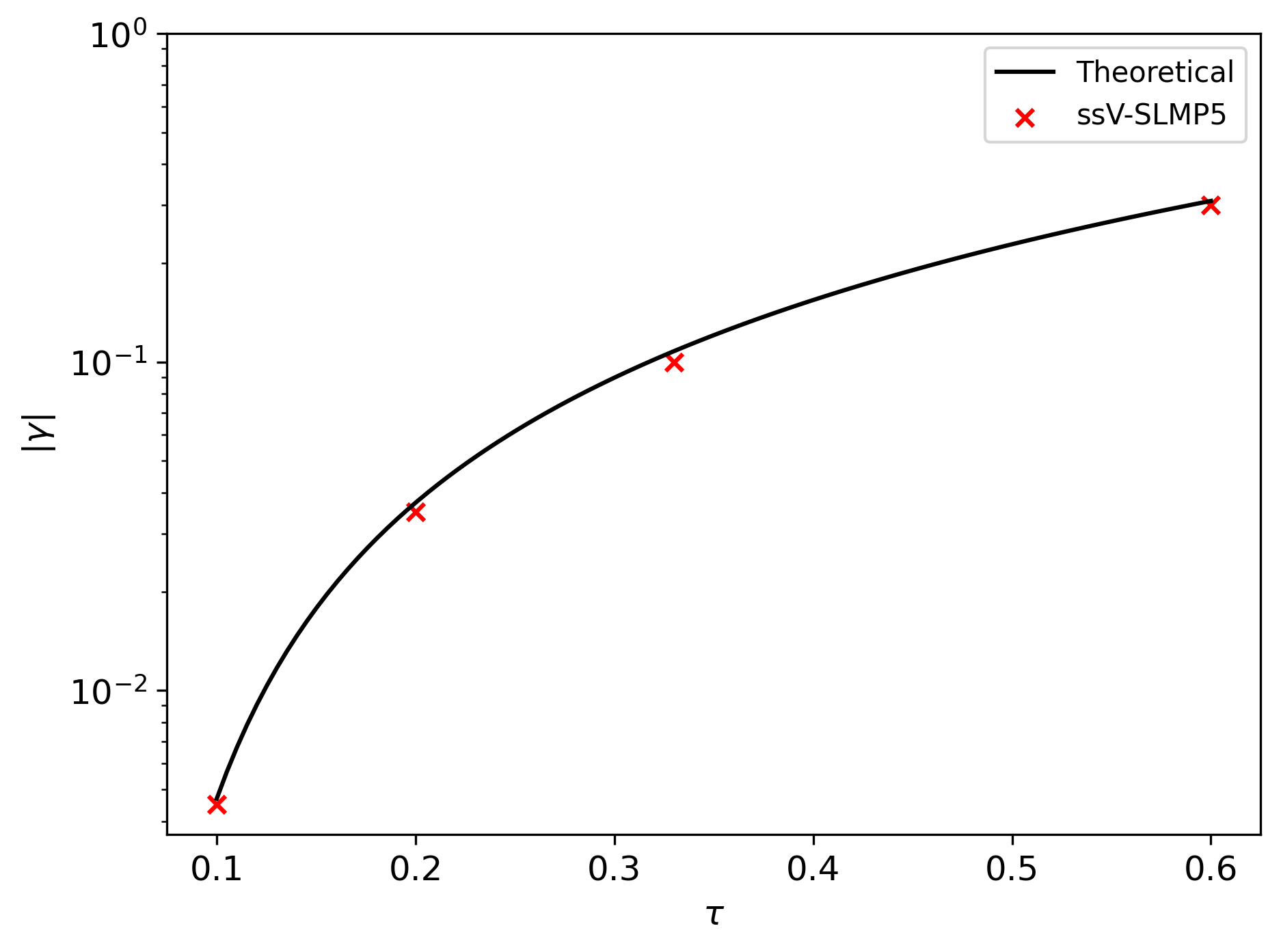}
}
\caption{Numerical verification of ion-acoustic wave dispersion and damping using SLMP5. (a) Electric energy evolution in log scale for different $\tau = 0.1,0.2,0.33,0.6 $, (b) Damping rate comparison with theoretical predictions.}
\label{fig:comparison_plots_ion}
\end{figure}

The SLMP5 scheme verifies the ion Landau damping rates and dispersion relations across $\tau =T_i/T_e $ regimes. The numerical results were consistent with the analytical predictions, as shown in Figure~\ref{fig:comparison_plots_ion}. In particular, in the cold ion limit, represented by $\tau =T_i/T_e = 0.1$, the numerical damping rate $\gamma_{\text{num}} = 0.004$ became negligible and remained within 1$\%$ of the theoretical value.

The semi-Lagrangian full kinetic ions and drift kinetic electron model efficiently capture quasineutrality while resolving important electron kinetic effects at a low computational cost.
The hybrid gyrokinetic semi-Lagrangian approach robustly verifies the ion-acoustic wave physics, bridging kinetic ion effects, and drift kinetic electron approximations. The results validated the capability of ssV code to simulate temperature-dependent damping and dispersion, which is critical for applications in plasma turbulence and wave-driven transport.  
\subsection{Ion Bernstein wave dynamics}  
Ion Bernstein waves (IBWs) are high-frequency electrostatic waves in magnetized plasmas that occur near the harmonics of ion Larmor frequency $\omega_{ci}$~\cite{Podesta2012,Schild2024}. The IBW's general dispersion relation $\omega(k)$, derived under the hybrid gyrokinetic formulation~\cite{Lopes2022}, defines the wave’s frequency as a function of wavenumber $k$. For $k_\parallel = 0$ (perpendicular propagation), the analytical dispersion relation is expressed via modified Bessel functions $I_n$ and plasma function Z:

\begin{equation}
k_\perp^2 + 2 \left( \frac{\omega_{pi}}{v_{ti}} \right)^2 e^{-\frac{1}{2} k_\perp^2 \rho_i^2} \sum_{n=-\infty}^{\infty} I_n\left( \frac{1}{2} k_\perp^2 \rho_i^2 \right) \left[ 1 + \zeta_0 Z(\zeta_n) \right] = 0
\end{equation}

where $\omega_{pi}$ is the ion plasma frequency and $v_{t,i}$ is the ion thermal speed. The SLMP5 scheme solves the ion Vlasov equation in the 4D phase space ($x, y, z, v$) with the following parameters~\cite{Schild2024}:
grid: $128 \times 82 \times 32 \times 162$ nodes in ($x, y, z, v$) with $x \in [0, 20]$, $y,z \in [0, 2]$ and 1D velocity space ($v \in [-v_{\text{max}}, v_{\text{max}}],v_{\text{max}}=4.5$).
Initialization with white-noise perturbations in the configuration space superimposed on a Maxwellian velocity distribution.

\begin{equation}
f_i(0, \mathbf{x}, \mathbf{v}) = \frac{1}{(2\pi)^{3/2}} e^{- |\mathbf{v}|^2/2 } \left(1 + \alpha \sum_m \cos(k_m x)\right),
\end{equation}

where $\alpha \ll 1$ ensures linearity.
\begin{figure}[h]
\centering
\includegraphics[width=0.8\textwidth]{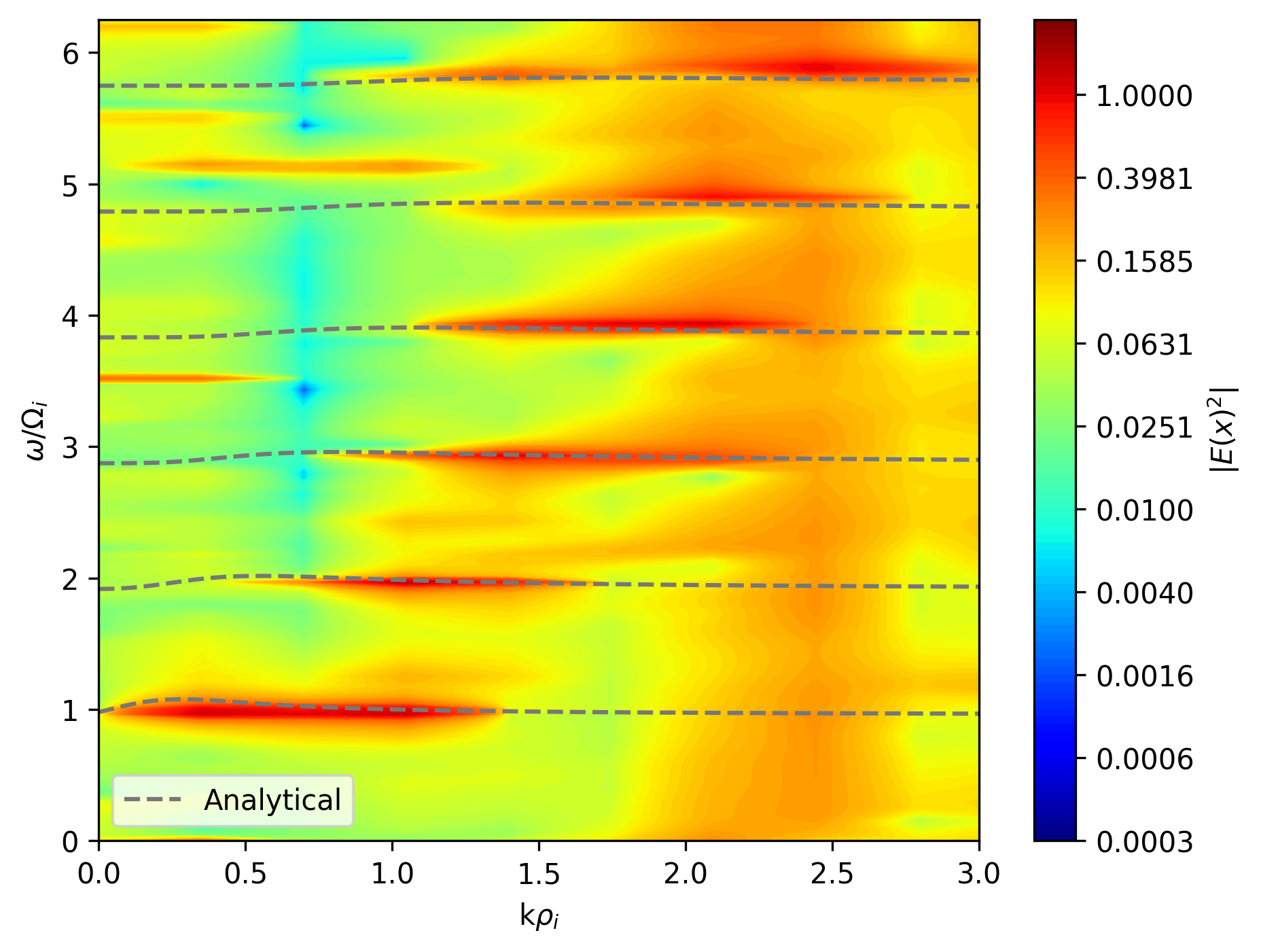}
\caption{Frequency wave number spectrum of Electrostatic potential. Dashed lines: analytical dispersion relation of IBW and the dashed lines show the theoretical Ion Bernstein waves solving the dispersion relation.}
\label{fig:ibw_spectrum}
\end{figure}
 The electrons were modeled as adiabatic species, which simplifies the electrostatic potential calculation. Time step: $\Delta t = 0.025$ resolves $\omega_{ci}$ and higher harmonics.
 
IBWs are found close to the frequencies $\omega \approx n\omega_{ci}$ ($n = 1, 2, ...$), with slight deviations from the Larmor harmonics showing up for finite $k_\perp$ in both the analytical results and the numerical simulations displayed in Figure~\ref{fig:ibw_spectrum}. At large $k_\perp$ the wave frequencies converge back to the exact harmonics, validating the code’s ability to resolve finite Larmor radius effects.
Despite the absence of a positivity limiter, the SLMP5 scheme does not exhibit unphysically negative values in $f_i$, ensuring stability in high-$\omega_{ci}$ regimes and resolving multiple IBW modes. 
The numerical results for $\omega(k)$ are consistent with the analytical predictions, as shown in Figure~\ref{fig:ibw_spectrum}. The SLMP5 scheme verifies the IBW dispersion relations across $k_\perp \rho_i$ regimes, demonstrating fidelity to analytical dispersion relations and efficient handling of high-frequency, magnetized plasma dynamics. This capability positions the method as a reliable tool for simulating instabilities, wave-driven transport, and magnetic reconnection in tokamak edge plasmas and astrophysical plasmas.
\subsection{Kinetic Alfven waves and the Ampere cancellation issue}

Kinetic Alfven waves (KAWs) are an essential component of low-beta plasma dynamics, in which the ion Larmor radius and electron kinetic effects modify the standard Alfven wave behavior. Unlike their MHD counterparts, KAWs account for separate ion/electron dynamics and finite parallel electric fields, which play a key role in electron Landau damping~\cite{Mandell2020,Pezzi2019,Dannert2004}. The correct numerical representation of these waves requires a proper balance between the current and charge density fluctuations to avoid artificial parallel electric fields, a problem commonly referred to as the Ampere cancellation issue. The effectiveness of ssV in addressing the Ampere cancellation problem was examined through a numerical test case simulating kinetic Alfven waves within the MHD limit. Here, we compare ssV with the gyrokinetic KAW dispersion relation shown in Ref.~\cite{Mandell2020}:

\begin{equation}
\omega^2 \left[ 1 + \omega \frac{\sqrt{2} k_\parallel v_{te} Z(\omega / (\sqrt{2} k_\parallel v_{te}))}{k_\parallel v_A} \right] = k_\parallel^2 v_{te}^2 \hat{\beta} \left[ 1 + k_\perp^2 \rho_s^2 + \omega \frac{\sqrt{2} k_\parallel v_{te} Z(\omega / (\sqrt{2} k_\parallel v_{te}))}{k_\parallel v_A} \right]
\label{eq:kawdisp}
\end{equation}

where $\omega$ is the frequency, $k_\parallel$ the parallel wave number, $v_{te}$ is the electron thermal speed, $\hat{\beta} = (\beta_e/2) m_i/m_e$, $\rho_s$ is the ion sound gyroradius, and $Z(x)$ the plasma dispersion function. Errors in the numerical calculations of integrals introduce corrections to Ampere’s law, resulting in a modified dispersion relation as given below~\cite{Mandell2020}.
\begin{equation}
\omega^2 = \frac{k_\parallel^2 v_A^2}{C_N + \frac{k_\perp^2 \rho_s^2}{\hat{\beta}}} \left[ 1 + \frac{(C_N - C_J) \hat{\beta}}{k_\perp^2 \rho_s^2} \right]
\end{equation}
As discussed by Mandell et al.~\cite{Mandell2020}, even in a linearized system, inconsistencies in the numerical evaluation or initialization of the charge and current densities can lead to non-canceling residual terms in the field equations. These residuals are represented by coefficients \( C_N \) and \( C_J \), which correspond to the charge and current moment responses, respectively. For the parallel electric field \( E_{\parallel} \) to remain zero at \( t = 0 \) and ensure correct linear wave dynamics, it is necessary that both \( C_N =  C_J = 1 \). This condition guarantees that the charge and current moments extracted from the distribution function are self-consistent and preserve the Ampere cancellation. 

For the following comparison, ssV evolves ions into the fully kinetic Vlasov equation, whereas electrons follow the drift-kinetic approximation. The coupled Poisson and Ampere equations are given by Equations ~\ref{eq:poissoneq} and ~\ref{eq:Ampereeq} respectively.

\begin{figure}[H]
    \centering
    \subfigure[Numerical dispersion relation \( \omega \) compared to theoretical KAW dispersion.]{
    \includegraphics[width=0.47\textwidth]{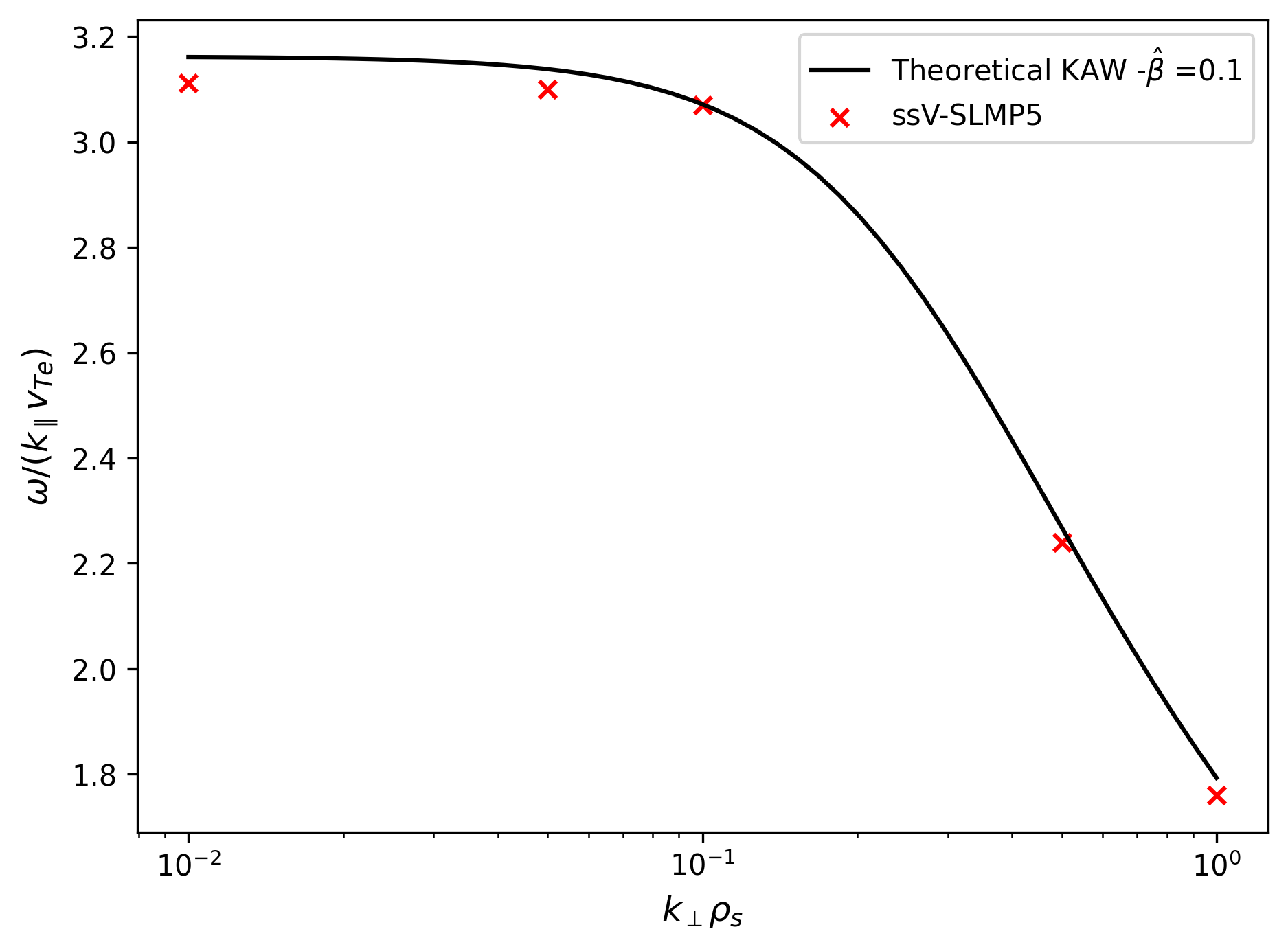}
    }
    \subfigure[Damping rate \( \gamma \) extracted from simulations compared to kinetic theory.]{
    \includegraphics[width=0.47\textwidth]{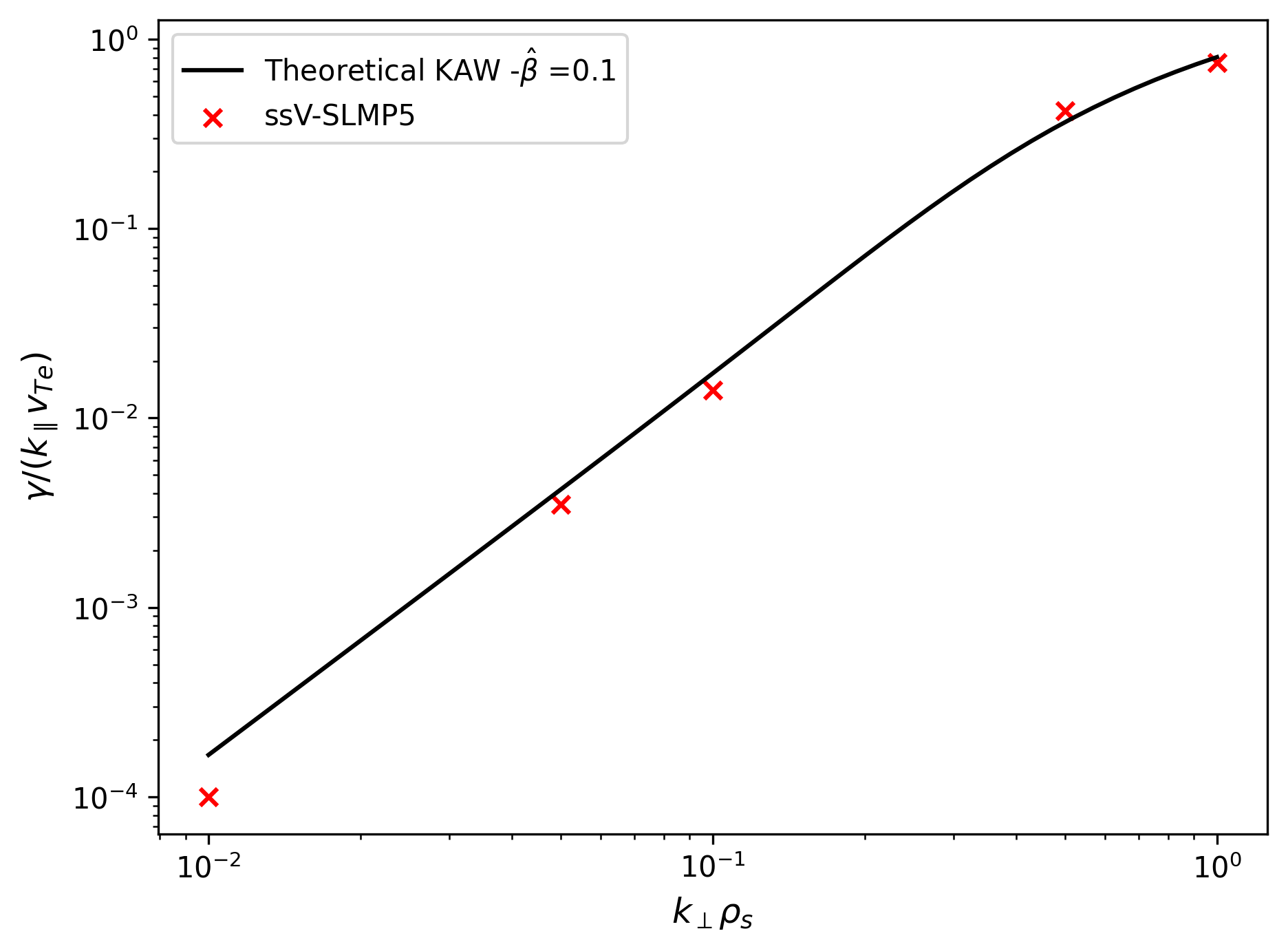}
    }
    \caption{(a) Numerical dispersion relation \( \omega \) (red color x marks) compared to theoretical KAW dispersion (solid black line). (b) Damping rate \( \gamma \) (red color x marks) extracted from simulations compared to the imaginary part of theoretical KAW dispersion (solid black line).}
    \label{fig:kaw_dispersion_damping}
\end{figure}

\begin{figure}[H]
    \centering
    \subfigure[ Magnetic energy evolution in log scale for $k_\perp \rho_s=0.01$, $\hat\beta=10$ ]{
    \includegraphics[width=0.48\textwidth]{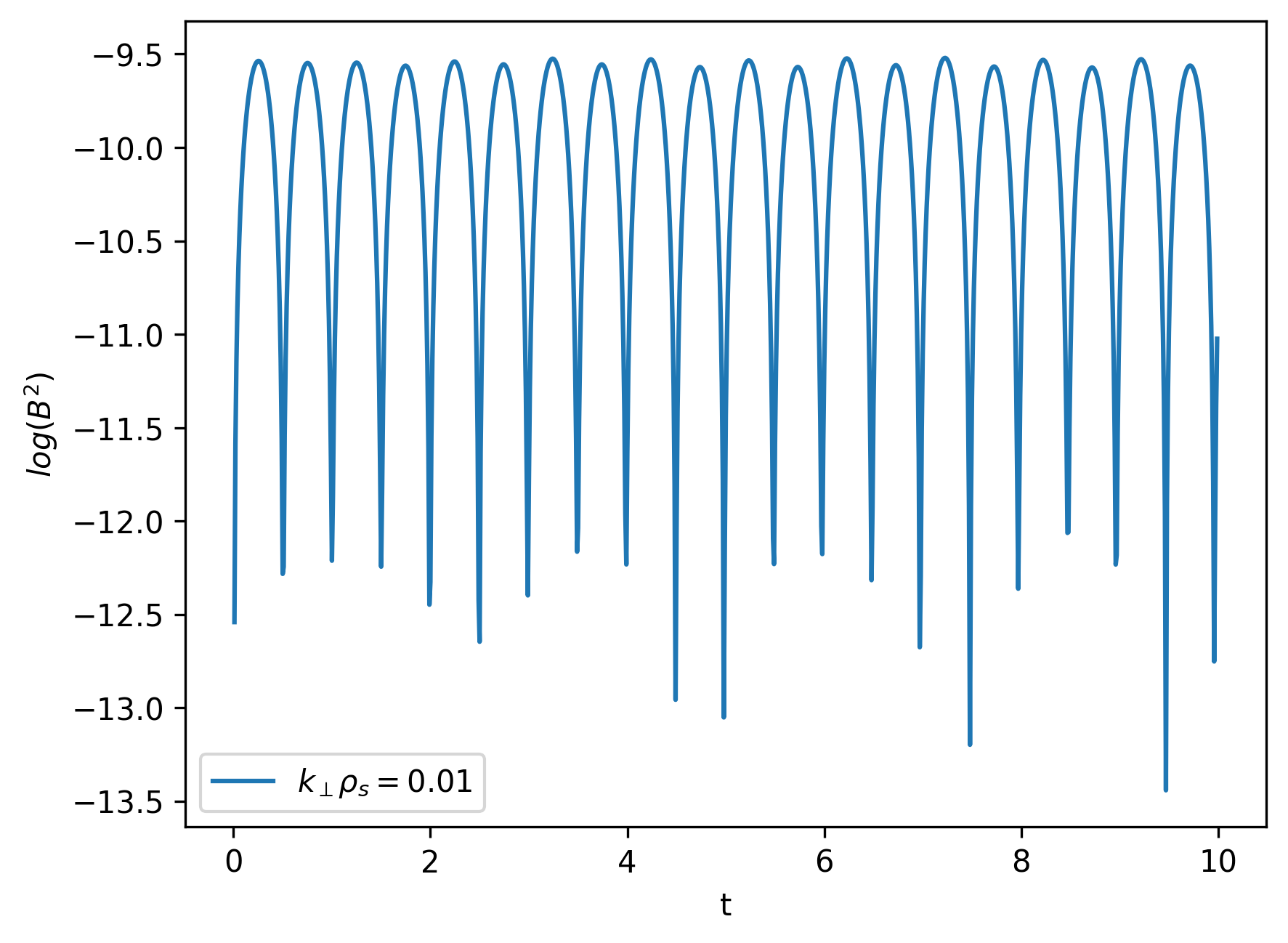}
    }
    \caption{ Numerical results from the EM branch of ssV align closely with analytical predictions for $\hat{\beta} / k_\perp^2 \rho_s^2 \gg 1$. Magnetic energy evolution in log scale shows no damping ($\gamma \approx 0$) 
    }
    \label{fig:amp_cancel}
\end{figure}

The charge and current densities in both equations are computed by taking the velocity space integrals (moments) of the distribution functions. Specifically, the normalized charge density is given by
\begin{equation}
\hat{\rho} = \sum_s \hat{q}_s \int \hat{f}_s \, d\hat{v},
\end{equation}
and the normalized parallel current density is computed as
\begin{equation}
\hat{J}_{\parallel} = \sum_s \hat{q}_s \int \hat{v}_{\parallel} \hat{f}_s \, d\hat{v},
\end{equation}
where \( \hat{f}_s \) is the normalized distribution function for species \( s \), and \( \hat{v} \) denotes the normalized velocity variable.

Figure~\ref{fig:kaw_dispersion_damping} shows the numerically obtained dispersion relation and damping rates compared to the theoretical prediction, which shows good agreement with the kinetic theory at moderate \( k_{\perp} \). At a larger \( k_{\perp} \), deviations may be expected once frequencies close to the cyclotron frequency are encountered, whose physics are not included in Eq.~\ref{eq:kawdisp}.

Numerical results from the EM branch of the ssV code align closely with analytical predictions for $\hat{\beta} / k_\perp^2 \rho_s^2 \gg 1$. A reduced dimensionality setup with one position and one velocity dimension was used, employing $(N_z, N_{v_\parallel}) = (32, 64)$ cells and periodic boundary conditions in $z$.In Figure~\ref{fig:amp_cancel}, we further examine the case with $k_\perp^2 \rho_s^2 =0.01$, $\hat{\beta}=10$, and $\hat{\beta} / k_\perp^2 \rho_s^2 = 1 e^5 $. The figure shows the magnetic energy evolution in log scale with no visible damping ($\gamma \approx 0$), which is consistent with MHD expectations~\cite{Pezzi2019,Dannert2004}.
These results demonstrate that ssV correctly captures the KAW dynamics, including dispersion, damping, and Ampere cancellation conditions. The omission of \( A_{\perp} \) and \( B_{\parallel} \) in our model may introduce deviations at high \( k_{\perp} \rho_i \) as well as high $\beta\gtrsim1$, but the essential KAW physics is retained for the parameter range of interest.
\section{Conclusion}\label{conclusion}
In this work, we presented a hybrid gyrokinetic model implemented in the ssV code, designed to simulate collisionless plasmas with fully kinetic ions and drift kinetic electrons. This framework bridges the gap between computational feasibility and physical fidelity, enabling the study of multi-scale plasma processes such as turbulence, wave-particle interactions, and reconnection. The combination of high-order semi-Lagrangian schemes and hybrid gyrokinetic modeling provides a flexible and accurate tool for exploring space and laboratory plasma phenomena.

The development and testing of the super simple Vlasov (ssV) code shows that it can handle complex plasma dynamics with both accuracy and efficiency. Using numerical schemes such as PFC, FCV, FCV-Umeda, and SLMP5, the code tackles common issues in plasma simulations, such as oscillations and numerical diffusion. The PFC scheme works well for step-like profiles, keeping the solution stable and positive, though it tends to introduce diffusion in the smooth regions. The FCV scheme improves the accuracy for sinusoidal cases but can cause oscillations near sharp gradients unless paired with limiters, such as those introduced by Umeda et al. Among all the schemes, SLMP5 stands out as the most balanced, delivering high accuracy while maintaining computational costs and monotonicity across a range of test cases.

Simulations confirm that ssV can reproduce key kinetic effects such as Landau damping, ion-acoustic waves, and ion Bernstein waves, with results that closely match the analytical predictions. The code also captures the kinetic Alfven wave behavior in the electromagnetic case by properly resolving the Ampere cancellation issue. 

The combination of semi-Lagrangian methods, high-order interpolation, and advanced limiters enables accurate and stable simulations without requiring fine grid resolution, making the ssV code well-suited for computationally efficient studies of kinetic plasma phenomena. The ssV code thus allows us to investigate a broad range of kinetic plasma phenomena that are challenging for other reduced-order models to capture efficiently. Future work will focus on applying this model to explore instabilities, magnetic reconnection, and turbulence at sub-ion scales across space and laboratory plasmas.
\section{Data availability statement}
The raw data take effort to interpret without expert knowledge. The data that support the findings of this study are available upon reasonable request from the authors.
\section{Acknowledgments}
The authors gratefully acknowledge support from the Helmholtz Young Investigator Group grant VH-NG-1239.
Computational resources were provided by the MPCDF Center, Garching. We also extend our gratitude to the Theoretical Physics Department at Ruhr University Bochum for their collaboration and for providing the base version of the MUPHY I code, which served as the foundation for ssV development.

\bibliographystyle{elsarticle-num} 
\bibliography{references}



\end{document}